\documentclass{aa}  

\usepackage{graphicx}
\usepackage{subfig}
\usepackage[varg]{txfonts}
\usepackage{color}

\usepackage{natbib}
\bibpunct{(}{)}{;}{a}{}{,} 

\begin{document}

   \title{Pan-STARRS1 variability of XMM-COSMOS AGN }

   \subtitle{I. Impact on photometric redshifts}

   \author{T. Simm\inst{1}
          \and R. Saglia\inst{1,2}
          \and M. Salvato\inst{1}
          \and R. Bender\inst{1,2}
          \and W. S. Burgett\inst{5}
          \and K. C. Chambers\inst{3} 
          \and P. W. Draper\inst{4}
          \and H. Flewelling\inst{3}
          \and N. Kaiser\inst{3}
          \and R.-P. Kudritzki\inst{3,2}
          \and E. A. Magnier\inst{3}
          \and N. Metcalfe\inst{4}
          \and J. L. Tonry\inst{3}
          \and R. J. Wainscoat\inst{3}
          \and C. Waters\inst{3}
          }

   \institute{Max Planck Institute for Extraterrestrial Physics, Giessenbachstrasse, Postfach 1312, 85741 Garching, Germany\\
   \email{tsimm@mpe.mpg.de} 
    \and University Observatory Munich, Ludwig-Maximilians Universitaet, Scheinerstrasse 1, 81679 Munich, Germany
        \and Institute for Astronomy, University of Hawaii at Manoa, Honolulu, HI 96822, USA
        \and Department of Physics, Durham University, South Road, Durham DH1 3LE, UK
        \and GMTO Corporation, 251 S. Lake Ave., Suite 300, Pasadena, CA 91101, USA
             }


 
  \abstract
   {}
   {Upcoming large area sky surveys like EUCLID and eROSITA, which are dedicated to studying the role of dark energy in the expansion history of the Universe and the three-dimensional mass distribution of matter, crucially depend on accurate photometric redshifts. The identification of variable sources, such as AGNs, and the achievable redshift accuracy for varying objects are important in view of the science goals of the EUCLID and eROSITA missions.}
   {We probe AGN optical variability for a large sample of X-ray-selected AGNs in the XMM-COSMOS field, using the multi-epoch light curves provided by the Pan-STARRS1 (PS1) 3$\pi$ and Medium Deep Field surveys. To quantify variability we employed a simple statistic to estimate the probability of variability and the normalized excess variance to measure the variability amplitude. Utilizing these two variability parameters, we defined a sample of varying AGNs for every PS1 band. We investigated the influence of variability on the calculation of photometric redshifts by applying three different input photometry sets for our fitting procedure. For each of the five PS1 bands $g_{\mathrm{P1}}$, $r_{\mathrm{P1}}$, $i_{\mathrm{P1}}$, $z_{\mathrm{P1}}$, and $y_{\mathrm{P1}}$, we chose either the epochs minimizing the interval in observing time, the median magnitude values, or randomly drawn light curve points to compute the redshift. In addition, we derived photometric redshifts using PS1 photometry extended by GALEX/IRAC bands.}
   {We find that the photometry produced by the 3$\pi$ survey is sufficient to reliably detect variable sources provided that the fractional variability amplitude is at least $\sim$3\%. Considering the photometric redshifts of variable AGNs, we observe that minimizing the time spacing of the chosen points yields superior photometric redshifts in terms of the percentage of outliers (33\%) and accuracy (0.07), outperforming the other two approaches. Drawing random points from the light curve gives rise to typically 57\% of outliers and an accuracy of $\sim$0.4. Adding GALEX/IRAC bands for the redshift determination weakens the influence of variability. Although the redshift quality generally improves when adding these bands, we still obtain not less than 26\% of outliers and an accuracy of 0.05 at best, therefore variable sources should receive a flag stating that their photometric redshifts may be low quality.}
   {}

   \keywords{   catalogs --
                                methods: data analysis -- 
                                techniques: photometric --
                                galaxies: active --
                                galaxies: distances and redshifts --
                                X-rays: galaxies
               }

   \maketitle
%

\section{Introduction}

Understanding the expansion history of the Universe is one of the fundamental questions of modern astrophysics. This is particularly true for the nature of dark energy and dark matter, the presumed agents behind cosmic acceleration and cosmological structure formation. Unveiling the dark Universe, which represents 96\% of the cosmic matter-energy content, allows setting major constraints on the past, present, and future evolution of the Universe and promises to provide insight into radically new physics. Significant progress in our understanding is expected to be delivered by current and upcoming surveys, such as the Dark Energy Survey (DES) \citep{2008SPIE.7014E..0ED,2008SPIE.7016E..0LM}, EUCLID \citep{2011arXiv1110.3193L}, and eROSITA \citep{2007SPIE.6686E..17P,2011MSAIS..17..159C}.  
 
The EUCLID mission aims to map the geometry of the dark Universe by accurately gauging distortions of galaxy shapes mediated by weak lensing effects and constraining the pattern of baryonic acoustic oscillations (BAO) from galaxy clustering measurements. Applying these two independent cosmological probes, EUCLID will survey the three-dimensional distribution of structures with unparalleled accuracy out to redshift $z\sim$2, thereby covering the entire period of the accelerated expansion of the Universe that is driven by dark energy. Observing 15\,000 $\mathrm{deg^{2}}$ of the extragalactic sky, EUCLID will probe the growth of cosmic structure in tomographic bins. This will be determined through photometric redshifts (photo-z's) that need to be as accurate as $\sigma_{z}/\left(1+z\right)<0.05$ at $I_{\mathrm{AB}}\leq 24.5$ \citep{2010MNRAS.406..881B,2012MNRAS.421.1671B} and as unbiased as possible. The mission will deliver photo-z's for an unprecedented large number of about two billion galaxies and a million AGNs \citep{2011arXiv1110.3193L,2013LRR....16....6A}. The photometric redshifts will be computed from optical and NIR photometry, with EUCLID providing the NIR Y, J, H bands and optical observations collected from ground-based deep wide area surveys.   

Complementary information about the large scale structure will be provided by the eROSITA all-sky survey, observing the hot X-ray Universe. The mission is expected to detect a very large sample of $\sim 10^{5}$ galaxy clusters, about three million obscured and un-obscured AGNs and $\sim$500\,000 stars \citep{2012arXiv1209.3114M}. This unique data set will allow the co-evolution of supermassive black holes and their host galaxies to be studied within the cosmic structure in unprecedented detail, provided that accurate redshifts can be obtained for the point-like and extended X-ray sources. This requires precise identification of the respective optical counterparts and sufficient multiband photometry for the photo-z computation, again to be supplied by deep wide area surveys. Various sophisticated methods have been developed to derive photo-z's, which either follow an empirical approach by exploring the possible color-redshift combinations of galaxies with the help of a spectroscopic training set \citep{2003AJ....125..580C,2004PASP..116..345C,2004A&A...423..761V,2010ApJ...715..823G,2009MNRAS.397..520W,2010ApJ...712..511C} or by applying template fitting \citep{1998AJ....115.2169G,2000A&A...363..476B,2000ApJ...536..571B,2001defi.conf...96B,2004MNRAS.353..654B,2005MNRAS.359..237P,       2006MNRAS.372..565F,2006A&A...457..841I,2007MNRAS.376.1861F,2007ApJS..172..117M,2008ApJ...686.1503B,    2008ApJ...676..286A,2009MNRAS.396..462K,2009A&A...508.1173P,2011ApJS..193...30B,2010ApJ...724..425D,2013ApJ...775...93D,2012ApJ...746..128S}. Although modern photo-z codes comfortably reach accuracies better than 5\% for inactive galaxies \citep{2004A&A...421...41G,2004A&A...421..913W,2006A&A...449..951G,2006A&A...457..841I,2009ApJ...690.1236I,2010ApJS..189..270C}, photometric redshifts of similar precision for AGNs require much more effort and are available solely for well-studied sky fields with extensive multiband coverage \citep{2009ApJ...690.1250S,2011ApJ...742...61S,2010ApJS..187..560L,2014ApJ...796...60H}. 

The difficulties related to SED fitting of AGNs are mainly driven by the fact that the spectrum is a superposition of the AGN core component and the host galaxy light, plus the strong intrinsic variability of AGNs across wavelength. While the former difficulty can be tackled by collecting high quality empirical templates of a representative subsample of the AGN population, the actual uncertainties introduced by multiband variability are currently not known. The latter may, however, introduce fatal biases into the photo-z accuracy for AGNs, which in turn affect the ability to study the evolution of the X-ray luminosity function and AGN clustering, as well as the star formation and stellar population properties of AGN host galaxies \citep{2010MNRAS.401.2531A,2013ApJ...763...59R}. 

For this reason, it is paramount to reliably detect variable sources in the entire extragalactic sky. That AGNs exhibit strong variability in a wide spectral range, covering radio, UV/optical, X-ray, and $\gamma$-ray wavelengths \citep{1997ARA&A..35..445U} allows identification of AGNs on the basis of their variability properties. The onset of wide-area massive time-domain optical imaging surveys triggered a multitude of AGN variability studies with the aim of characterizing the optical variability and establishing a method for quasar selection \citep{2009ApJ...698..895K,2011ApJ...730...52K,2013ApJ...779..187K,2010ApJ...708..927K,2011ApJS..194...22K,2012ApJ...746...27K,2013ApJ...775...92K,2010ApJ...721.1014M,2011ApJ...728...26M,2012ApJ...753..106M,2010ApJ...714.1194S,       2012ApJ...744..147S,2011A&A...530A.122P,2011AJ....141...93B,2011ApJ...735...68K,2012ApJ...760...51R,2012ApJ...758..104Z,        2013A&A...554A.137A,2013ApJ...765..106Z,2014ApJ...784...92M,2014ApJ...792...54S,2014MNRAS.439..703G,2015A&A...574A.112D,2015A&A...579A.115F,    2014ApJ...783...46K,2015MNRAS.449...94K}. These investigations confirmed the general picture that AGNs show non-periodic, stochastic flux variability occurring on timescales of several months to several years with a fractional amplitude of typically $\sim$10\%--20\%.         

The surveys carried out with the Pan-STARRS1 instrument deliver multi-epoch light curves for $3\pi$ of the sky and ten additional fields observed with higher cadence in five optical and NIR bands, thus providing variability information for millions of AGNs. Motivated by the aforementioned issues we used the light curves of the PS1 $3\pi$ and Medium Deep Field surveys to define a sample of variable AGNs in each PS1 band. The sample is drawn from the well-characterized source list of X-ray-selected AGNs from the XMM-COSMOS survey. Utilizing this sample of variable objects, we study how multiband variability affects the quality of photometric redshifts in detail and assess the achievable redshift accuracy using solely PS1 photometry and PS1 photometry plus GALEX/IRAC bands. In a follow-up paper we will then use the same sample to search for correlations between optical variability and physical parameters, such as black hole mass, luminosity, Eddington ratio, and redshift.                
        
This work is organized as follows. In section \ref{sec:obsdata} we describe the observations and characterize our data set, the sample definition and the detection of variability for our AGNs is depicted in sections \ref{sec:sampledef} and \ref{sec:defvarsample}, the fitting technique and the methods of studying the effects of variability on photo-z calculations are introduced in section \ref{sec:photoz}, the photometric redshift results for variable AGNs are presented in section \ref{sec:results}, and section \ref{sec:conclu} summarizes the results.  Throughout the paper we use AB magnitudes and assume a $\Lambda$CDM cosmology with $H_{0}=70\,\mathrm{km\,s^{-1} Mpc^{-1}}$, $\Omega_{\mathrm{m}}=0.3,$ and $\Omega_{\Lambda}=0.7$. 

\section{Observational data set}
\label{sec:obsdata}
\subsection{The Pan-STARRS1 $3\pi$ and Medium Deep Field surveys}

The observational data used in this work are based on the surveys carried out by the Pan-STARRS1 Science Consortium covering a period of about four years from November 2009 to March 2014. The Pan-STARRS1 instrument is a single wide-field telescope designed for survey mode operation and is located at the Haleakala Observatory on the island of Maui in Hawaii. The f/4.4 optical system, comprising a 1.8 m primary mirror and a 0.9 m secondary, generates a $3.3\degr$ field of view in combination with the PS1 gigapixel camera (GPC1). The 1.4 Gpixel detector is composed of a mosaic of 60 CCD chips each of 4800 x 4800 pixels with one 10 $\mu$m pixel mapping 0.258 arcsec of the sky. The PS1 system performs imaging through five main broadband filters denoted as $g_{\mathrm{P1}}$, $r_{\mathrm{P1}}$, $i_{\mathrm{P1}}$, $z_{\mathrm{P1}}$, $y_{\mathrm{P1}}$ covering optical to near infrared spectral regimes with respective "pivot" wavelengths of 481, 617, 752, 866, and 962 nm and a "wide" filter $w_{\mathrm{P1}}$, used for large depth solar system observations
\citep{2004SPIE.5489..667H,2010SPIE.7733E..12K,2009amos.confE..40T}. The PS1 photometric system is described in \citet{2012ApJ...750...99T}, whereas passband shapes are detailed in \citet{2010ApJS..191..376S}.   

Among the several surveys that PS1 accomplished, two major ones, the 3$\pi$ survey and the Medium Deep Field (MDF) survey, are of primary importance for extragalactic studies. The 3$\pi$ survey observed the three-quarter of the sky north of $-30\degr$ declination in the five main filters officially starting in May 2010 and lasting until March 2014. By completion of the survey mission, each observable field should ideally be imaged 12 times per filter in six different observing nights with typical exposure times of 30--60 s. Based on the requirements of the various science projects, the observations follow a complicated operating schedule dictating that each individual field is visited twice per observing night in a single filter with a temporal gap of 20--30 min. This enables the detection of moving objects like asteroids and near earth objects \citep{2013ApJS..205...20M,2014AAS...22311601C}.  

The MDF survey provides deeper multi-epoch data by repeatedly exposing a set of ten selected fields, with observations of each field distributed throughout the period of the year that allows for 1.3 airmass pointings at least. The scheduled cadence comprises observations in each night periodically running through the five PS1 bands. One cycle starts with 8 $\times$ 113 s in the $g_{\mathrm{P1}}$ and $r_{\mathrm{P1}}$ bands in the first night, followed by 8 $\times$ 240 s in the $i_{\mathrm{P1}}$ band the second night, and finishing with 8 $\times$ 240 s in the $z_{\mathrm{P1}}$ band the third night. Afterwards the next cycle begins by again integrating 8 $\times$ 113 s in the  $g_{\mathrm{P1}}$ and $r_{\mathrm{P1}}$ bands. Additionally for each of the three nights on either side of full Moon, 8 $\times$ 240 s in the $y_{\mathrm{P1}}$ band are obtained \citep{2012ApJ...746..128S,2012ApJ...745...42T}. The large number of exposures taken in the course of the MDF survey deliver very deep stack images and the observing strategy produces light curves permitting extensive variability investigations.  

The raw science frames exposed with the PS1 telescope are reduced by the PS1 Image Processing Pipeline (IPP) conducting standard procedures of image calibration, source detection, astrometry, and photometry. The resulting object catalogues can be accessed via the Published Science Products Subsystem (PSPS) database \citep{2008AIPC.1082..352H}. Amongst the various data products stored in the PSPS database in view of variability studies, the object and detection tables are very important. The object table lists the collected information about all sources identified as an astronomical object in multiple detections, such as sky coordinates, mean and stack magnitudes in all bands, and summary properties obtained from model fits like the PS1 star/galaxy separator. The detection table contains all available information about the individual detections of each object comprising instrumental fluxes, zeropoints, exposure times, and the PSF model fit parameters, to name but a few. Magnitudes in the "AB system" \citep{1983ApJ...266..713O} for each bandpass can be obtained from the instrumental flux $F_{\mathrm{instr}}$ in the considered filter and the respective zeropoint $zp$ stored in the detection table under the terms of $mag_{\mathrm{AB}}=-2.5\log_{10} (F_{\mathrm{instr}})+zp$. The PS1 IPP provides instrumental fluxes computed from PSF model fits suitable for point sources and Kron fluxes \citep{1980ApJS...43..305K}, giving a meaningful flux estimation for extended sources like galaxies. The Kron flux is defined as the flux within the Kron radius, with the latter given by 2.5 times the first radial moment of the flux in the PS1 IPP. From the AB magnitudes calibrated fluxes in units of $3631\,\mathrm{Jy}$ may be obtained according to
\begin{flalign}
        \label{eq:fluxcal}
        F=10^{-0.4 mag_{\mathrm{AB}}}=\frac{\int f_{\nu}\left(h\nu\right)^{-1}A\left(\nu\right)\mathrm{d\nu}}{\int 3631\,\mathrm{Jy}\left(h\nu\right)^{-1}A\left(\nu\right)\mathrm{d\nu}},
\end{flalign}  
with $1\,\mathrm{Jy}=10^{-23}\,\mathrm{erg\,s^{-1}cm^{-2}Hz^{-1}}$.
The right part of Eq. \ref{eq:fluxcal} follows from the definition of the "bandpass AB magnitude", where $\nu$ denotes the photon frequency, $f_{\nu}\,\left(\mathrm{erg\,s^{-1}cm^{-2}Hz^{-1}}\right)$ the flux density, $h$ the Planck constant, and $A\left(\nu\right)$ the capture cross section \citep{1983ApJ...266..713O,2012ApJ...750...99T}. The capture cross section measures the probability of releasing an electron per incoming photon within the detector. In the course of this work, the variability parameters defined in section \ref{sec:varmethod} are computed from the fluxes calculated after equation \ref{eq:fluxcal} and the corresponding flux errors. Throughout this work, we use PS1 data from two processing versions, PV1.2 for the 3$\pi$ survey and PV2 for the MDF survey.    

\subsection{XMM-COSMOS}

The initial sample of objects building the starting point of our studies is a catalogue of 1674 X-ray selected point sources \citep{2010ApJ...716..348B} from the XMM-COSMOS survey, which have been observed in the 0.5--2 keV, 2--10 keV, and 5--10 keV energy bands for a total of $\sim$1.5 Ms in 55 XMM-Newton pointings \citep{2007ApJS..172...29H,2009A&A...497..635C}. The survey reaches a depth of $\sim 5\times 10^{-16}$, $\sim 3\times 10^{-15}$, and $\sim 7\times 10^{-15}\,\mathrm{erg\,s^{-1}cm^{-2}Hz^{-1}}$ in these mentioned bands.

In this work we want to focus on QSOs, so we first limited the sample to those 495 X-ray detected sources that have a secure optical counterpart \citep{2010ApJ...716..348B} and that are classified as pointlike on the basis of the morphological analysis performed by \citet{2007ApJS..172..219L} using deep HST/ACS images.
In addition, to ascertain that our photometry is not influenced by blending effects, we cross-matched the COSMOS-ACS catalogue \citep{2007ApJS..172..219L} on the positions of our objects and removed every source from our sample that has a nearby object within 1.5 arcsec\footnote{This is a reasonable value considering that 75\% of the PS1 frames have a FWHM below 1.51, 1.39, 1.34, 1.27, 1.21 arcsec for $g_{\mathrm{P1}}, r_{\mathrm{P1}}, i_{\mathrm{P1}}, z_{\mathrm{P1}}, y_{\mathrm{P1}}$ \citep{2013ApJS..205...20M}.}. This reduced the final sample to 384 sources. Throughout this work we use PSF photometry for these objects.

Out of the 384 sources, 249 have reliable spectroscopic redshifts \citep{2009ApJS..184..218L,2007ApJS..172..383T}. For the rest, the availability of deep and homogeneous photometry in 31 bands, including intermediate- and narrow-band filters \citep{2007ApJS..172....9T}, allowed computing high quality photometric redshifts with an accuracy of 0.015 with only a handful of outliers \citep{2011ApJ...742...61S}. Our final sample also contains 47 objects that are classified as stars by their spectral features. In the following analysis we include both the AGNs and the stars in order to be able to compare the PS1 observational data for these different object types.

\section{Sample definition}
\label{sec:sampledef}

\subsection{Limiting magnitudes}
\label{sec:ps1phot}
 
To identify these objects within the PS1 3$\pi$ and MDF surveys again, with the latter including XMM-COSMOS in the MDF04 field, we matched the positions of the counterparts of the X-ray sources to the PS1 catalogues and recovered 285 sources within the 3$\pi$ survey and 313 within the MDF04 survey, hereafter referred to as the "3$\pi$ sample" and the "MDF04 sample". The angular separation of the XMM-COSMOS and PS1 coordinates is less than 0.25 arcsec for all of these sources. We note that within the photometry errors, none of the 285 objects of the 3$\pi$ sample has a median magnitude exceeding the 5$\sigma$ median limiting magnitudes for individual 3$\pi$ survey exposures of 22.1 ($g_{\mathrm{P1}}$), 21.9 ($r_{\mathrm{P1}}$), 21.6 ($i_{\mathrm{P1}}$), 20.9 ($z_{\mathrm{P1}}$), and 19.9 ($y_{\mathrm{P1}}$) by more than $\sim$0.1 mag \citep{2014ApJ...784...92M}. Therefore we do not apply a further magnitude cut within the 3$\pi$ sample. However, among the 313 AGNs of the MDF04 sample, we find a number of sources that are considerably fainter than the expected 5$\sigma$ limiting magnitudes for MDF single exposures. Since we observe from the MDF04 detection table that the individual MDF04 exposure times are on average a factor of two longer than the single 3$\pi$ exposure times and since the signal-to-noise ratio is proportional to the square root of the exposure time $S/N\propto\sqrt{t_{\mathrm{exp}}}$, we expect an average increase in the 5$\sigma$ limiting magnitudes of the MDF04 survey of $|-2.5\log_{10}\sqrt{2}|\sim 0.4$ magnitudes compared to the 3$\pi$ survey. Adding this correction of 0.4 magnitudes to the respective 3$\pi$ survey values quoted above, the approximate 5$\sigma$ median limiting magnitudes for single detections of the MDF04 survey become 22.5 ($g_{\mathrm{P1}}$), 22.3 ($r_{\mathrm{P1}}$), 22.0 ($i_{\mathrm{P1}}$), 21.3 ($z_{\mathrm{P1}}$), and 20.3 ($y_{\mathrm{P1}}$). We applied a magnitude cut in each band by discarding every object with median magnitude larger than these limiting magnitudes. 

\subsection{Removal of false detections}
\label{sec:cleanup}

Since the PS1 GPC1 is a prototype camera constructed for fast readout consisting of almost 4000 CCD cells there are several different defects and a huge number of detector edges that can lead to false detections \citep{2013MNRAS.435.1825M}. Moreover, there can be reflections that lead to ghost images, diffraction spikes of bright stars, or masked pixels, potentially resulting in spurious detections and misleading photometric measurements. To reduce the contamination by "bad" and "poor" detections we downloaded only those detections from the PSPS database with none of the following flags set: FITFAIL, SATSTAR, BADPSF, DEFECT, SATURATED, CR LIMIT, MOMENTS FAILURE, SKY FAILURE, SKYVAR FAILURE, SIZE SKIPPED, POORFIT, PAIR, BLEND, MOMENTS SN, BLEND FIT, ON SPIKE, ON GHOST, and OFF CHIP. In addition we removed detections suffering from very bad seeing or focus shifts by excluding PSF model fits with psfWidMajor $>$ 6 arcsec and extremely elliptic model fits with $\mathrm{psfWidMinor/psfWidMajor}<0.65$. To minimize the effects of pixel masking on the measurements, we only kept detections with psfQf $>$ 0.85 and psfQfPerfect $>$ 0.85, i.e. PSF model fits with fewer than 15\% masked pixels weighted by the PSF. Finally to exclude very faint measurements, we worked with 5$\sigma$ detections according to $\mathrm{psfFlux/psfFluxErr}>5$.     

We note that the vast majority of the detections in the 3$\pi$ and MDF04 samples have zeropoint errors that are more accurate than 10 millimag from the "Ubercal" photometric calibration \citep{2012ApJ...756..158S}. All detections of the MDF04 sample have specified zeropoint errors, but a substantial fraction of the detections of the 3$\pi$ sample have $\Delta zp=-999$. For these unspecified zeropoint errors, we have assumed a conservative value of $\Delta zp=0.07$ in the calculation of the photometric errors. Saturation for individual detections with typical 3$\pi$ survey exposure times sets in at $g_{\mathrm{P1}}$, $r_{\mathrm{P1}}$, $i_{\mathrm{P1}}\sim 13.5$, $z_{\mathrm{P1}}\sim 13.0,$ and $y_{\mathrm{P1}}\sim 12.0$ \citep{2013ApJS..205...20M}. Although we do not expect any of our objects to be affected by saturation in any of the PS1 bands, since none of our sources has $z$ (Subaru) $<$ 17, few PS1 detections exist that are significantly brighter than these saturation limits, and even negative magnitude values occur. Because these bright detections have very likely been wrongly associated with our sources, we excluded every detection with $g_{\mathrm{P1}}, r_{\mathrm{P1}}, i_{\mathrm{P1}}, z_{\mathrm{P1}}, y_{\mathrm{P1}} < 14.0$. Furthermore, even after applying a 5$\sigma$ cut, we observed a few magnitude values of single exposures amongst our sample detections that lie well above the 5$\sigma$ limiting magnitudes of two-year stack images from the MDF04 survey (for reference $mag_{\mathrm{lim}}\left(g_{\mathrm{P1}}\right)\sim 24.5$ \citep{2012ApJ...746..128S}), with some being as faint as $g_{\mathrm{P1}}\sim 40$. For this reason we additionally removed every detection with $g_{\mathrm{P1}}, r_{\mathrm{P1}}, i_{\mathrm{P1}}, z_{\mathrm{P1}}, y_{\mathrm{P1}}$ $>$ 23.5, 23.5, 23.5, 22.5, 21.5, thereby discarding these extremely faint measurements.

\subsection{Light curve treatment}
\label{sec:lc}

The light curves of the 3$\pi$ survey suffer from extreme sparse sampling because they consist of pairs of observations carried out within $\sim$30 min in one night, followed by large temporal gaps of several months until the next observation. Only the data acquired at the end of 2009 comprises up to eight observations within one night. In contrast, a typical MDF04 light curve is divided into several observing blocks lasting about three to four months with a high sampling rate of  about eight observations per night taken approximately every one to three days. The individual observing blocks are separated by gaps of  around seven to nine months with no observations. The full light curve covers a period of about four years.

\begin{figure*}
\centering
\textbf{\hspace*{6.0mm}AGNs\hspace*{68.5mm}Stars}

\subfloat{%
        \includegraphics[width=.40\textwidth]{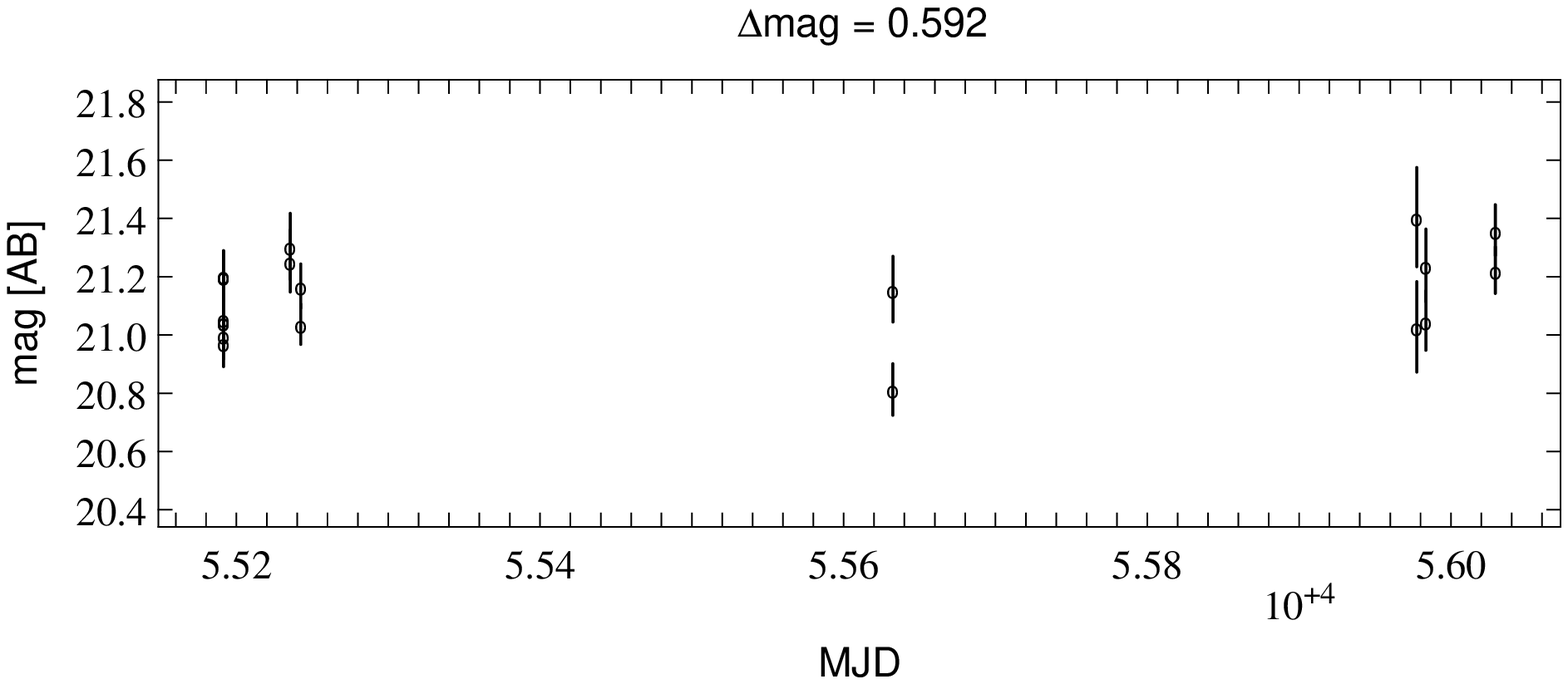}}
\quad
\subfloat{%
        \includegraphics[width=.40\textwidth]{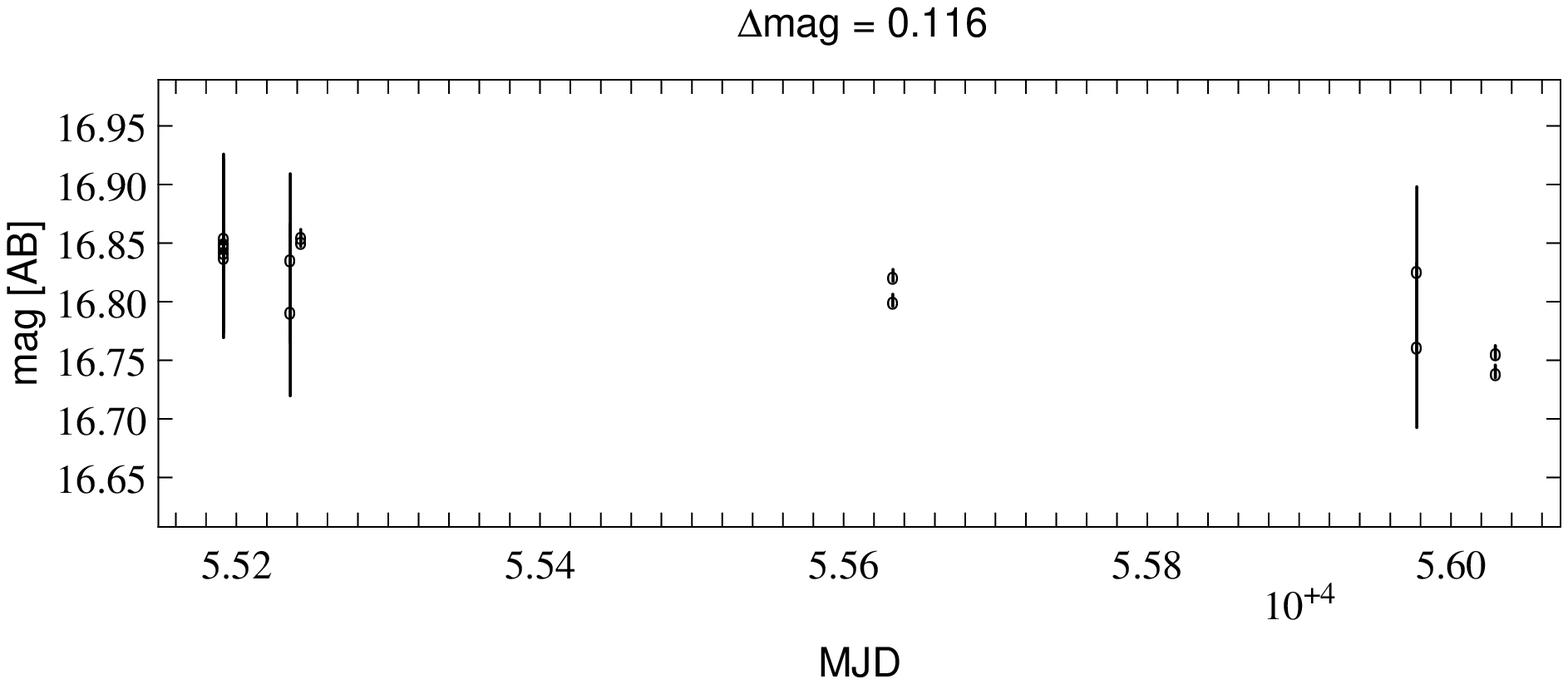}}
        
\subfloat{%
        \includegraphics[width=.40\textwidth]{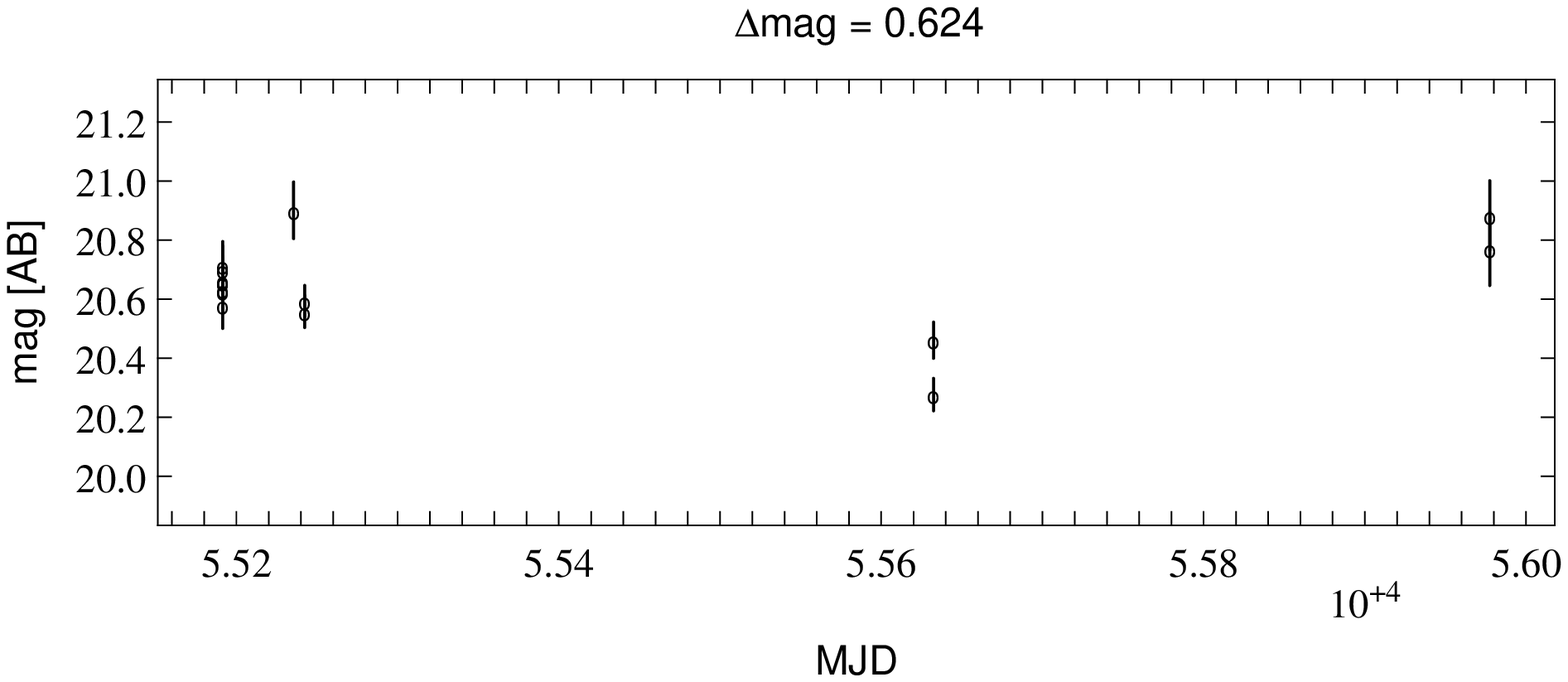}}
\quad
\subfloat{%
        \includegraphics[width=.40\textwidth]{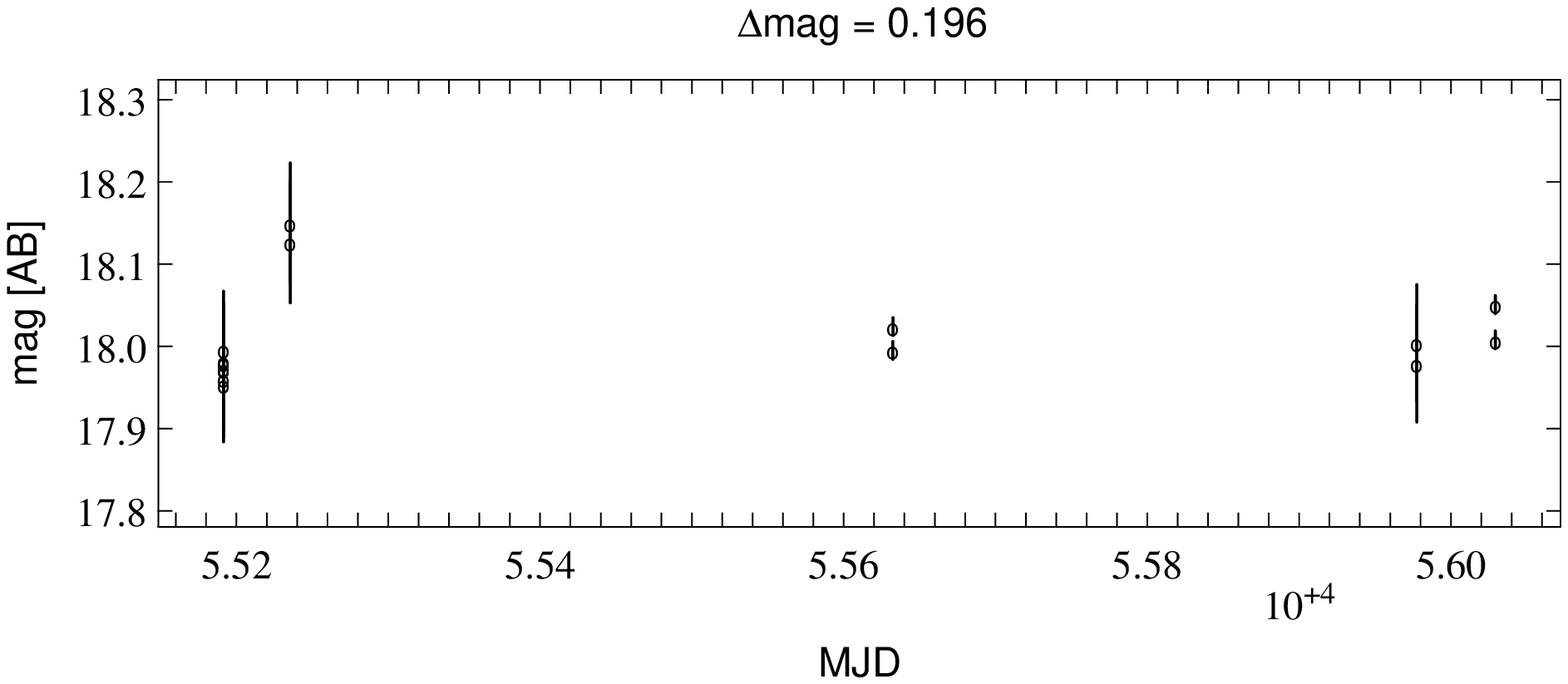}}
        
\subfloat{%
        \includegraphics[width=.40\textwidth]{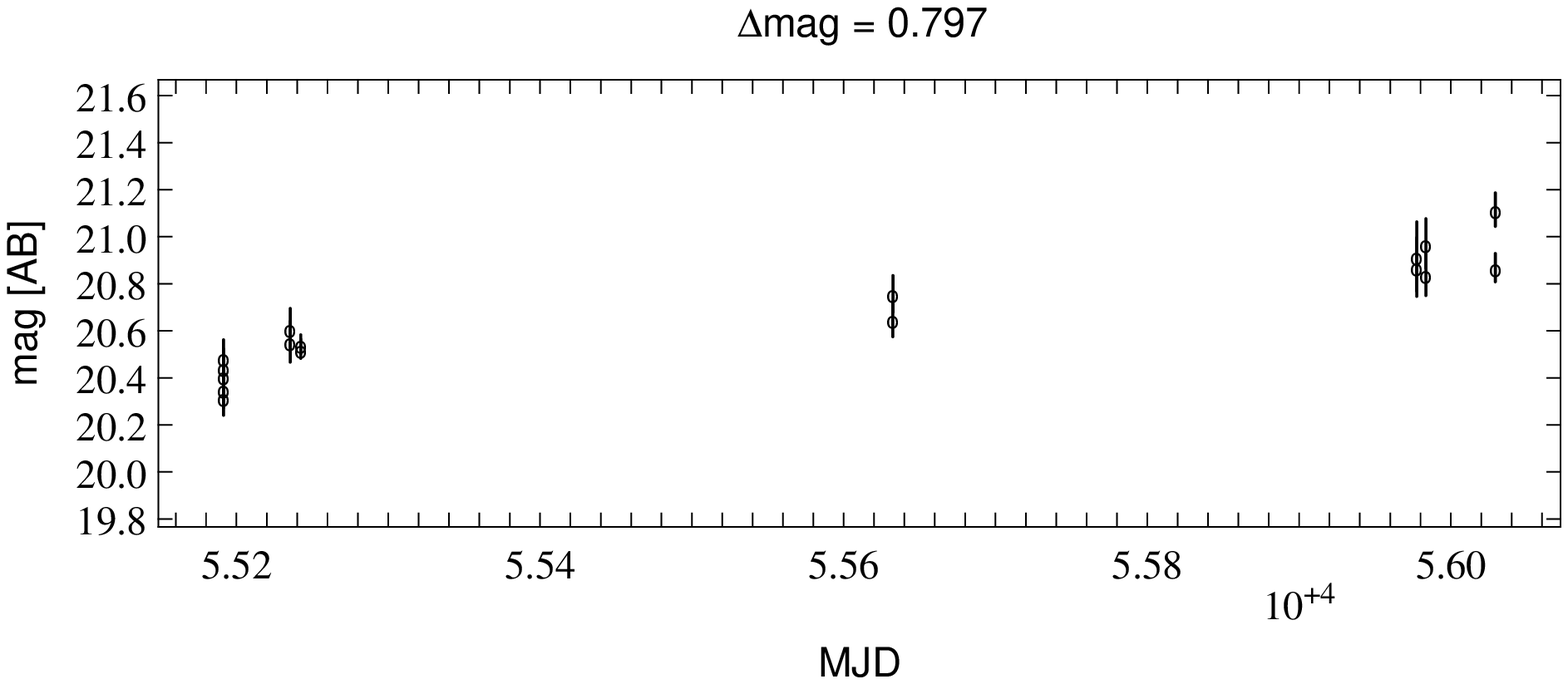}}
\quad
\subfloat{%
        \includegraphics[width=.40\textwidth]{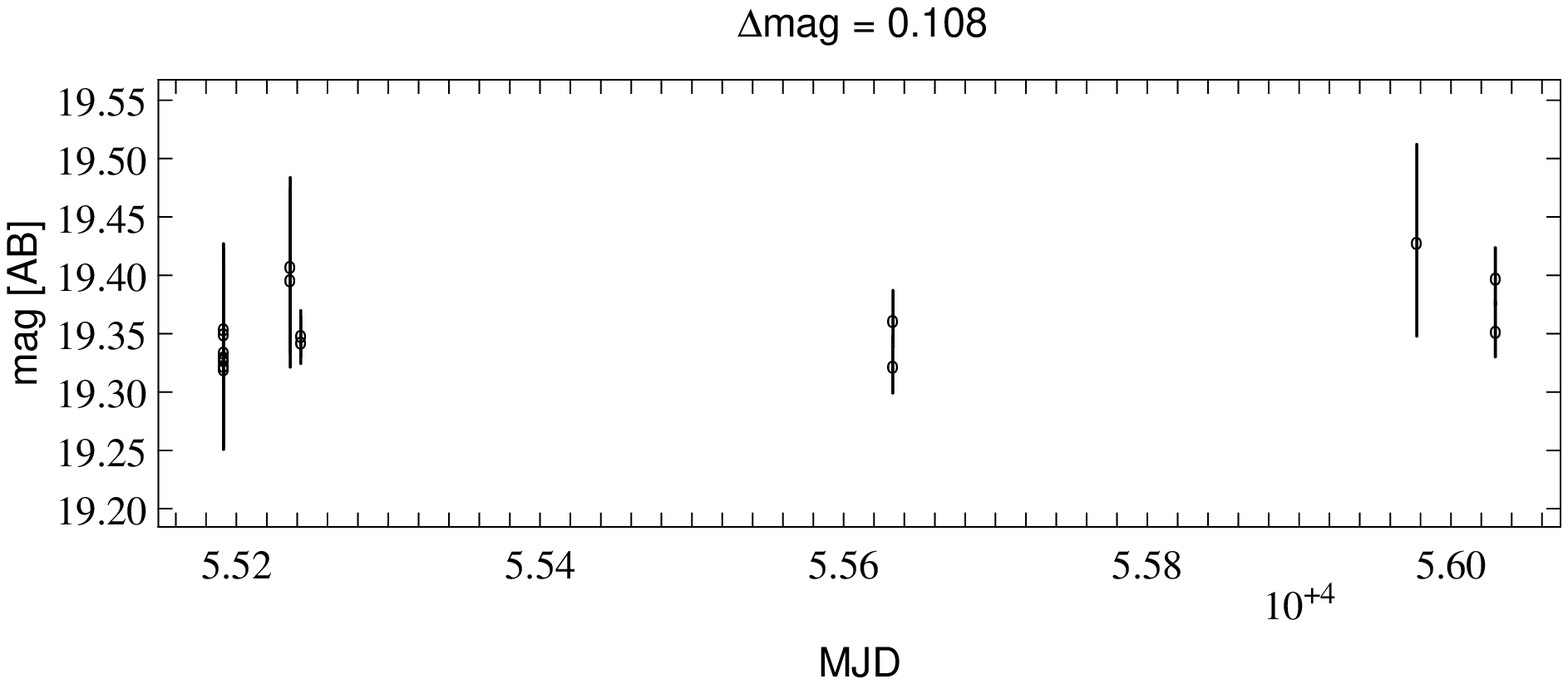}}
\quad

\subfloat{%
        \includegraphics[width=.40\textwidth]{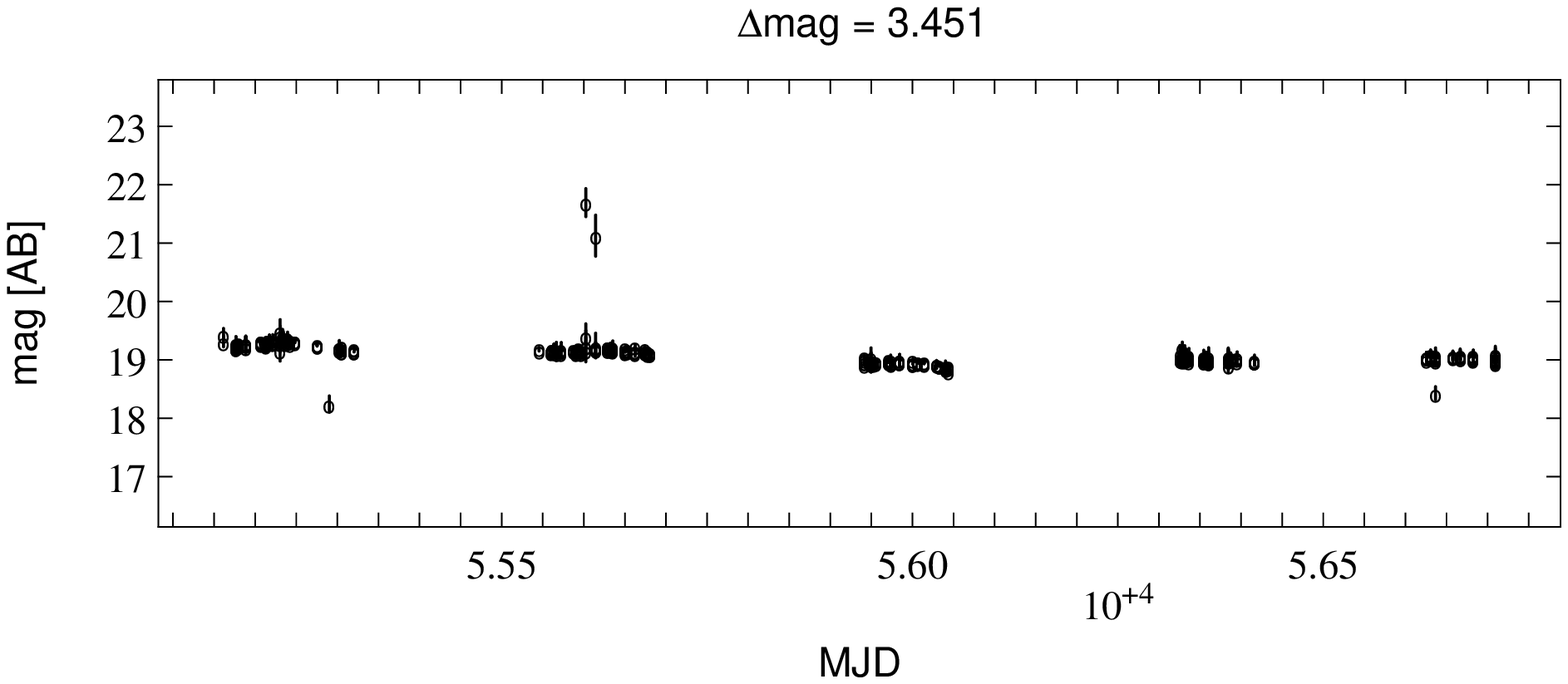}}
\quad
\subfloat{%
        \includegraphics[width=.40\textwidth]{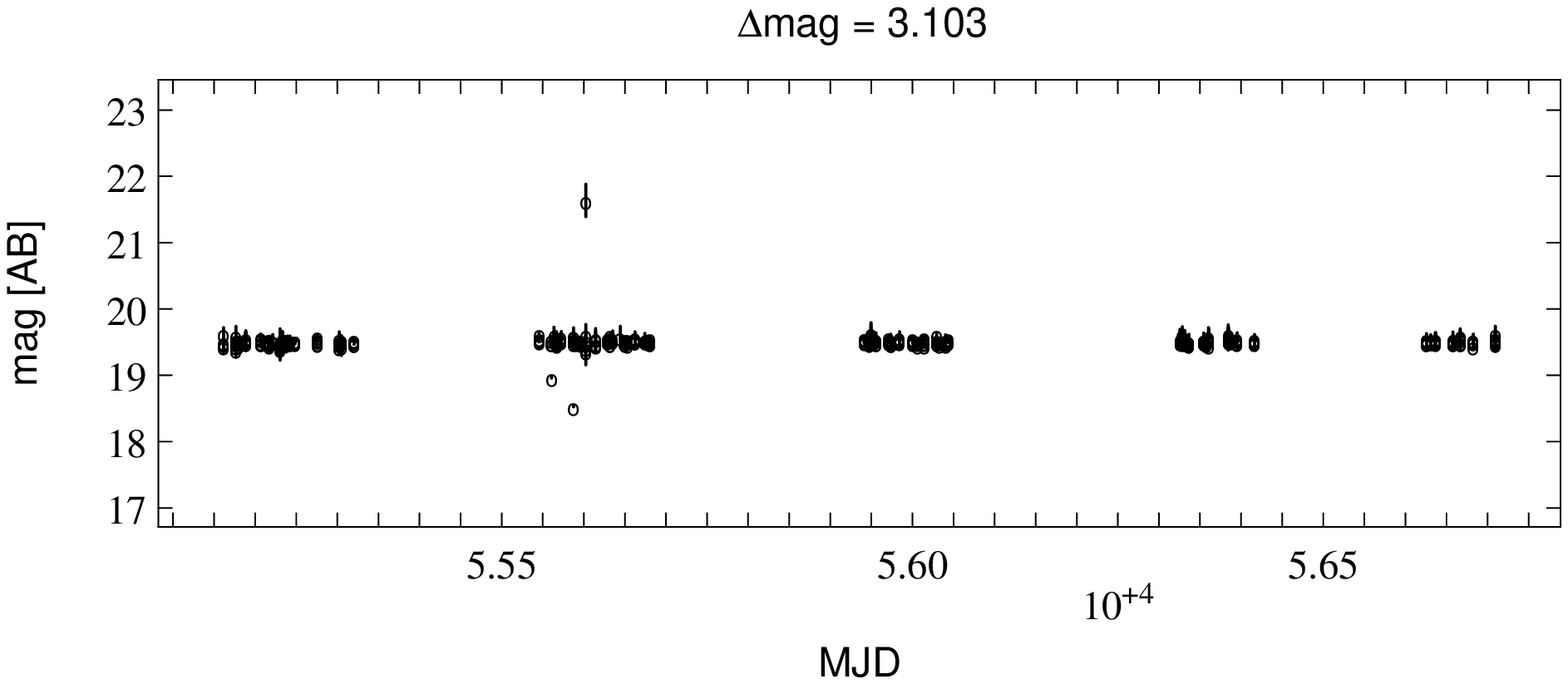}}

\subfloat{%
        \includegraphics[width=.40\textwidth]{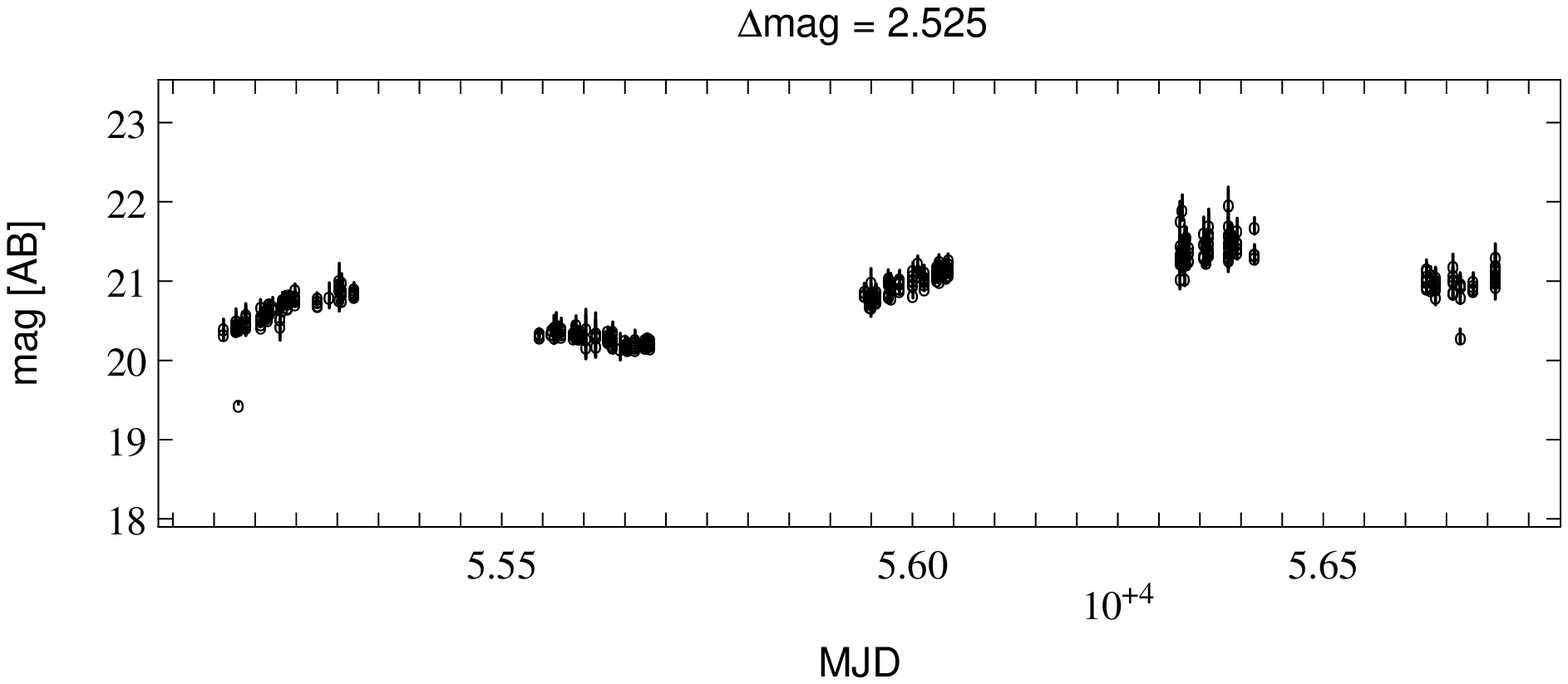}}    
\quad
\subfloat{%
        \includegraphics[width=.40\textwidth]{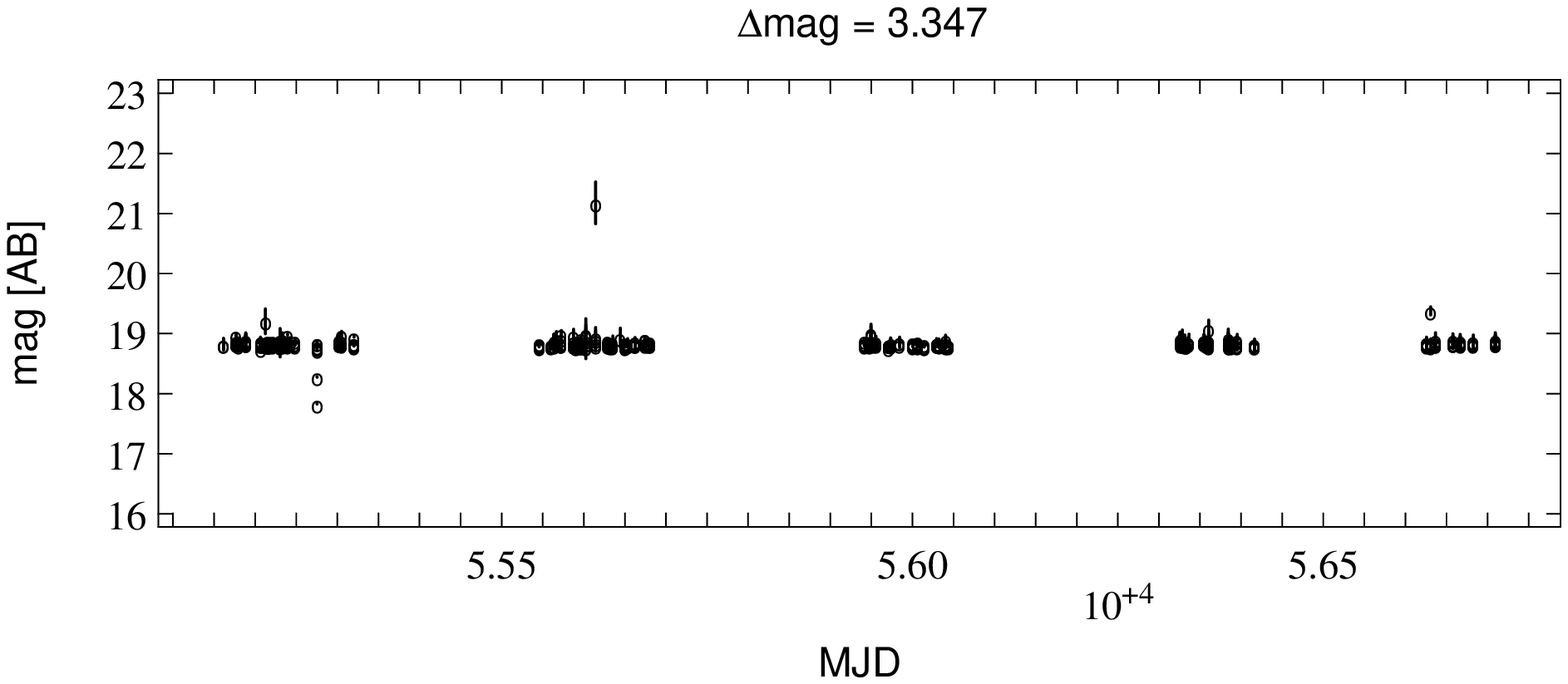}}

\subfloat{%
        \includegraphics[width=.40\textwidth]{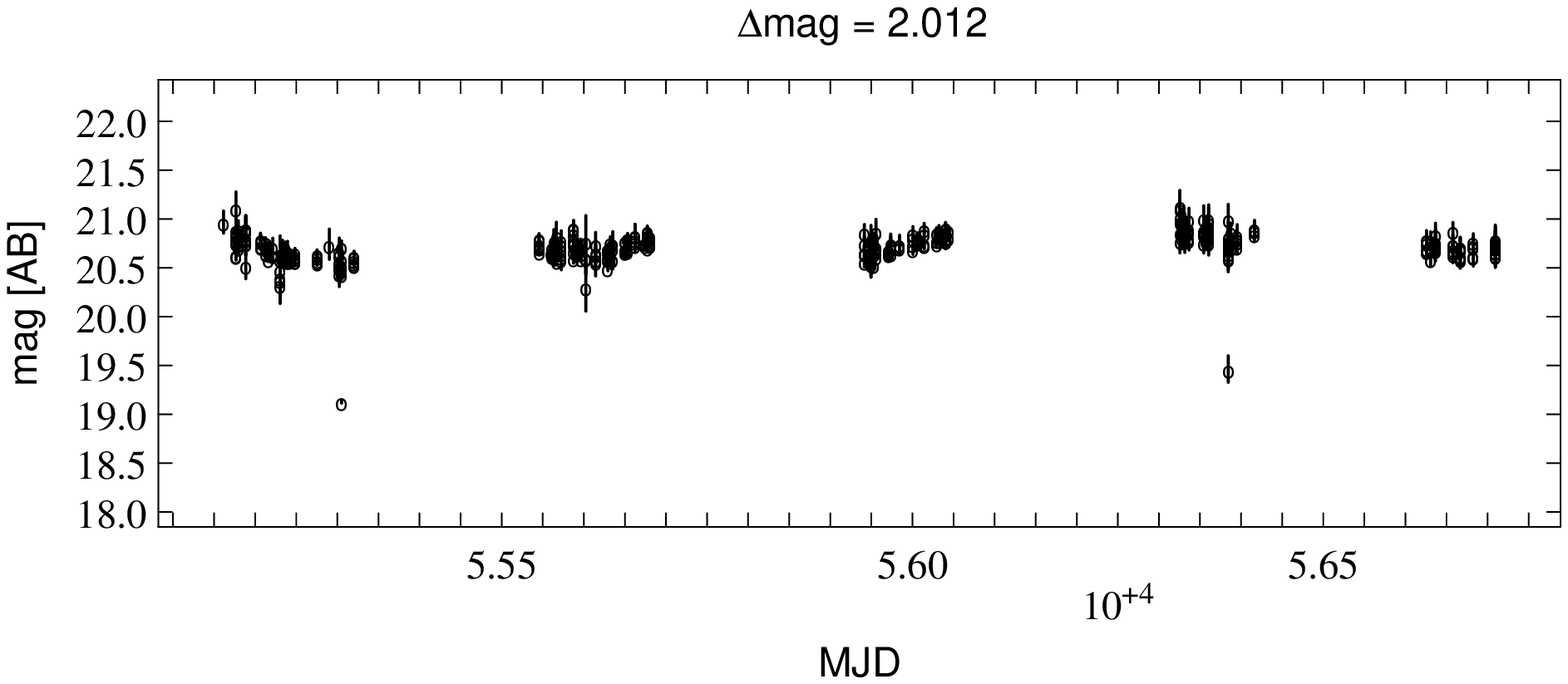}}    
\quad
\subfloat{%
        \includegraphics[width=.40\textwidth]{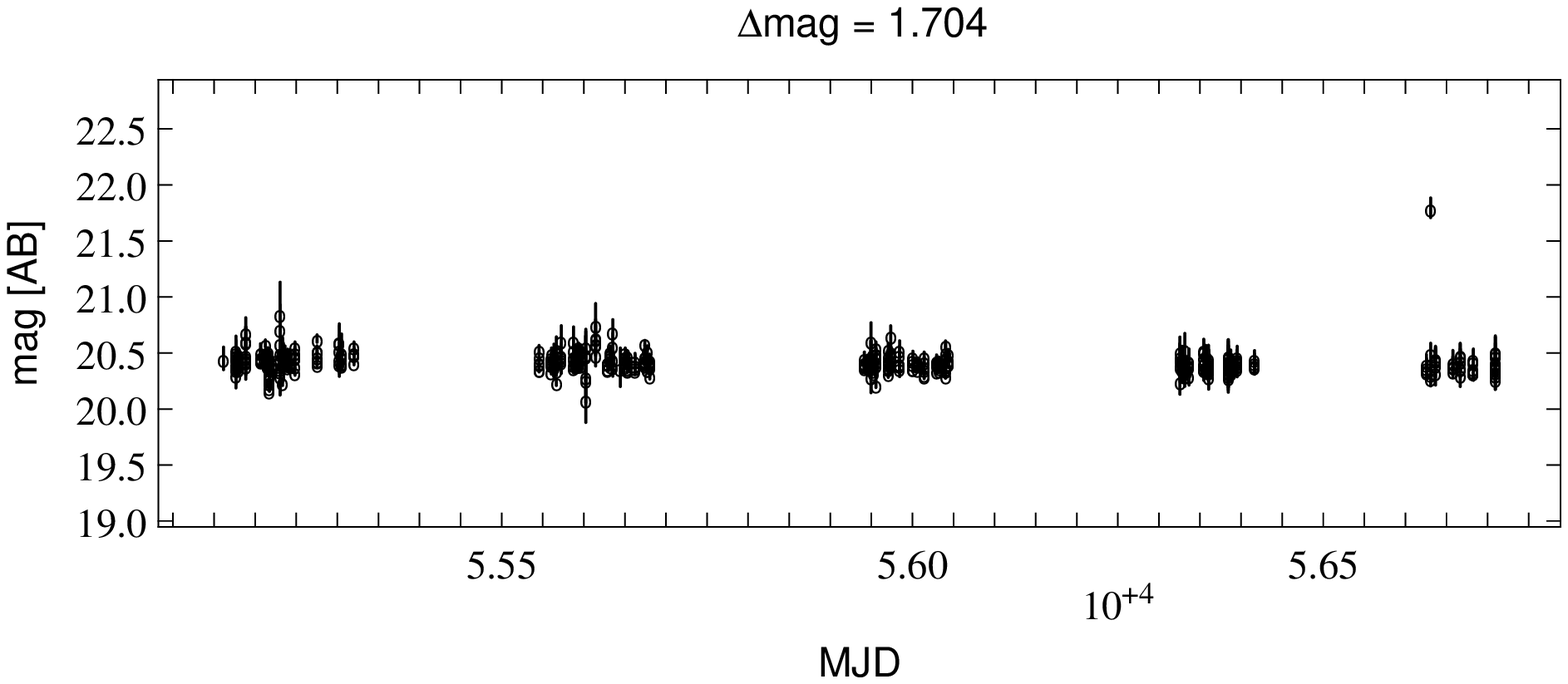}}
\caption{Raw light curves ($g_{\mathrm{P1}}$ band) of six AGNs (\textit{left column}) and six stars (\textit{right column}). The top three panels show data from the 3$\pi$ survey, the bottom three panels from the MDF04 survey, respectively. The value $\Delta mag=\mathrm{max}\left(mag\right)-\mathrm{min}\left(mag\right)$ quotes the maximum amount of variability in each light curve.}
\label{fig:rawlc3pimdf} 
\end{figure*}

Prior to performing a variability analysis, it is instructive to visually inspect the light curves in order to identify possible problems. Looking at a large number of light curves from the 3$\pi$ and MDF04 surveys, we discovered a significant number of measurements that imply variability of up to several tenths of a magnitude within one observing night. This must be compared with typical intra-night optical variability, termed as micro-variability, of $\sim$0.01--0.1 mag for normal AGNs. Only extreme objects like blazars or optically-violently variable (OVV) objects may show micro-variability of a few 0.1 mag within one night \citep{2003ApJ...586L..25G,2003A&AT...22..661G,2004MNRAS.350..175S,2005MNRAS.356..607S,2005A&A...440..855G,2007AJ....133..303C}. Considering that the nightly observations of the 3$\pi$ and MDF04 surveys are separated by $\sim$30--60 min at most, which corresponds to even shorter time intervals in the AGN rest frame, it is very likely that the observed micro-variability is not physically founded, but rather stems from low quality measurements with underestimated error bars. Moreover, we detect this short-term variability not only in AGN light curves, but also in the stellar light curves of our sample. In addition a number of light curves of the MDF04 survey exhibit few fatal outlier measurements, sometimes deviating from the bulk of data points by several magnitudes. These problems are illustrated in Fig. \ref{fig:rawlc3pimdf}, showing the raw light curves of six AGNs in the left-hand column and of six objects classified as stars in the right-hand column for both the 3$\pi$ and MDF04 surveys. The light curves of the AGNs exhibit clear signs of variability on timescales of months to years, whereas the stellar light curves are comparably flat. However, the intra-night variations in both the stellar and AGN light curves are essentially undistinguishable. This is also true considering the occurrence of catastrophic outliers in the MDF04 light curves.   

To probe the quality of the bulk of the measurements in the MDF04 light curves, we re-computed the photometry from the raw images for a $0.4\degr\times 0.4\degr$ field using the Munich Difference Imaging Analysis (MDIA) pipeline described in \cite{2013ExA....35..329K}. Comparing the resulting light curves with the ones created by the PS1 IPP, we observe that the overall trends of variability, visible in the light curve segments, are the same for both pipelines. This implies that the vast majority of the retained PS1 IPP detections exhibit sufficient quality. However, the occurrence of fatal outliers is much higher in the PS1 IPP light curves, suggesting that these measurements are indeed not credible. The origin of the fatal outliers may be spurious detections in the vicinity of our sources, which have been wrongly associated with the latter\footnote{This is supported by over-plotting all single detections remaining after the steps described in section \ref{sec:cleanup} onto a much deeper MDF04 stack image, revealing the presence of a number of false detections that cannot be associated with an optical counterpart.}. It is clear that the presence of these catastrophic outliers means that any variability measurement would be significantly biased towards very large variability amplitudes. On these grounds we decided to remove the few fatal outliers from the (PS1 IPP) MDF04 light curves. 

Finally, in view of the non-negligible number of detections showing considerable short-term variability on timescales less than one hour, we calculated nightly averages from the observations of the 3$\pi$ and MDF04 surveys. To assign a conservative and meaningful error to each averaged flux or magnitude value, considering both the presence of points with small error bars showing large scatter about the mean, as well as points with large error bars but negligible scatter, we take as error the larger of the two estimates
\begin{flalign}
        \label{eq:errmean}
        \sigma\left(\bar{f}\right)=\sqrt{\frac{1}{n\left(n-1\right)}\sum_{i=1}^{n}\left(f_{i}-\bar{f}\right)^2}
\end{flalign}  
\begin{flalign}
        \label{eq:errgauss}
        \Delta \bar{f}_{\mathrm{gauss}}=\frac{1}{n}\sqrt{\sum_{i=1}^{n}\left(\Delta f_{i}\right)^2}.
\end{flalign}   
Here $\sigma\left(\bar{f}\right)$ denotes the standard error of the arithmetic mean calculated from $n$ values $f_{i}$ observed in one night, and $\Delta \bar{f}_{\mathrm{gauss}}$ the uncertainty following Gaussian error propagation of the individual errors $\Delta f_{i}$.  

\section{Detection of variability}
\label{sec:defvarsample}

\subsection{Statistical methods to characterize variability}
\label{sec:varmethod}

Considering the tremendous data volumes provided by the PS1 surveys and future massive time-domain optical surveys, e.g. LSST \citep{2006AAS...209.8602I}, identifying large numbers of variable sources requires using variability estimators that can be obtained with low computational effort. To quantify variability we therefore utilize two variability parameters that possess well known statistical properties, and owing to their fast evaluation, they can be easily applied to very large samples. 

To estimate the probability that an object is actually varying, we first calculate the observed $\chi^{2}_{\mathrm{obs}}$ given by 
\begin{flalign}
        \label{eq:chiobs}
        \chi^{2}_{\mathrm{obs}}=\sum_{i=1}^{N}\frac{\left(f_{i}-\bar{f}\right)^{2}}{\sigma^{2}_{\mathrm{err},i}}
\end{flalign}    
from the light curve consisting of N measured fluxes $f_{i}$ with individual errors $\sigma_{\mathrm{err},i}$ and arithmetic mean $\bar{f}$. Then assuming an intrinsically non-varying source, we compute the probability that a $\chi^{2}$ larger than the observed one could just emerge by chance due to Poisson noise following 
\begin{flalign}
        \label{eq:probchi}      P\left(\chi^{2}\geq\chi_{\mathrm{obs}}^{2}\right)=\int_{\chi^{2}_{\mathrm{obs}}}^{\infty}f\left(\chi^{2},\mathrm{N}-1\right)\mathrm{d}\chi^{2}
\end{flalign}  
where $f\left(\chi^{2},d.o.f\right)$ is the probability density function of the $\chi^{2}$-distribution with $\mathrm{N}-1$ degrees of freedom ($d.o.f$). Subsequently we define the variability index V according to 
\begin{flalign}
        \label{eq:V}
        \mathrm{V}=-\log_{10}P\left(\chi^{2}\geq\chi_{\mathrm{obs}}^{2}\right),
\end{flalign}           
providing a measure of the strength of the evidence of variability. This method is depicted in \citet{1996ApJ...473..763M} and was subsequently applied by \citet{2004ApJ...611...93P}, \citet{2012ApJ...748..124Y}, and \citet{2014ApJ...781..105L}. Values of $\mathrm{V}=1.0$, $\mathrm{V}=1.3$,  and $\mathrm{V}=2.0$ correspondingly express that we reject the null hypothesis of an intrinsically non-variable source with 90\%, 95\%, and 99\% confidence. The V parameter is a useful tool for pre-selecting variable objects, yet beyond that it contains no information about the magnitude of the flux variations.

To evaluate the variability amplitude, we employ the normalized excess variance \citep{1997ApJ...476...70N} defined by 
\begin{flalign}
        \label{eq:nev}  \sigma_{\mathrm{rms}}^{2}=\left(s^{2}-\overline{\sigma_{\mathrm{err}}^{2}}\right)/\left(\bar{f}\right)^{2}=\frac{1}{\left(\bar{f}\right)^{2}}\left(\sum_{i=1}^{N}\frac{\left(f_{i}-\bar{f}\right)^{2}}{\left(N-1\right)}-\sum_{i=1}^{N}\frac{\sigma^{2}_{\mathrm{err},i}}{N}\right)
\end{flalign}     
with N, $f_{i}$, and $\sigma^{2}_{\mathrm{err},i}$ describing the same quantities as in equation \ref{eq:chiobs}. The normalized excess variance (hereafter just excess variance) depicts the residual variance after subtracting the average statistical error $\overline{\sigma_{\mathrm{err}}^{2}}$ from the sample variance $s^{2}$ of the light curve flux. Since the excess variance is normalized to the squared mean of the flux, it essentially specifies the squared fractional variability $F_{\mathrm{var}}=\sqrt{\sigma_{\mathrm{rms}}^{2}}$ \citep{1990ApJ...359...86E}. The excess variance is an estimator of the intrinsic fractional variance of a source and provides a meaningful measure of the variability amplitude even for sparsely sampled light curves. The excess variance has been frequently used in X-ray variability studies and was found to be correlated with the black hole mass, X-ray luminosity, and X-ray spectral index of AGNs \citep{1997ApJ...476...70N,1999ApJ...524..667T,1999ApJS..125..297L,2000ApJ...531...52G,2004MNRAS.348..207P,2005MNRAS.358.1405O,   2010ApJ...710...16Z,2011A&A...526A.132G,2012A&A...542A..83P,2014ApJ...781..105L}. In this work we adopt the excess variance as an estimator of the optical variability amplitude calculated from the PS1 light curves in each of the five bands.   

The uncertainty of $\sigma_{\mathrm{rms}}^{2}$ caused by Poisson noise alone has been determined by \citet{2003MNRAS.345.1271V}. They performed Monte Carlo simulations thereby generating a random "red noise" light curve (power spectral density with logarithmic slopes between $-1$ and $-2$, see Appendix \ref{sec:appendixa} for details), adding Poisson noise by drawing fluxes from the Poisson distribution and then measuring the excess variance of the smeared light curve. The width of the $\sigma_{\mathrm{rms}}^{2}$ distribution resulting from $10^{4}$ "observations" of the light curve is found to be well fitted by
\begin{flalign}
        \label{eq:errnev}
        err\left(\sigma_{\mathrm{rms}}^{2}\right)=\sqrt{\left(\sqrt{\frac{2}{N}}\cdot\frac{\overline{\sigma_{\mathrm{err}}^{2}}}{\left(\bar{f}\right)^{2}}\right)^{2}+\left(\sqrt{\frac{\overline{\sigma_{\mathrm{err}}^{2}}}{N}}\cdot\frac{2 F_{\mathrm{var}}}{\left(\bar{f}\right)}\right)^{2}}.
\end{flalign}  
In the case of a low intrinsic variability amplitude or very faint sources $s^{2}\sim\overline{\sigma_{\mathrm{err}}^{2}}$,  the excess variance will thus be small and can even be negative. In this situation the first term of equation \ref{eq:errnev} dominates. In contrast, if the variability signal is strong, $s^{2}>>\overline{\sigma_{\mathrm{err}}^{2}}$, and the second term of equation \ref{eq:errnev} dominates \citep{2003MNRAS.345.1271V}. It is well known that there are additional uncertainties connected to an excess variance measurement. These are related to the stochastic nature of AGN variability and the light curve sampling pattern. However, as discussed in Appendix \ref{sec:appendixa}, the bias factor associated with these uncertainties is expected to be close to one for our observational data, so we do not correct for these errors.

Following \citet{2014ApJ...781..105L}, when using solely the excess variance to quantify variability, we regard it as a detection of variability if 
\begin{flalign}
        \label{eq:detectnev}
        \sigma_{\mathrm{rms}}^{2}-err\left(\sigma_{\mathrm{rms}}^{2}\right)>0.
\end{flalign} 
To be able to define robust samples of varying sources, we take advantage of the information provided by both of the introduced variability parameters. Throughout this work, unless quoted differently, we consider it as a detection of variability when the probability for spurious variability is less than 5\% and the excess variance is greater than zero within its error, expressed by the condition
\begin{flalign}
        \label{eq:varboth}
        \mathrm{V}>1.3\wedge\sigma_{\mathrm{rms}}^{2}-err(\sigma_{\mathrm{rms}}^{2})>0.
\end{flalign}  
We apply these two variability parameters to identify variable objects using the light curves of the 3$\pi$ and MDF04 surveys by testing for condition \ref{eq:varboth} in each of the five PS1 band light curves.

\subsection{Catalogues of variable objects}
\label{sec:varsample}

To investigate variability in the 3$\pi$ and MDF04 samples, we only considered objects with more than two detections ($N>2$). From the nightly averaged flux light curves of the point-like and isolated sources, we calculated the V parameter and the normalized excess variance. The numbers of AGNs satisfying $\mathrm{V}>1.3$, $\sigma_{\mathrm{rms}}^{2}-err(\sigma_{\mathrm{rms}}^{2})>0$, or both of these conditions are listed for each filter in Table \ref{tab:pi3varstat} for the 3$\pi$ sample and Table \ref{tab:mdfvarstat} for the MDF04 sample, respectively. The numbers reveal that, when estimating the
probability of variability, the V parameter has a tendency to select more objects as variable than the excess variance, with the latter quantifying the net amplitude of variability. Nevertheless, the intersection of the two variability detection methods, given in the last column of Tables \ref{tab:pi3varstat} and \ref{tab:mdfvarstat}, is large; i.e., both methods are consistent for identification of variable sources. Considering the numbers in this last column, 59.6\% ($g_{\mathrm{P1}}$), 46.6\% ($r_{\mathrm{P1}}$), 28.0\% ($i_{\mathrm{P1}}$), 38.9\% ($z_{\mathrm{P1}}$), and 22.2\% ($y_{\mathrm{P1}}$) of the AGNs with $N>2$ are detected as variable in the 3$\pi$ survey and 98.4\% ($g_{\mathrm{P1}}$), 98.4\% ($r_{\mathrm{P1}}$), 98.2\% ($i_{\mathrm{P1}}$), 97.0\% ($z_{\mathrm{P1}}$), and 97.4\% ($y_{\mathrm{P1}}$) in the MDF04 survey, respectively.
\begin{table}
\caption{Number of variable AGNs from the 3$\pi$ sample.}
\centering
\begin{tabular}{ccccc}
\hline\hline
Filter & $N>2$ & $V>1.3$ & $\sigma_{\mathrm{rms}}^{2}-err(\sigma_{\mathrm{rms}}^{2})>0$ & (1) $\wedge$ (2) \\ 
& & (1) & (2) & \\
\hline
$g_{\mathrm{P1}}$ & 151 & 107 & 92 & 90 \\
$r_{\mathrm{P1}}$ & 116 & 76 & 55 & 54 \\
$i_{\mathrm{P1}}$ & 50 & 27 & 14 & 14 \\
$z_{\mathrm{P1}}$ & 95 & 38 & 39 & 37 \\
$y_{\mathrm{P1}}$ & 36 & 12 & 8 & 8 \\
\hline
\end{tabular}
\tablefoot{Number that fulfil the conditions given in the column headings. N: number of detections}
\label{tab:pi3varstat}
\end{table}
\begin{table}
\caption{Number of variable AGNs from the MDF04 sample.}
\centering
\begin{tabular}{ccccc}
\hline\hline
Filter & $N>2$ & $V>1.3$ & $\sigma_{\mathrm{rms}}^{2}-err(\sigma_{\mathrm{rms}}^{2})>0$ & (1) $\wedge$ (2) \\ 
& & (1) & (2) & \\
\hline
$g_{\mathrm{P1}}$ & 187 & 187 & 185 & 184 \\
$r_{\mathrm{P1}}$ & 184 & 183 & 182 & 181 \\
$i_{\mathrm{P1}}$ & 165 & 165 & 163 & 162 \\
$z_{\mathrm{P1}}$ & 135 & 135 & 132 & 131 \\
$y_{\mathrm{P1}}$ & 76 & 76 & 74 & 74 \\
\hline
\end{tabular}
\tablefoot{Number that fulfil the conditions given in the column headings.
N: number of detections}
\label{tab:mdfvarstat}
\end{table}

After comparing the different PS1 filters, the fraction of variable AGNs is found to be larger for the "bluer" bands, and yet the $i_{\mathrm{P1}}$ band of the 3$\pi$ sample comprises fewer varying sources than the "redder" $z_{\mathrm{P1}}$ band. We stress, however, that one should be careful when comparing these fractions since the 3$\pi$ light curves suffer from extreme sparse sampling, so that the ability to detect variability crucially depends on the number of observations. In fact the $i_{\mathrm{P1}}$ band has on average the fewest detections in the 3$\pi$ sample in our data set with the mean number of observations of 4.5 ($g_{\mathrm{P1}}$), 3.9 ($r_{\mathrm{P1}}$), 3.1 ($i_{\mathrm{P1}}$), 3.4 ($z_{\mathrm{P1}}$), and 3.5 ($y_{\mathrm{P1}}$), which might explain the observed lack of variability. Furthermore, considering the corresponding fractions of variable sources from the MDF04 sample, we find more variable objects in the $i_{\mathrm{P1}}$ band than in the $z_{\mathrm{P1}}$ band. Given the sampling rate of the MDF04 survey, essentially all AGNs in our sample show some amount of variability during the  nearly four years of repeated monitoring, and the vast majority are variable in multiple bands. After averaging the MDF04 light curves, the mean number of observations for all bands is given by 69.1 ($g_{\mathrm{P1}}$), 70.5 ($r_{\mathrm{P1}}$), 83.6 ($i_{\mathrm{P1}}$), 88.2 ($z_{\mathrm{P1}}$), and 51.9 ($y_{\mathrm{P1}}$). The MDF04 survey produced significantly fewer observations in the $y_{\mathrm{P1}}$ band, while the $g_{\mathrm{P1}}$ and $r_{\mathrm{P1}}$ bands suffered the most from fatal outliers. This catalogue of well characterized XMM-COSMOS sources therefore allows us to study the variability properties of $\sim$180 AGNs in the "blue" bands and more than 100 AGNs in the "red" bands. These catalogues of variable AGNs are available in the online journal (see Appendix \ref{sec:appendixc} for details). 

\begin{figure*}
\centering
\textbf{\hspace*{12mm}AGN (XID 1)}
\subfloat{%
        \includegraphics[width=0.83\textwidth]{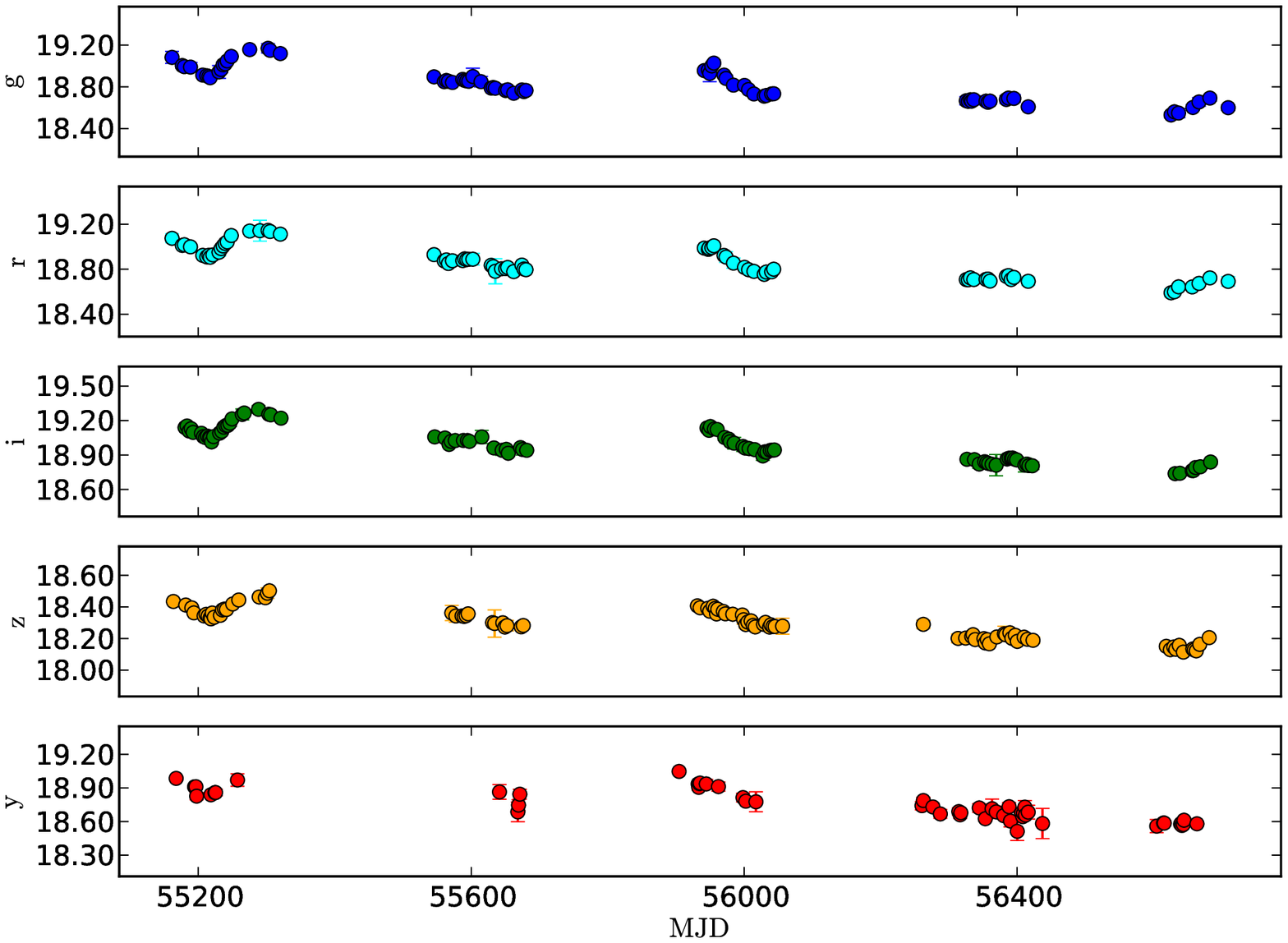}}

\textbf{\hspace*{11mm}Star (XID 60462)} 
\subfloat{%
        \includegraphics[width=0.83\textwidth]{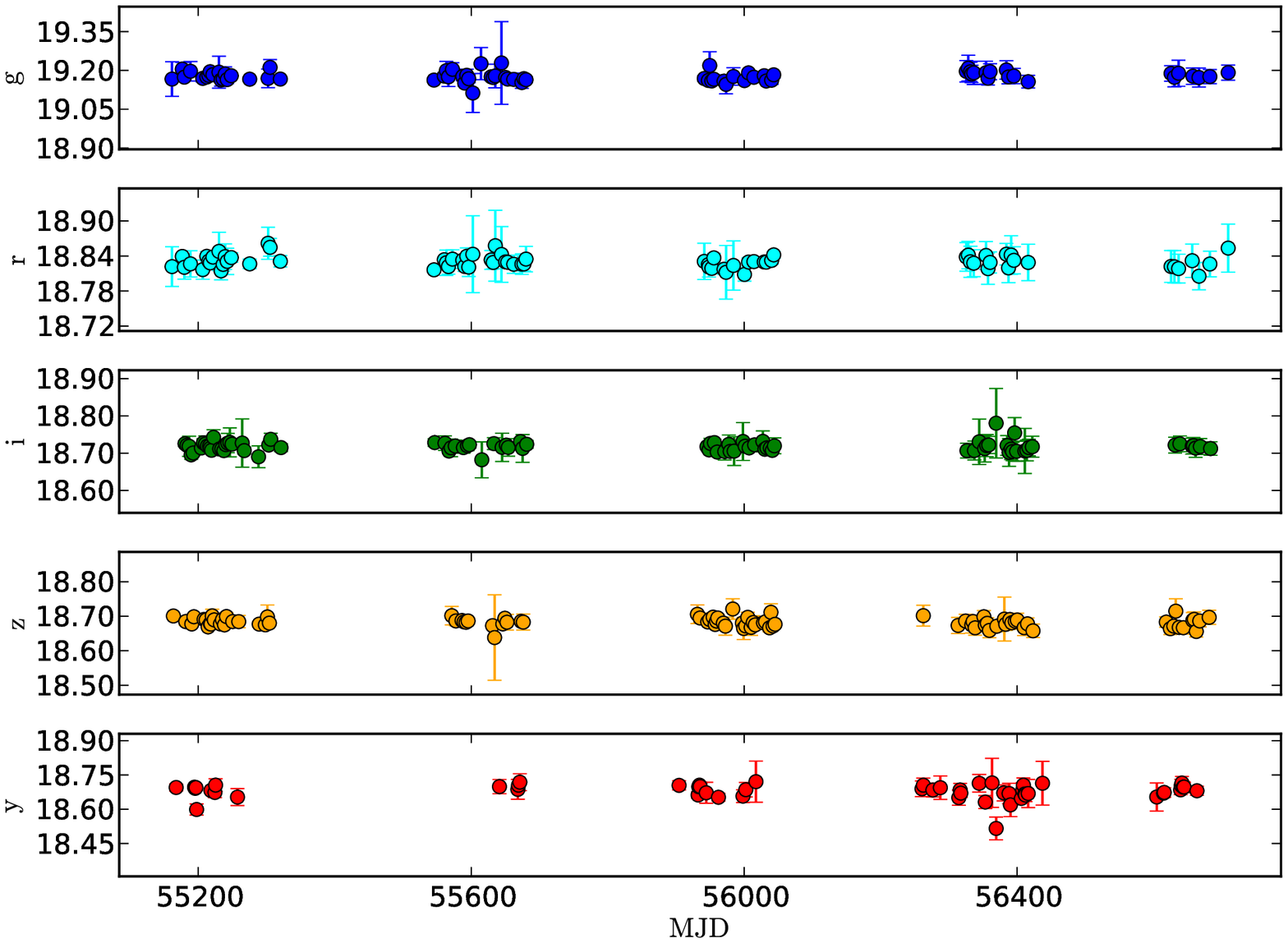}}
\caption{Nightly averaged MDF04 light curves showing all five PS1 bands of the AGN with XID 1 (\textit{top}) and the star with XID 60462 (\textit{bottom}).}
\label{fig:mdf5bandlc}
\end{figure*}

As a summarizing example, Fig. \ref{fig:mdf5bandlc} shows the nightly averaged light curves of one AGN that is varying in all five PS1 bands, along with the light curves of one star that does not vary in any band. Whereas the AGN light curves exhibit approximately simultaneous variations in all five PS1 bands with significant amplitudes of about $\sim$0.5 magnitudes, the stellar light curves are constant within the photometric errors with only two outliers appearing in the $y_{\mathrm{P1}}$ band, which however do not cause a detection of variability according to condition \ref{eq:varboth}.

\subsection{Comparison of the 3$\pi$ and MDF04 variability}
\label{sec:cmp3pimdf}

\begin{figure}
\centering
\subfloat{%
        \includegraphics[width=0.40\textwidth]{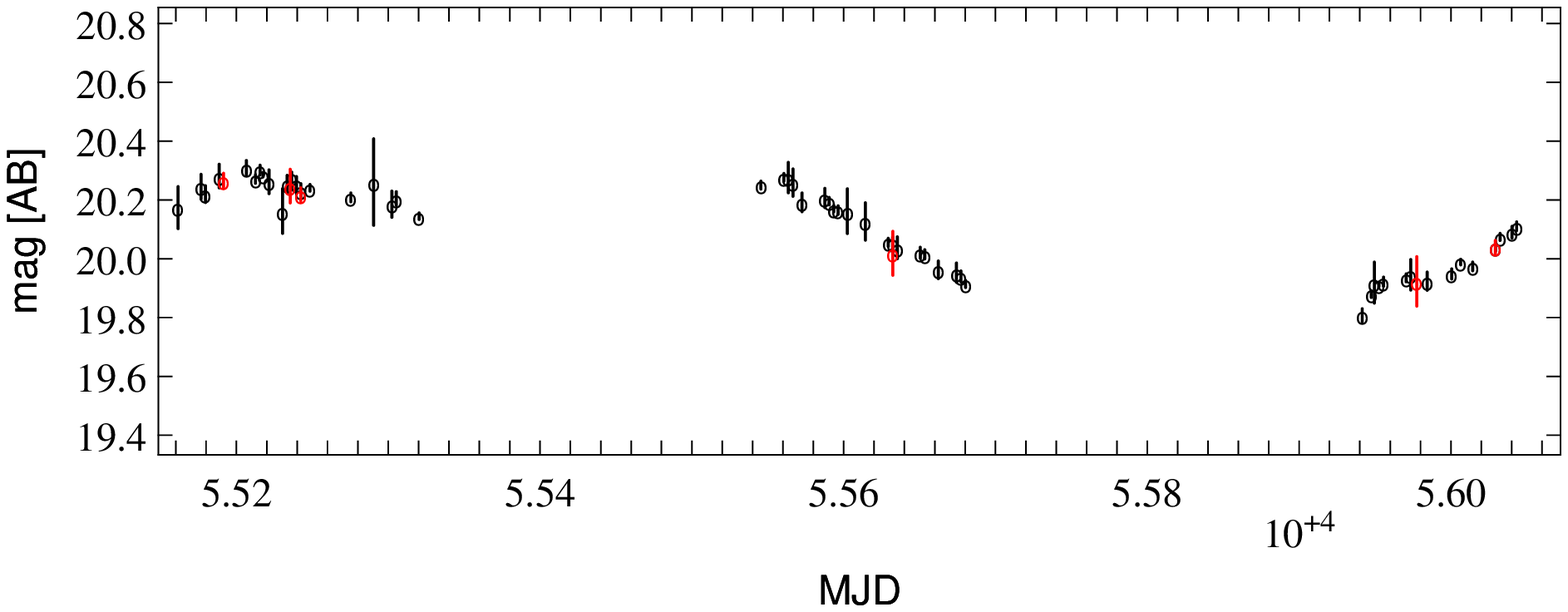}}
        
\subfloat{%
        \includegraphics[width=0.40\textwidth]{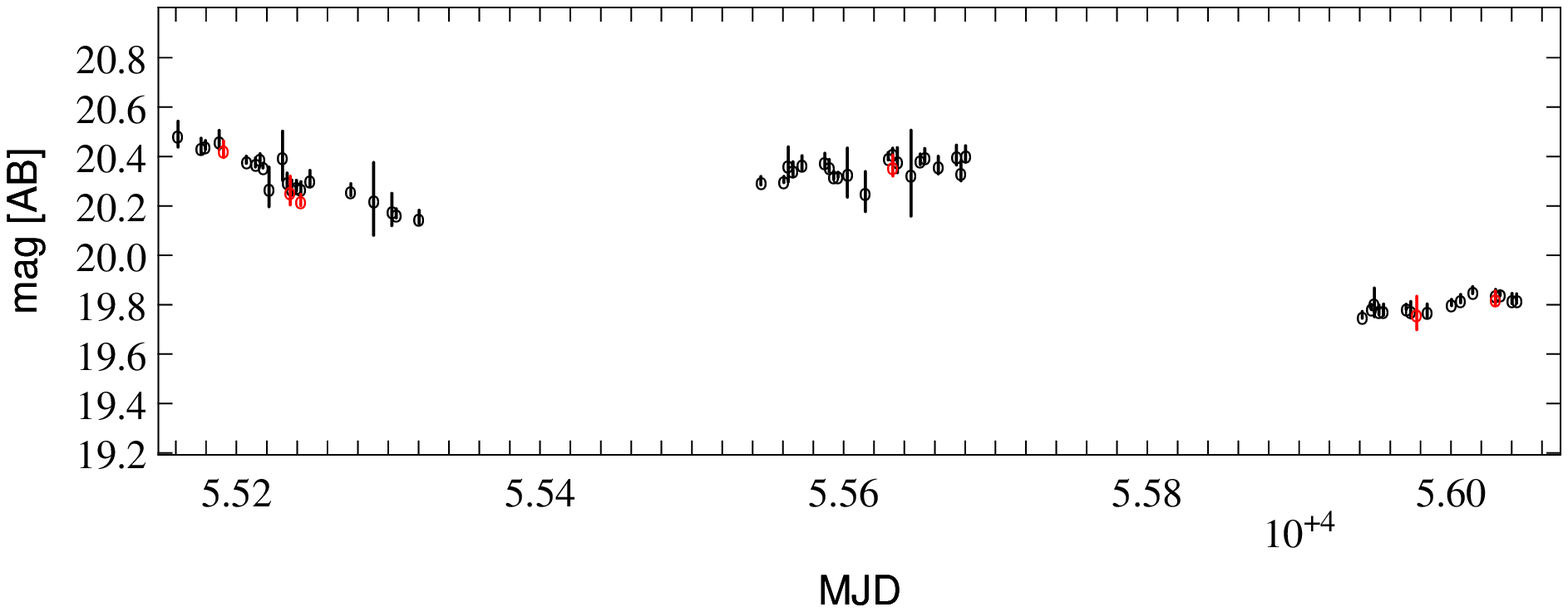}}
        
\subfloat{%
        \includegraphics[width=0.40\textwidth]{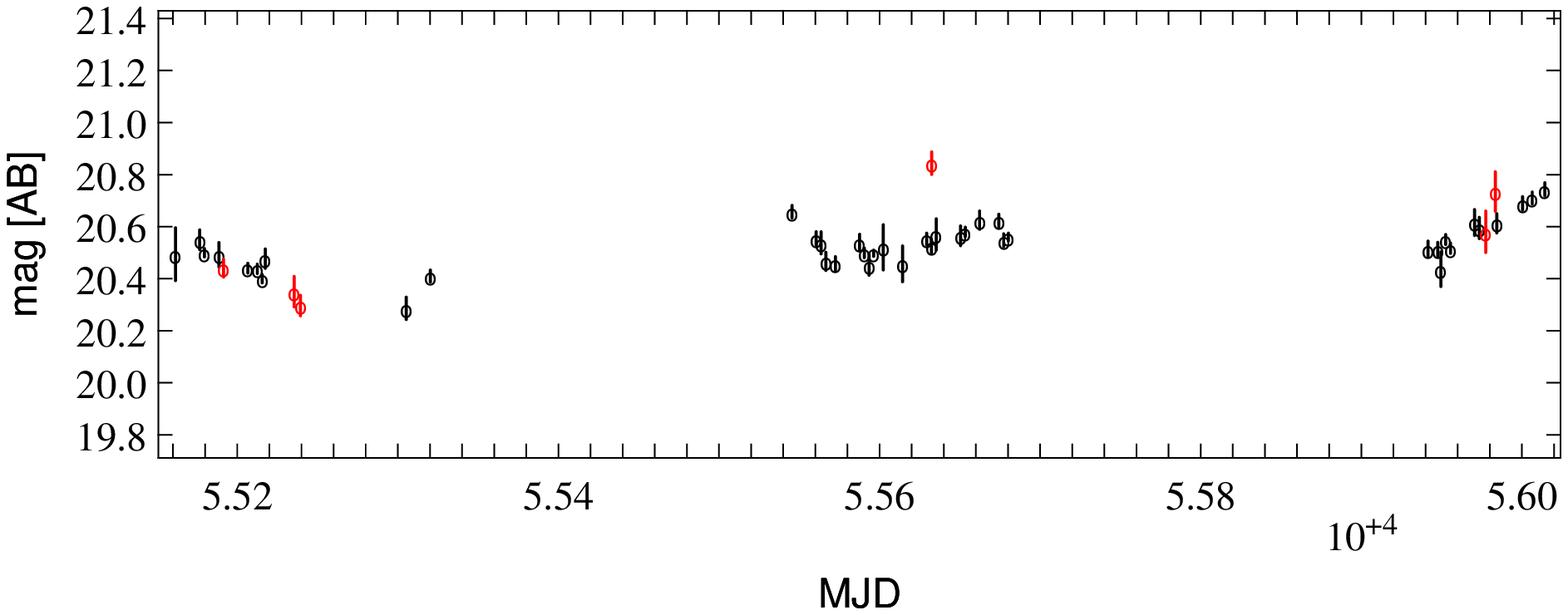}}
\caption{MDF04 light curves ($g_{\mathrm{P1}}$ band) of three AGNs over-plotted with the corresponding 3$\pi$ light curves in red.}
\label{fig:cmpmdf3pi}
\end{figure}

Since the 3$\pi$ survey covers three-quarters of the sky, it allows the properties of millions of AGNs to be
investigated and provides optical photometry for the sources to be observed with EUCLID and eROSITA. It is therefore important to understand to what extent we can probe variability using the sparsely sampled 3$\pi$ light curves as compared to the much better sampled MDF light curves, which are however only available for ten selected sky fields. To address this question we performed a visual comparison of our AGN light curves of the two surveys. Examination of a large number of all light curves reveals that the vast majority of the detections of both surveys yield very similar magnitude values; i.e., the nightly averaged light curves of the 3$\pi$ survey fit almost perfectly in the corresponding ones of the MDF04 survey. That is illustrated in Fig. \ref{fig:cmpmdf3pi}, showing the light curves of the MDF04 survey in black, over-plotted with the respective 3$\pi$ observations in red for three AGNs. However, some of the 3$\pi$ survey light curves are still contaminated by fatal outliers that can only be identified as such with the additional information provided by the MDF04 light curves also showing the long-term trends in variability. Such a case is visible in the bottom panel of Fig. \ref{fig:cmpmdf3pi}, whereas the two other AGN light curves of the same plot agree very well.  

\begin{figure}
\centering
\subfloat{%
        \includegraphics[width=0.40\textwidth]{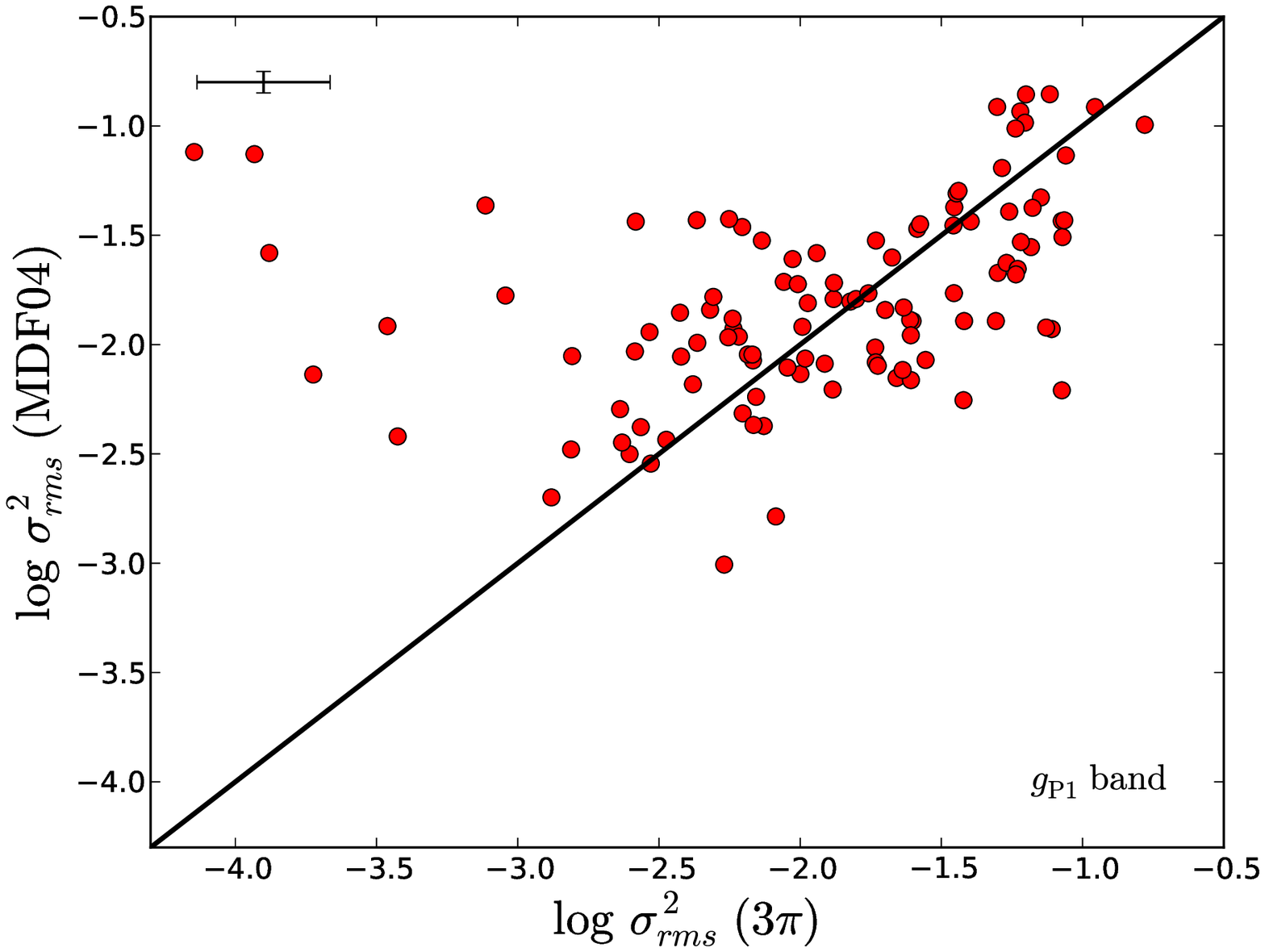}}
        
\subfloat{%
        \includegraphics[width=0.40\textwidth]{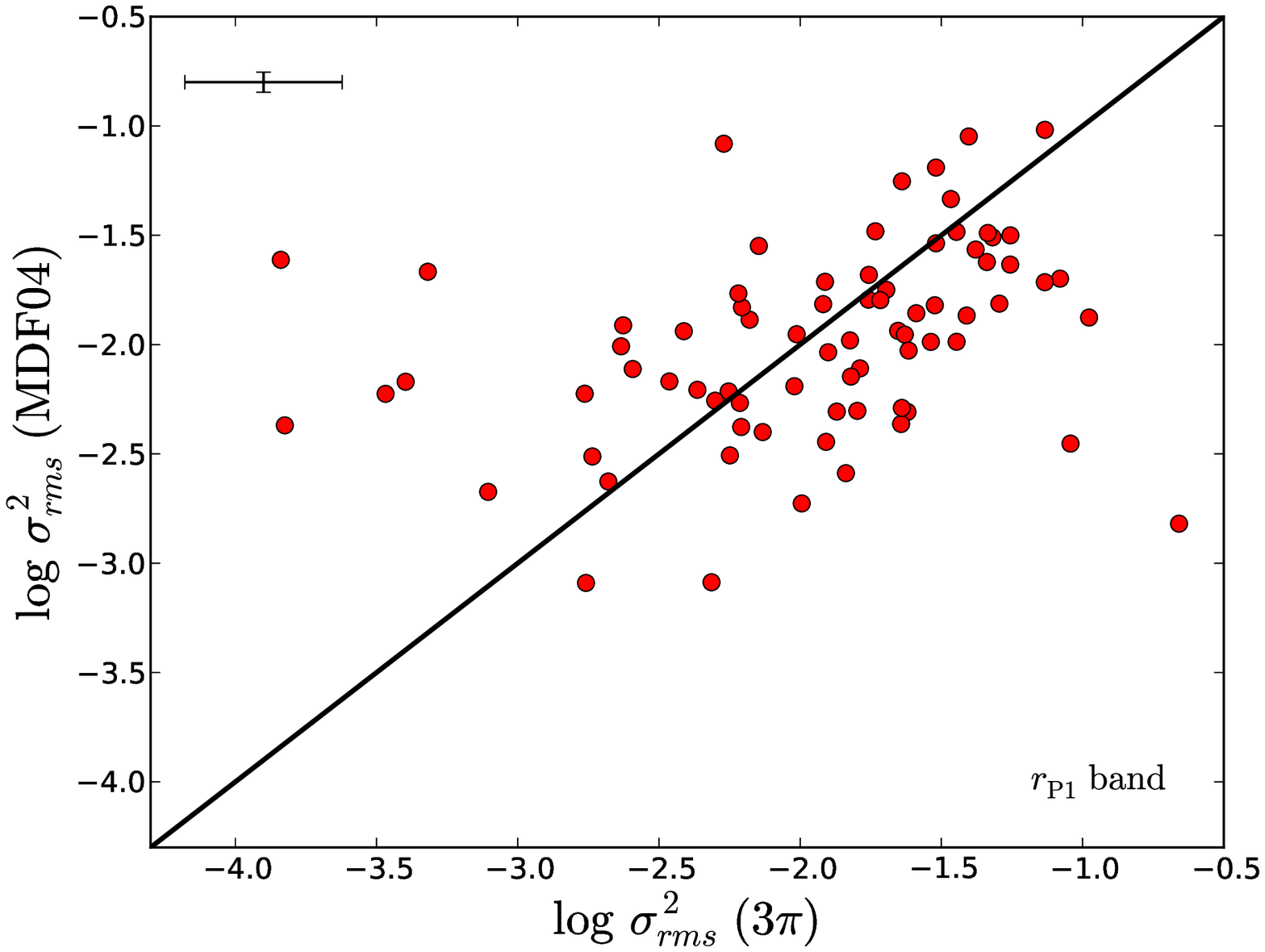}}
\caption{Excess variance calculated from the MDF04 light curves versus the respective value computed from the 3$\pi$ light curves for all AGNs with $\sigma_{\mathrm{rms}}^{2}>0$ in the $g_{\mathrm{P1}}$ band (\textit{top}) and $r_{\mathrm{P1}}$ band (\textit{bottom}). The black line represents the one-to-one relation, and the error bars show the average value of $err(\sigma_{\mathrm{rms}}^{2})$.}
\label{fig:nevmdfvs3pi}
\end{figure}

A more quantitative comparison of the two surveys in view of the variability amplitude can be done by contrasting the excess variances as measured from the light curves of the 3$\pi$ and MDF04 survey. This is displayed in Fig. \ref{fig:nevmdfvs3pi} for those AGNs with a positive $g_{\mathrm{P1}}$ and $r_{\mathrm{P1}}$ band excess variance. Even though the excess variance is calculated from only about six points at best in the case of the 3$\pi$ survey, as opposed to typically $\sim$70 points in the MDF04 survey, the two measurements yield similar estimates for a large portion of the tested sample. Nevertheless, the error of the excess variance is considerably larger for the 3$\pi$ sample, so the variability signal cannot be detected as well as within the MDF04 survey.  

The differences in the $\sigma_{\mathrm{rms}}^{2}$ measurements are particularly large for objects whose 3$\pi$ light curves suggest a constant source just because they miss the variability occurring in between the observations, which is however visible in the MDF04 light curves. Such light curves give rise to the dramatic outliers apparent in the top left region within each panel of Fig. \ref{fig:nevmdfvs3pi}. For these reasons the fractions of variable AGNs reported in Table \ref{tab:pi3varstat} are significantly lower than the corresponding fractions obtained using the MDF04 survey. For example, we lose 39\% in the variability detection for the $g_{\mathrm{P1}}$ band and 52\% for the $r_{\mathrm{P1}}$ band as compared to the MDF04 survey. Nonetheless, we point out that although the uncertainties in the excess variance measurements are large, it is possible to obtain a reasonable variability amplitude estimation by utilizing the light curves of the 3$\pi$ survey for a large number of our sources. Considering Fig. \ref{fig:nevmdfvs3pi}, we may assume that all objects with $\log\sigma_{\mathrm{rms}}^{2}\left(\mathrm{MDF04}\right)>-3$ and $\log\sigma_{\mathrm{rms}}^{2}\left(3\pi\right)>-3$, i.e., all sources varying at least at the 3\% level, have a well-estimated variability amplitude even when using 3$\pi$ survey light curves. When assuming this variability cut, the excess variance values of both surveys are similar for 91\% of the $g_{\mathrm{P1}}$ band objects and 89\% of the $r_{\mathrm{P1}}$ band objects. This means that the 3$\pi$ sample of variable objects is pure but not complete at the 3\% level of variability, therefore the observations provided by the 3$\pi$ survey allow variable objects to be selected for three-quarters of the sky, at least as long as the intrinsic variability amplitude is large.  

\subsection{Definition of the photo-z sample}
\label{sec:photozsample}

\begin{figure}
        \centering
        \includegraphics[width=.45\textwidth]{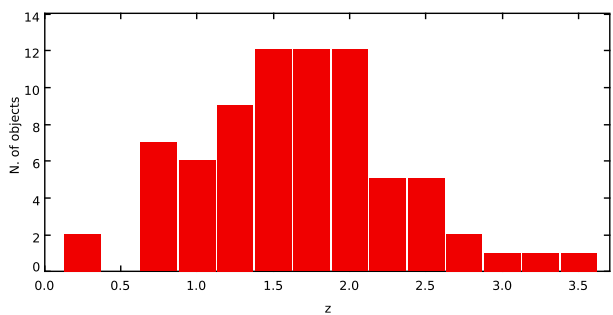}
        \caption{Redshift distribution of the 75 AGNs from the MDF04 sample used in the photo-z analysis.}
        \label{fig:histozmdf}
\end{figure}

The sample of varying sources for our photo-z analysis is drawn by selecting only those AGNs from the 3$\pi$ and MDF04 samples, which are detected as variable according to condition \ref{eq:varboth} in at least one of the PS1 bands and have at least one observation in each band. In this way we are unaffected by different numbers of available bands per object. This means that from the samples defined in section \ref{sec:varsample}, we are left with 40 type-1 AGNs from the 3$\pi$ survey and 75 type-1 AGNs from the MDF04 survey, for which we can compute photometric redshifts from the five PS1 bands. In the following photo-z analysis, we focus on the results obtained with the MDF04 sample, since it is almost twice as large as the 3$\pi$ sample, and the sampling pattern of the MDF04 light curves allows for a more thorough investigation of the effects of variability on photo-z calculations. Amongst the 75 AGNs from the MDF04 sample, 72 sources vary in all five PS1 bands, with three sources varying in only three bands. We point out that although our variability detection threshold defined in equation \ref{eq:varboth} corresponds to a $1\sigma$ detection regarding the excess variance, 72 ($g_{\mathrm{P1}}$), 72 ($r_{\mathrm{P1}}$), 72 ($i_{\mathrm{P1}}$), 71 ($z_{\mathrm{P1}}$), and 61 ($y_{\mathrm{P1}}$) of the 75 sources satisfy $\sigma_{\mathrm{rms}}^{2}-3 err(\sigma_{\mathrm{rms}}^{2})>0$. The redshift distribution of these sources is shown in Fig. \ref{fig:histozmdf}. For comparison reasons a clean sample of non-varying AGNs would be very useful. We stress, however, that we are unable to define such a sample because the vast majority of our AGNs vary in at least one band, and the few non-varying sources lacking photometry in several bands.

\section{Photometric redshifts of variable AGNs}
\label{sec:photoz}

\subsection{Multiband data}
\label{sec:bands}

To study the effects of AGN optical variability on the calculation of photometric redshifts in detail, we used the PSF photometry of the five broad band PS1 filters for which we have variability information. We determined photometric redshifts obtained with just these five PS1 bands to estimate the achievable photo-z accuracy for a photometry set consisting exclusively of wavelength bands that show strong variability. 

In addition, we derived photometric redshifts by extending the photometry set with the near-UV (NUV) and far-UV (FUV) bands of the Galaxy Evolution Explorer (GALEX) and the IRAC1/IRAC2 mid-infrared (MIR) Spitzer bands. We used the GALEX-COSMOS catalogue of \citet{2007ApJS..172..468Z}, which provides de-blended, PSF-fitted NUV, and FUV magnitudes in the AB system, to find the nearest object within 0.25 arcsec to the COSMOS coordinates of each of our sources. Among the 75 AGNs from the MDF04 sample, five objects lack GALEX photometry. Regarding the IRAC photometry \citep{2007ApJS..172...86S,2010ApJ...709..644I} we searched for the closest counterpart within 1.0 arcsec to the optical coordinates of each of our objects. However, only five IRAC counterparts deviate by more than 0.25 arcsec from the corresponding optical coordinates. From the COSMOS-IRAC catalogue, we then extract the 1.9 arcsec aperture fluxes of the IRAC1/IRAC2 bands and obtained total fluxes by dividing the aperture fluxes by 0.765 (IRAC1) and 0.740 (IRAC2), following the instructions given in the readme file attached to the catalogue (see also \citet{2005AAS...207.6301S}). The total fluxes were finally transformed from $\mu$Jy to AB magnitudes according to $mag_{\mathrm{AB}}=-2.5\log F_{\mathrm{tot}}+23.9$. The total wavelength coverage of the used bands is listed in Table \ref{tab:multibands}.  
\begin{table}
\caption{Photometric coverage used for the redshift computation.}
\centering
\begin{tabular}{cccc}
\hline\hline
Filter & Telescope & $\lambda_{\mathrm{eff}}$ & FWHM \\
& & $\left(\mathrm{\AA}\right)$ & $\left(\mathrm{\AA}\right)$ \\
\hline
FUV & GALEX & 1546      & 234 \\   
NUV & GALEX & 2345      & 795 \\
$g_{\mathrm{P1}}$ & PS1 & 4900  & 1149 \\   
$r_{\mathrm{P1}}$ & PS1 & 6241  & 1398 \\
$i_{\mathrm{P1}}$ & PS1 & 7564  & 1292 \\
$z_{\mathrm{P1}}$ & PS1 & 8690  & 1039 \\
$y_{\mathrm{P1}}$ & PS1 & 9645  & 665 \\
IRAC1 & Spitzer & 35634 & 7412 \\
IRAC2 & Spitzer & 45110 & 10113 \\
\hline
\end{tabular}
\tablefoot{Values calculated from the transmission curves with the LePhare code.}
\label{tab:multibands}
\end{table}          

\subsection{Fitting technique}
\label{sec:prep}

The SED fitting is realized with the publicly available LePhare code \citep{1999MNRAS.310..540A,2006A&A...457..841I}, which performs a $\chi^{2}$ minimization, comparing the observed flux with the template flux in each band to determine the most likely redshift, SED template, and intrinsic extinction. When aiming to calculate redshifts for AGNs, it is of primary importance to utilize a library of SED templates covering the variety of possible superpositions of the AGN and host galaxy emission components. To account for this, we use the well-tested model set employed in \citet{2009ApJ...690.1250S,2011ApJ...742...61S} (hereafter termed S09 and S11, respectively). This library comprises hybrid templates with varying contributions (90:10, 80:20,...,20:80, 10:90) of several host galaxy types and different types of AGNs (type-1, type-2, QSO1, QSO2). These templates are described in detail in S09. Since all of our sources are luminous point-like type-1 AGNs, we only consider the nine templates with a significant QSO-fraction, listed in Table \ref{tab:sedtempl}. This comparably small number of models helps to reduce degeneracy between templates and redshifts. Moreover, since we are driving the fitting routine towards QSO-dominated templates, we minimize the color-redshift degeneracy that is produced by AGNs and galaxies occupying similar regions in color space for certain redshifts \citep{2002AJ....123.2945R,2004A&A...421..913W}. 

\begin{table}
\caption{Template SEDs used in this work.}
\centering
\begin{tabular}{cc}
\hline\hline
Model ID & Model name \\
\hline
1 & I22491{\_}60{\_}TQSO1{\_}40 \\
2 & I22491{\_}50{\_}TQSO1{\_}50 \\
3 & I22491{\_}40{\_}TQSO1{\_}60 \\
4 & pl{\_}I22491{\_}30{\_}TQSO1{\_}70 \\ 
5 & pl{\_}I22491{\_}20{\_}TQSO1{\_}80 \\
6 & pl{\_}I22491{\_}10{\_}TQSO1{\_}90 \\
7 & pl{\_}QSOH \\
8 & pl{\_}QSO \\
9 & pl{\_}TQSO1 \\    
\hline
\end{tabular}
\tablefoot{Same model names as in Table 2 of S09.}
\label{tab:sedtempl}
\end{table} 
       
To account for Galactic extinction, we corrected each of our photometric measurements in the optical PS1 bands by the corresponding total absorption $A_{\lambda}$ in magnitudes. The extinction values were obtained from the NASA/IPAC Extragalactic Database (NED) and are based on the extinction maps of \citet{2011ApJ...737..103S}. The GALEX photometry was corrected for Galactic extinction by subtracting $8.612\times 0.0167$ from the NUV magnitudes and $8.290\times 0.0167$ from the FUV magnitudes, respectively. These $A_{\lambda}$ values were calculated with the LePhare code using the Galactic extinction law of \citet{1989ApJ...345..245C} as a function of color excess $E\left(B-V\right)$, assuming $A_{\mathrm{V}}=R_{\mathrm{V}}\times E\left(B-V\right)$ with $R_{\mathrm{V}}=3.1$. We did not perform a Galactic extinction correction for the IRAC bands, because the extinction in the MIR wavelength range is typically much less than the photometric errors of the observations. 

To obtain a representative library of expected intrinsic SEDs in the AB photometric system of the PS1 bands, we performed the following steps. First we multiply the template SEDs with the filter transmission curves of the used bands and integrate over the wavelength range covered by the latter. Then the SEDs are redshifted within a range of $z=0.02$--5, applying a bin size of $\Delta z=0.01$. Subsequently, we create a grid of redshift and host extinction values by allowing for a range of $E\left(B-V\right)$ values between 0 and 0.5 with steps of 0.05 to take care of the intrinsic reddening caused by the AGN host galaxy. For the latter we apply the SMC extinction law of \citet{1984A&A...132..389P}, which was found to produce the best photo-z results for the XMM-COSMOS sources in S09. The $\mathrm{Ly_{\alpha}}$ absorption produced by the intergalactic medium (IGM) is considered to depend on redshift according to \citet{1995ApJ...441...18M}. 

Following S09 we additionally use a luminosity prior by allowing only absolute magnitudes within $-20>M_{\mathrm{g_{P1}}}>-30$ to prevent unreasonable combinations of luminosity and redshift. Considering that quasars typically have absolute magnitudes of $M_{\mathrm{B}}\leq -23$, this prior is suitable for our sample of luminous type-1 AGNs \citep{2009ApJ...690.1250S}. Finally we add 0.05 mag in quadrature to the individual errors of the optical PS1 band measurements to avoid values with underestimated errors getting too much weight during the fitting process.

Throughout this work we quantify the accuracy of the photo-z in terms of the normalized median absolute deviation (NMAD) \citep{1983ured.book.....H}, defined as    
\begin{flalign}
        \label{eq:sigmanmad}
        \sigma_{\mathrm{NMAD}} = 1.48\times\mathrm{median}\,\frac{|z_{\mathrm{phot}}-z_{\mathrm{spec}}|}{1+z_{\mathrm{spec}}}.
\end{flalign}  
Here, $z_{\mathrm{phot}}$ is the newly computed photometric redshift and $z_{\mathrm{spec}}$ the known spectroscopic redshift (or high quality photometric redshift from S09, S11), respectively. Since this statistic is based on the median, catastrophic outliers do not strongly affect the quoted accuracy. The fraction of objects we consider as not fitted with the correct redshift is evaluated according to  
\begin{flalign}
        \label{eq:eta}
        \eta = \mathrm{fraction}\,\mathrm{of}\,\mathrm{objects}\,\mathrm{with}\,\,\frac{|z_{\mathrm{phot}}-z_{\mathrm{spec}}|}{1+z_{\mathrm{spec}}} > 0.15.
\end{flalign}  
The quantity $\eta$, defined in this way, is usually referred to as the fraction (or percentage) of outliers.
 
\subsection{Selection of input photometry} 
\label{sec:inputphot}

Considering that we are dealing with strongly varying sources, with the vast majority showing variability in multiple PS1 bands, it is paramount to appropriately select the photometry of the different bands for the fitting procedure. Ideally the set of multiband photometry should be obtained from simultaneous observations in each band, allowing the spectrum to be fit with a snapshot SED. However, simultaneous multiband observations are often not available and thus the photometry for a specific object must be collected from several epochs, possibly introducing biases due to variability. The non-simultaneous five-band observations of the PS1 surveys provide an ideal test bed for studying the effects of multiband variability on the calculation of photometric redshifts in view of the achievable accuracy and fraction of outliers.

To address this question we apply three different kinds of input photometry for our fitting routine. First we try to get as close as possible to the realization of a snapshot SED by choosing those photometric measurements from the five PS1 band light curves with minimum relative distance in observing time. To determine this set of light curve points, a combinatoric procedure is adopted, which is described in detail in Appendix \ref{sec:appendixb}. In the following this set of input photometry is referred to as Case A. We point out that due to the extreme sparse sampling of the $3\pi$ light curves, the minimal time interval between the observations, $\Delta T_{\mathrm{min}}$, is typically four to five months but can be even longer than one year for our sources, which represents a very bad approximation of a snapshot SED. In contrast, the MDF04 light curves of our sample allow collection of the five-band photometry within 2.5 days at least, and for the majority of our objects $\Delta T_{\mathrm{min}}<1.2$ days. 

Another reasonable approach is to choose the median magnitude of each filter light curve as input photometry, giving a "typical" light curve value that is insensitive to outliers. Taking the median resembles the common procedure for using stack photometry for the photo-z computation in order to obtain deeper data. In this case (hereafter termed Case B), we assign a "typical" photometric error from the light curve, given by $median\left(err\left(mag\right)\right)$ to each of the five $median\left(mag\right)$ values. Finally, to mimic the situation where the set of input photometry can only be obtained by collecting the filter observations from different epochs with large temporal gaps, we randomly select the light curve points in each band. For this purpose we draw integer values from the uniform distribution out of the $i=1, 2,..., N$ light curve points for each band. To allocate a representative photometry set for this case (hereafter Case C), we create ten different random realizations. Table \ref{tab:3cases} summarizes the different kinds of input photometry used in this work.
\begin{table}
\caption{Different cases of input photometry studied in this work.}
\centering
\begin{tabular}{cc}
\hline\hline
Case & Description \\
\hline
A & values with minimum temporal distance  \\
B & median magnitude of each light curve  \\
C & randomly chosen values (10 realizations)  \\     
\hline
\end{tabular}
\label{tab:3cases}
\end{table}     
 
\section{Results} 
\label{sec:results}
\subsection{Redshift accuracy using PS1 photometry}
\label{sec:ps1only}

\begin{figure*}
\centering
\subfloat{%
        \includegraphics[width=0.4\textwidth]{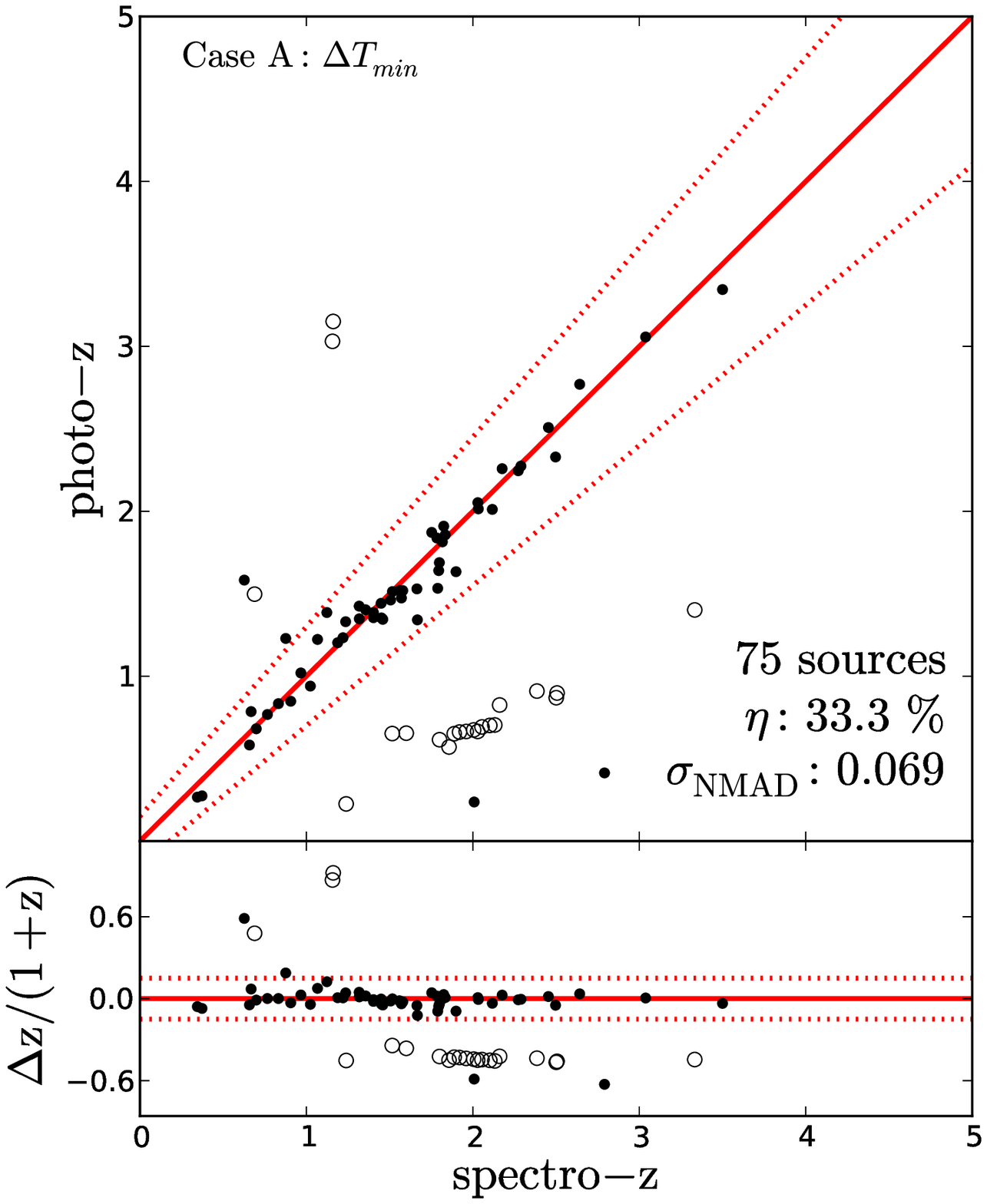}}
\quad
\subfloat{%
        \includegraphics[width=0.4\textwidth]{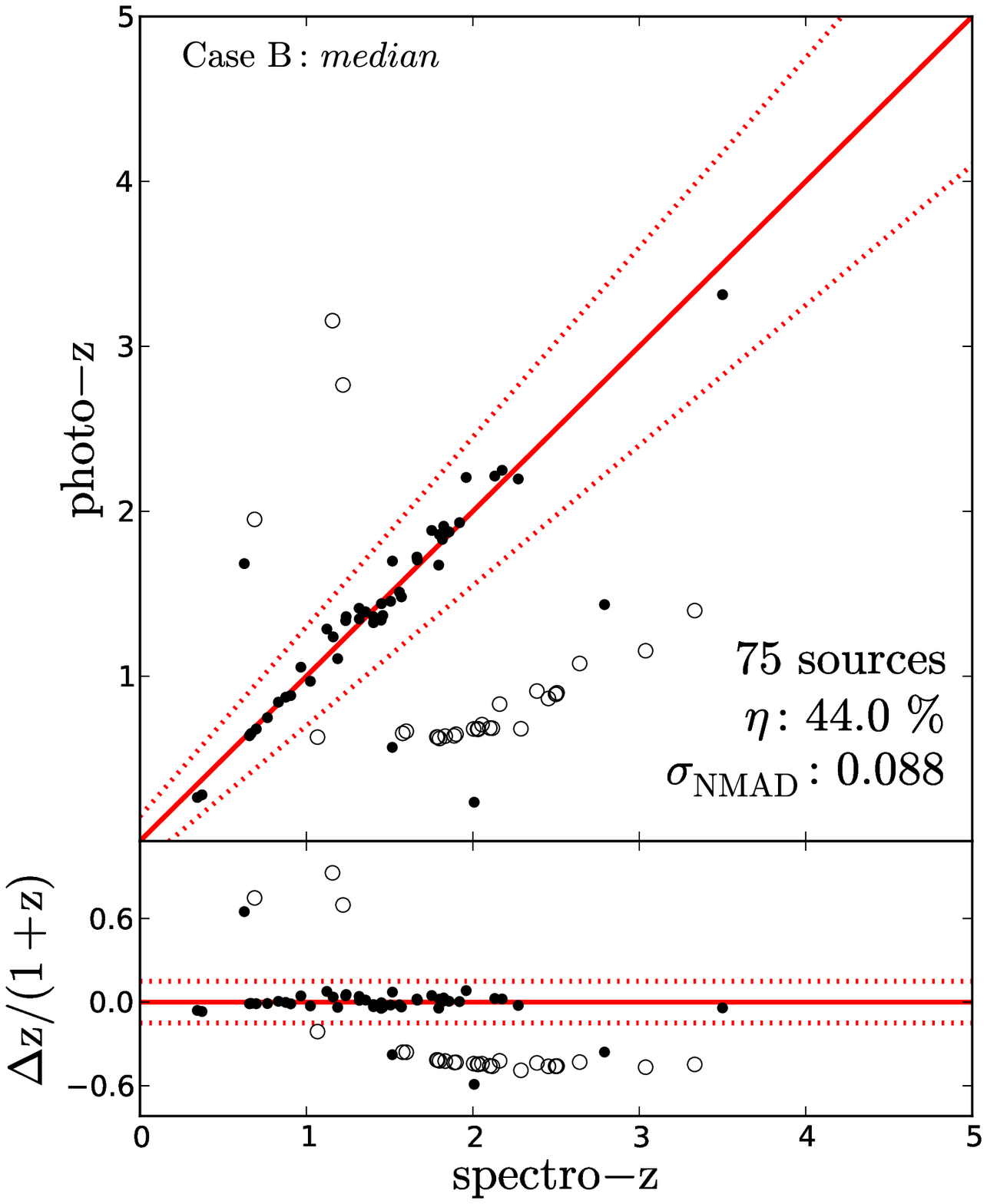}}

\subfloat{%
        \includegraphics[width=0.4\textwidth]{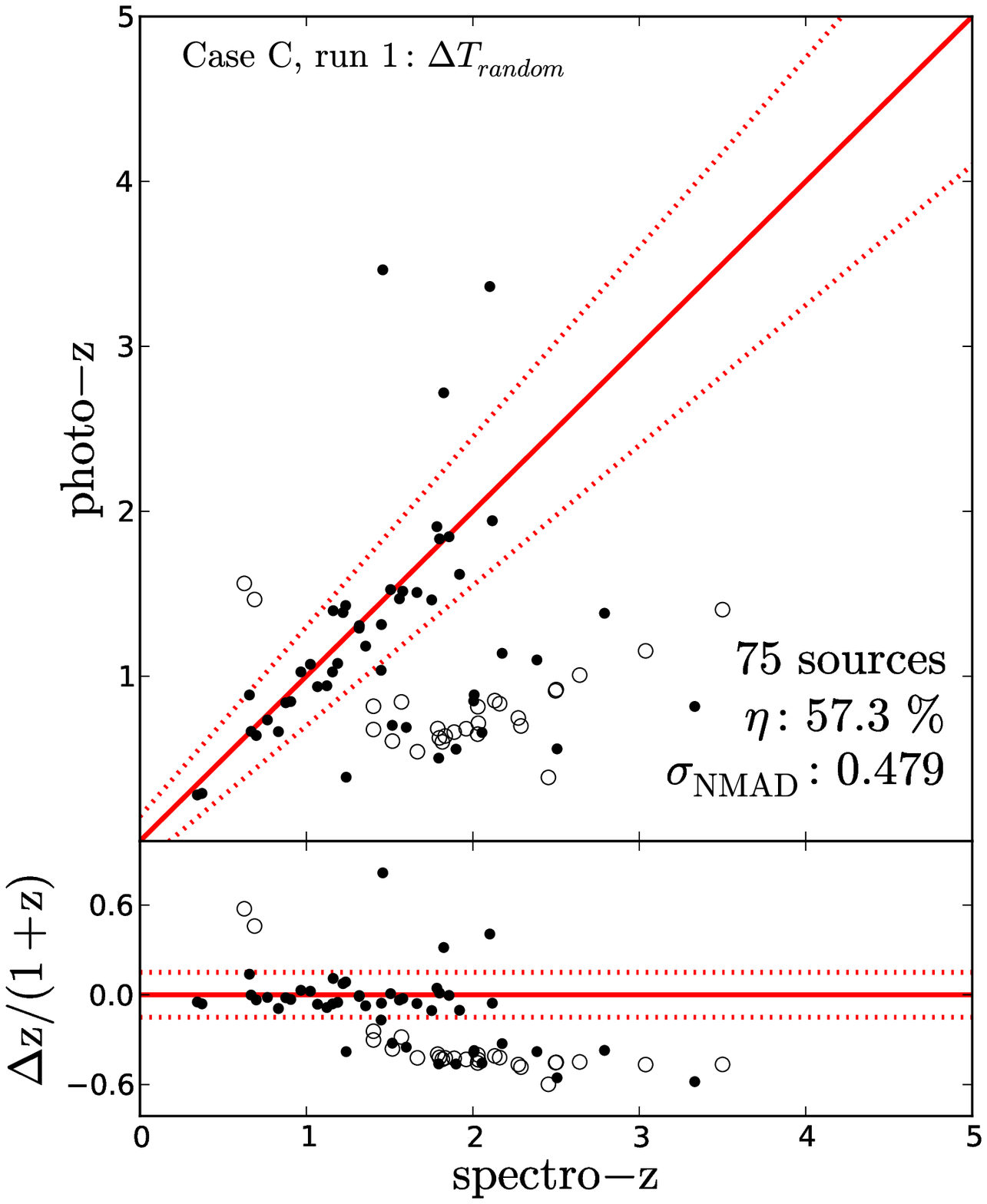}}
\quad
\subfloat{%
        \includegraphics[width=0.4\textwidth]{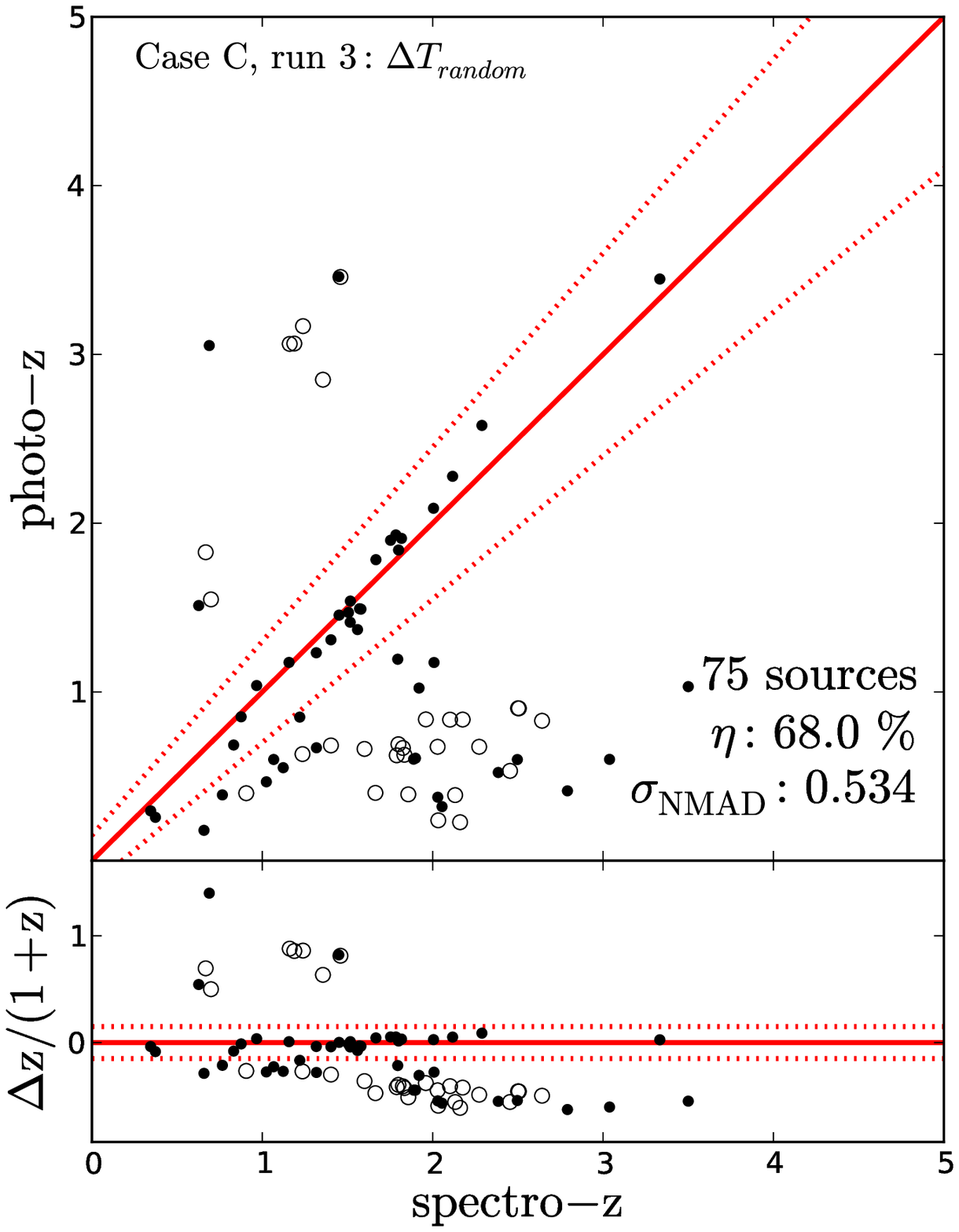}}

\caption{Comparison of photometric and spectroscopic redshifts for the 75 AGNs of the MDF04 sample. Empty circles represent sources for which the second peak of the redshift probability distribution agrees with the correct redshift. The solid line represents the one-to-one relation, and the dotted lines correspond to $z_{\mathrm{phot}}=z_{\mathrm{spec}}\pm 0.15\left(1+z_{\mathrm{spec}}\right)$.}
\label{fig:zresultps1}
\end{figure*}

Following the procedure outlined in section \ref{sec:prep} we determine photometric redshifts for the input photometry sets of Case A, Case B, and Case C using only the five-band PS1 photometry. The results in terms of accuracy and fraction of outliers for the MDF04 sample are summarized in Table \ref{tab:zresultps1} and Fig. \ref{fig:zresultps1}, plotting for each case the photometric redshift versus the redshift listed in the catalogues of \citet{2010ApJ...716..348B}, which was updated in \citet{2011ApJ...742...61S}. The results for each of the ten runs of Case C are listed in Table \ref{tab:caseCruns}. It is clear that irrespective of the AGN variability, we do not expect to obtain very accurate photo-z results using only the five PS1 bands. However, the setup is the same for all input photometry sets and regarding the comparison of the three cases, only the relative photo-z quality is important for dissecting variability effects. The numbers show that Case A outperforms the other cases in terms of accuracy and fraction of outliers. This result is perhaps not totally surprising considering the fact that Case A resembles a snapshot SED  and should therefore be the least affected by multiband variability. What is more, we stress that we find strong evidence that the randomly selected input photometry of Case C produces by far the worst photo-z results for a set of variable AGNs. As can be seen from Table \ref{tab:caseCruns}, the fraction of outliers and $\sigma_{\mathrm{NMAD}}$ is much larger than the related values of Cases A and B for every of the ten runs of Case C.
\begin{table}
\caption{Assessing photo-z quality for the MDF04 sample using PS1 photometry.}
\centering
\begin{tabular}{cccccccc}
\hline\hline
& \multicolumn{2}{c}{Case A} & \multicolumn{2}{c}{Case B} & \multicolumn{2}{c}{Case C} \\

        & $\eta\,(\%)$ & $\sigma_{\mathrm{NMAD}}$
        & $\eta\,(\%)$ & $\sigma_{\mathrm{NMAD}}$
        & $\eta\,(\%)$ & $\sigma_{\mathrm{NMAD}}$ \\
        \hline
        & 33.3 & 0.069 & 44.0 & 0.088 & 57.2 & 0.400 & \\ 
        \hline
\end{tabular}
\tablefoot{Quoted values of Case C are the average values of the ten realizations.}
\label{tab:zresultps1}
\end{table}
\begin{table}
\caption{Results for each of the ten realizations of Case C using PS1 photometry.}
\centering
\begin{tabular}{ccccc}
\hline\hline
 & \multicolumn{3}{c}{Case C} \\
 
    & Run ID & $\eta\,(\%)$ & $\sigma_{\mathrm{NMAD}}$ \\
    \hline
    & 1 & 57.3  &   0.479 & \\
        & 2 & 61.3  &   0.510 & \\
        & 3 & 68.0  &   0.534 & \\
        & 4 & 62.7  &   0.544 & \\
        & 5 & 57.3  &   0.329 & \\      
        & 6 & 58.7  &   0.429 & \\
        & 7 & 54.7  &   0.460 & \\
        & 8 & 49.3  &   0.218 & \\
        & 9 & 52.0  &   0.246 & \\
        & 10 & 50.7  &   0.235 & \\
    \hline
\end{tabular}
\label{tab:caseCruns}
\end{table}

\begin{figure}
\centering
\text{\hspace*{6mm}Case A: $\Delta T_{\mathrm{min}}$}
\subfloat{%
        \includegraphics[width=0.4\textwidth]{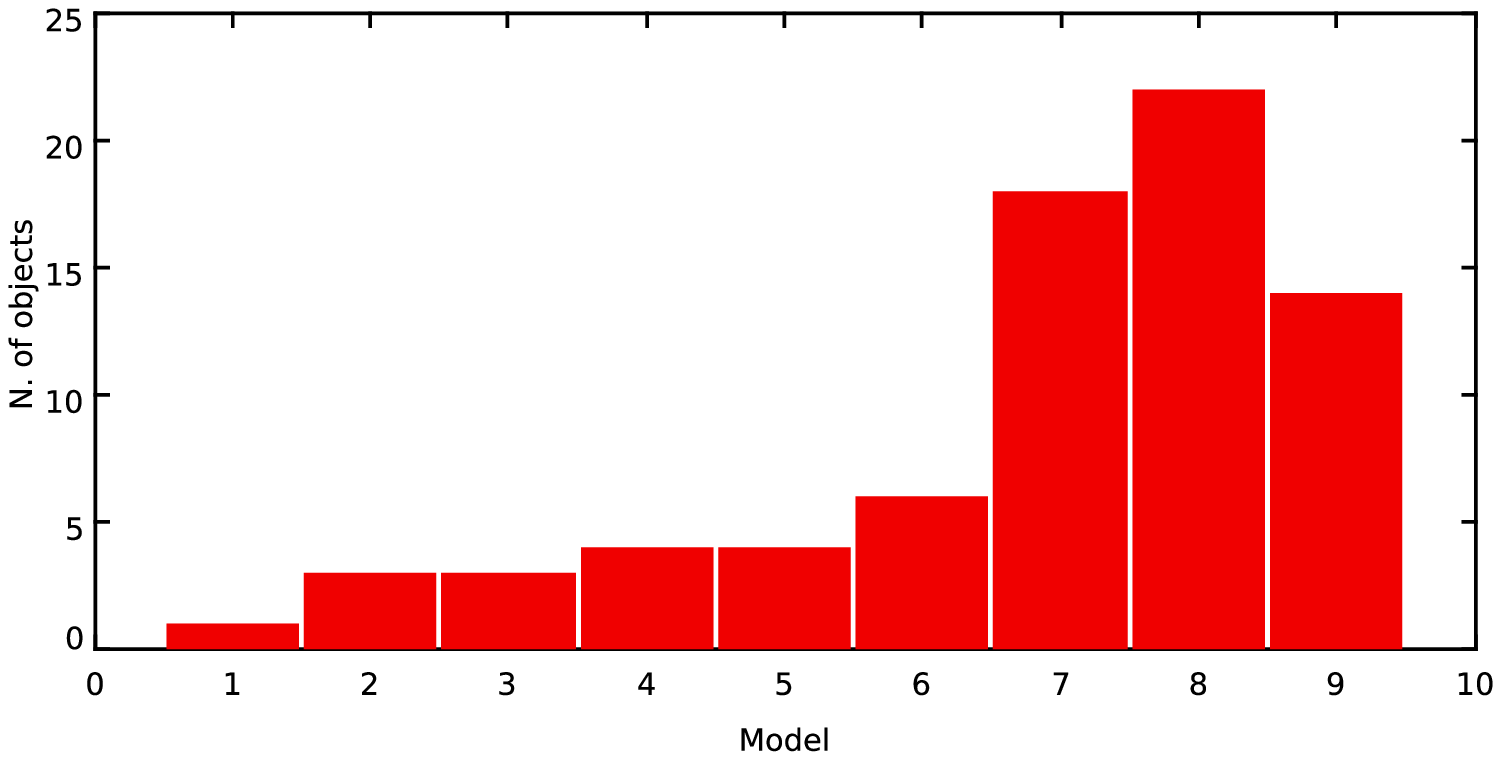}}

\text{\hspace*{6mm}Case B: median}      
\subfloat{%
        \includegraphics[width=0.4\textwidth]{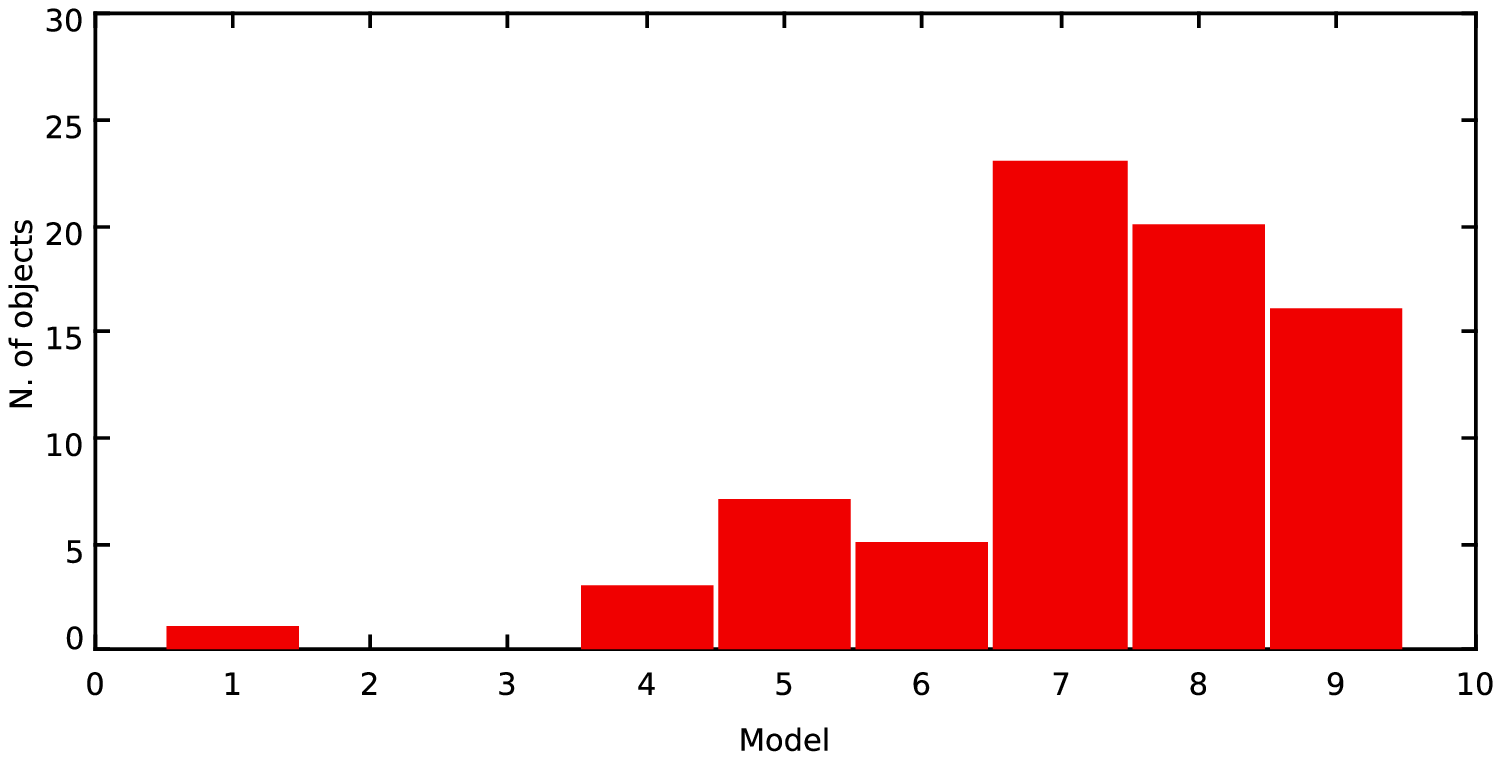}}

\text{\hspace*{6mm}Case C: $\Delta T_{\mathrm{random}}$, run 1} 
\subfloat{%
        \includegraphics[width=0.4\textwidth]{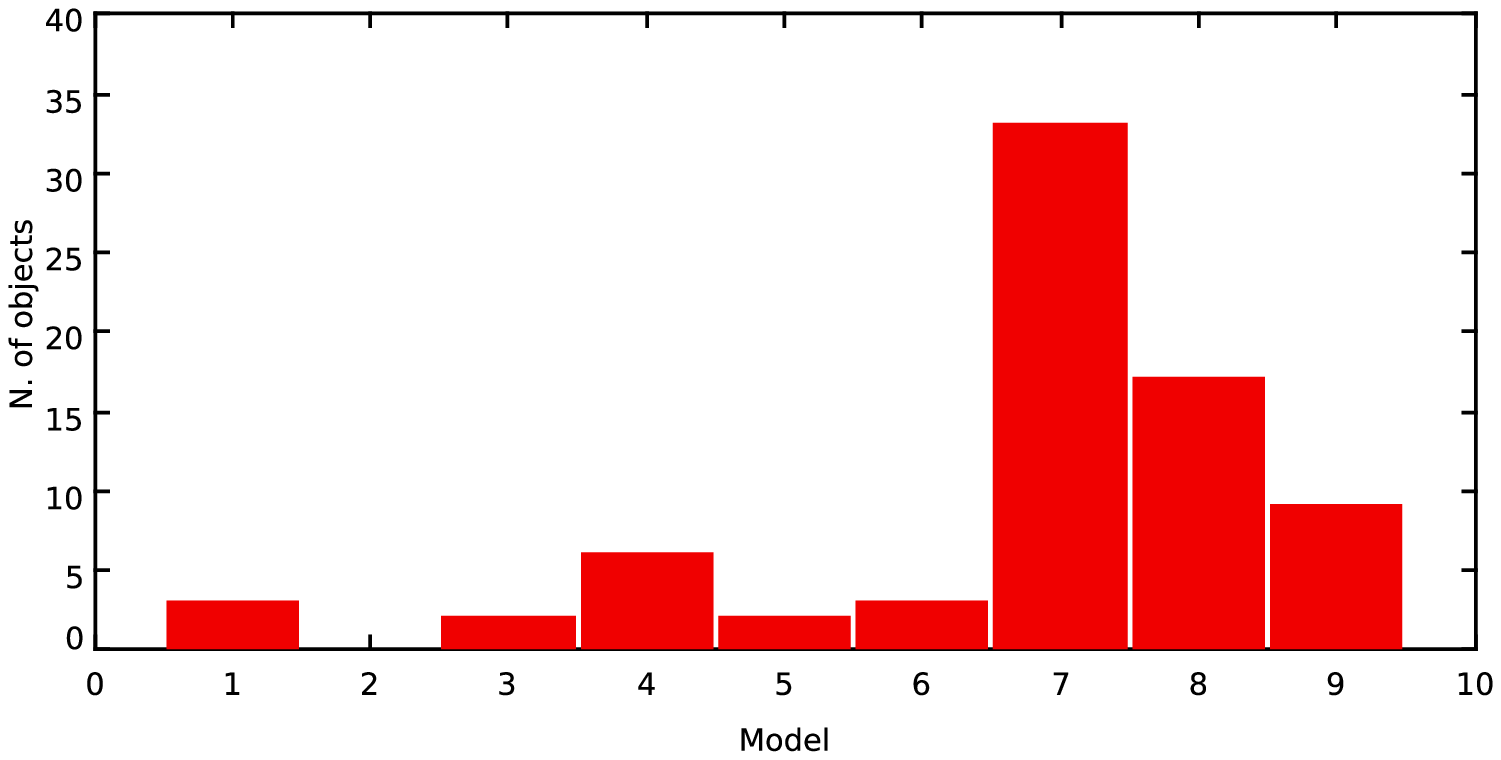}}
\caption{Histograms of the model ID (see Table \ref{tab:sedtempl}) chosen by the fitting routine for Cases A and B and for Run 1 of Case C, respectively.}
\label{fig:histmodelps1}
\end{figure}

Contrasting to this, we also observe similarities between the considered cases. For one, each panel of Fig. \ref{fig:zresultps1} exhibits an exceedingly large number of outliers with greatly underestimated redshifts in the redshift range $1.8<z<2.2$. This systematic failure suggests that our fitting routine is not able to correctly differentiate between the continua of our SEDs with just the five PS1 bands and without any redshift prior in this particular redshift range. To uncover any further dependency with redshift, we adopted a cumulative approach by sorting our sample in ascending order of redshift and subsequently derived redshifts for the first 20, 25, 30, 35, ..., 75 objects. Apart from the strong increase in the outlier fraction in the range $1.8<z<2.2$, however, we find no dramatic evolution of $\eta$ and $\sigma_{\mathrm{NMAD}}$ with redshift significantly biasing the comparison of our photometry sets. Moreover, we stress that the second best model agrees with the correct redshift for 84\%, 88\%, and 60\% of the outliers of Cases A and B and Run 1 of Case C, respectively. Therefore we emphasize that the photometric redshift assigned to a source should always be reviewed considering the value of a possible secondary peak in the redshift probability distribution. Finally, as displayed in Fig. \ref{fig:histmodelps1}, the distribution of the best-fit model templates is very similar for each set of input photometry. The QSO-dominated Models 7, 8, and 9 are preferably chosen for our sources, which is in good agreement with the fact that our sample consists of luminous point-like type-1 AGNs.  

Performing the same experiment with the 40 AGNs of the $3\pi$ sample, we are not able to state significant differences in the photo-z quality by comparing the three studied cases. For all three photometry sets, the percentage of outliers is fairly high with values of about 60\%--65\%, and the typical accuracy is not much better than $\sigma_{\mathrm{NMAD}}\sim 0.400$. As outlined above, the light curves of the $3\pi$ sample do not allow defining a value of $\Delta T_{\mathrm{min}}$ that is less than several months. Since our sources show strong variability on these timescales, the photometry of Case A is biased by multiband variability and therefore gives comparable results to Case C. Although the median photometry of Case B can smooth out variability to some degree, it is clear that a median value calculated from only three to six light curve points might still give a rather poor estimation of the actual light curve "average", hence leading to a much lower photo-z quality than for the MDF04 sample.  

\subsection{Dissecting the photo-z quality differences}

\begin{figure}
\centering
\text{\hspace*{6mm}Case A: $\Delta T_{\mathrm{min}}$}
\subfloat{%
        \includegraphics[width=0.45\textwidth]{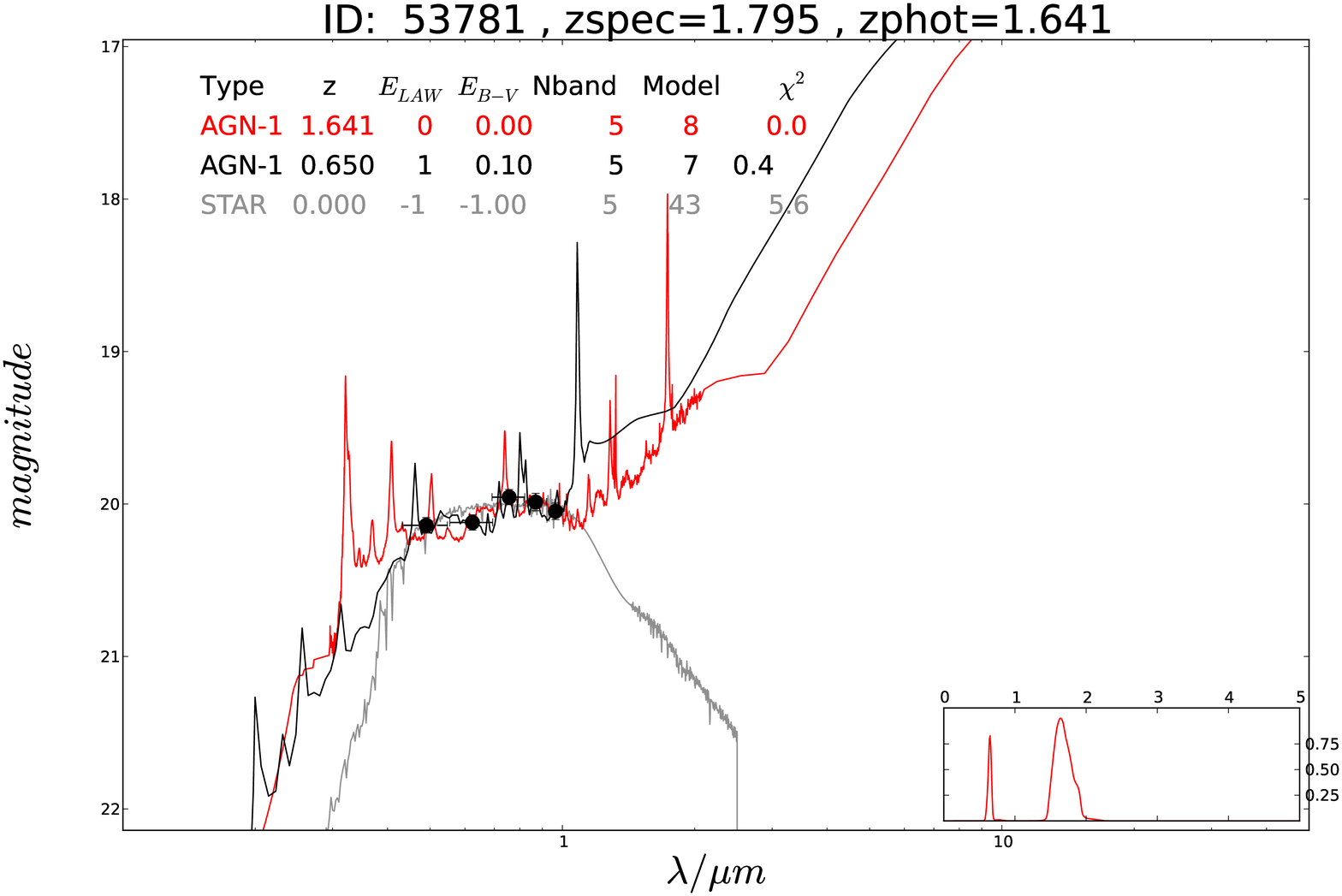}}

\text{\hspace*{6mm}Case B: median}      
\subfloat{%
        \includegraphics[width=0.45\textwidth]{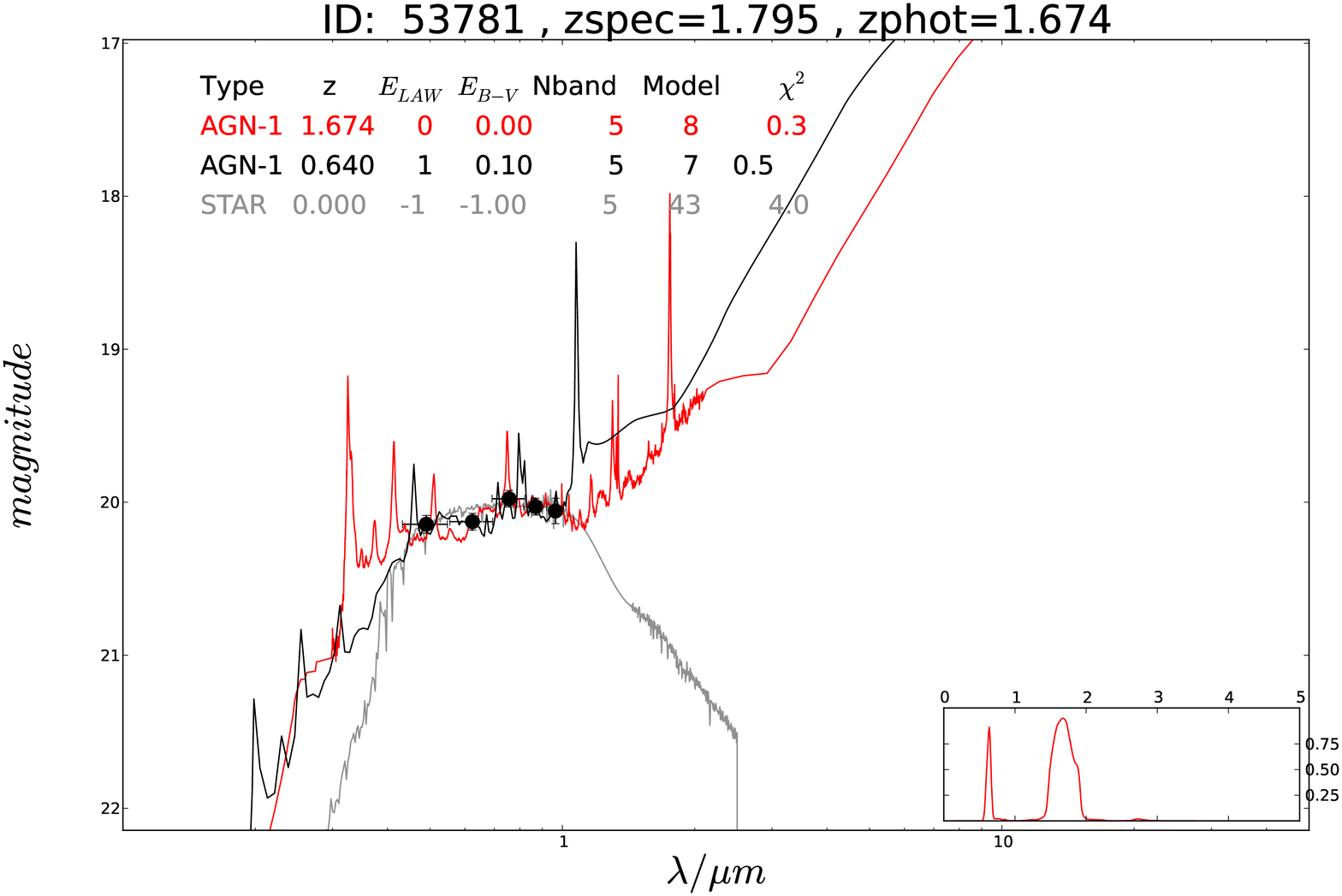}}

\text{\hspace*{6mm}Case C: $\Delta T_{\mathrm{random}}$, run 1} 
\subfloat{%
        \includegraphics[width=0.45\textwidth]{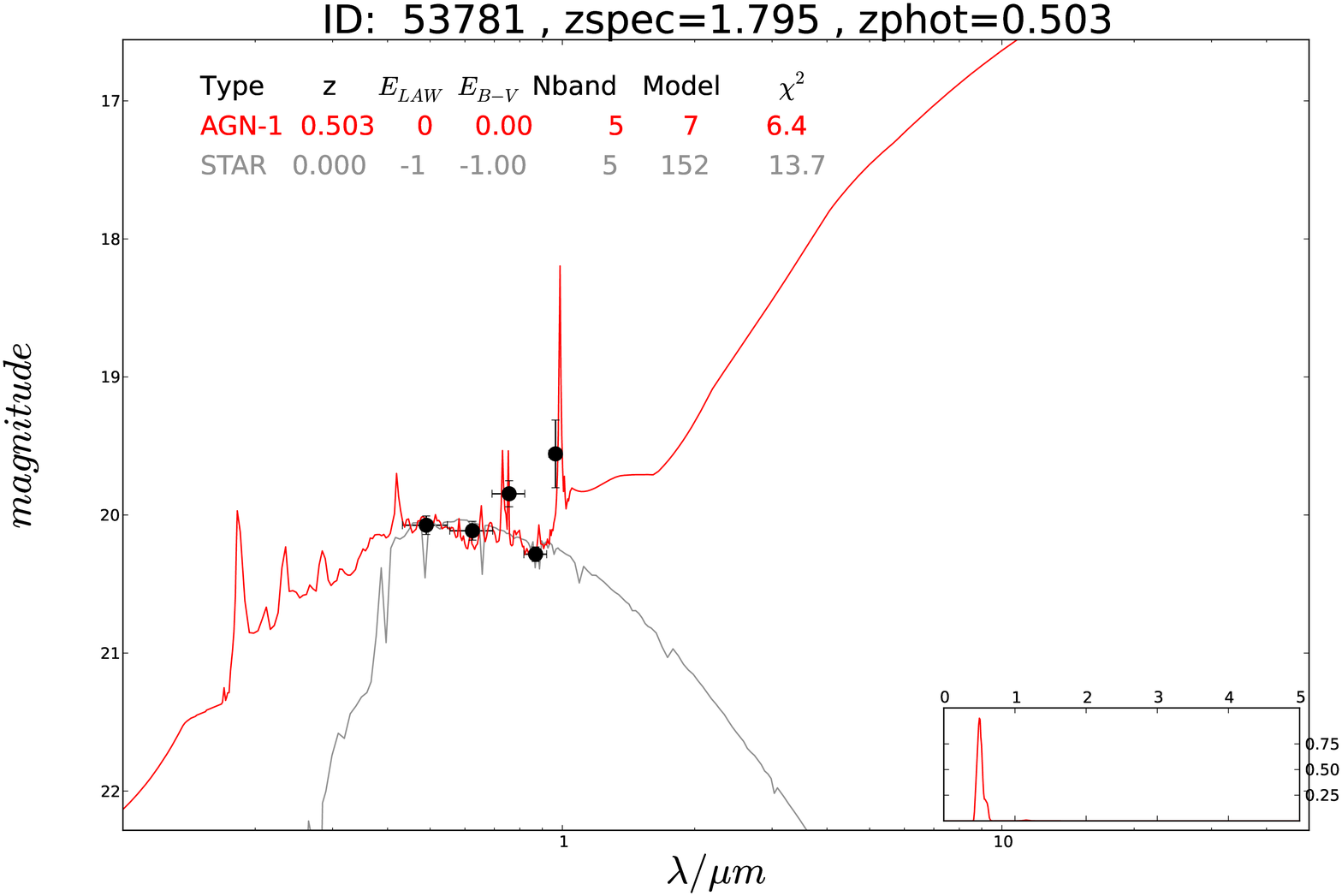}}
\caption{Best fit SED in red and input photometry of the AGN with XID 53781 for Case A (\textit{top panel}) and Case B (\textit{middle panel}) and Run 1 of Case C (\textit{bottom panel}). Some basic information about the model fit is listed in each panel. The redshift probability distribution is shown in the inserts. The black and grey curves are the second-best AGN and best stellar SEDs, respectively.}
\label{fig:trandvstmin}
\end{figure}

To dissect the observed performance differences of our considered photometry sets, we visually inspected the SED fits of those sources that were correctly fitted in one case, but among the outliers in another case. Comparing  Cases A and C in this way, it turns out that the multiband variability of our AGNs causes fatal outliers predominantly for Case C. Owing to the random selection of light curve points, the SED shape implied by the relative positions of the five PS1 bands in magnitude space is deformed in such a way that either the continuum cannot be correctly described, or variability in individual bands mocks "wrong" emission lines. The latter case is visible in Fig. \ref{fig:trandvstmin}, with variability in the $i_{\mathrm{P1}}$ and $y_{\mathrm{P1}}$ bands leading to fatal emission line hits, whereas Case A is not affected by the variability. However, comparing Case A with Case B, it is not possible to state an unambiguous reason that Case A yields slightly better results in terms of accuracy and fraction of outliers than Case B. The detailed inspection of the respective SED fits reveals that the relative positions of the chosen five band photometry values of Cases A and B are very similar and, for many objects, almost indistinguishable. For these sources taking the median magnitudes was sufficient to smooth out multiband variability. 

It seems natural to associate the observed differences in the photo-z quality with the strength of variability of our objects. To test this idea, we selected two subsamples of sources with very large variability amplitude in at least one band and two subsamples of objects showing considerable variability amplitudes in all bands simultaneously. Then we checked separately for Cases A, B, and C, whether these most variable AGNs tend to be among the fraction of outliers. In detail we chose the 17 AGNs with fractional variability $F_{\mathrm{var}}>0.15$ and the 13 AGNs with $\Delta mag>0.8$ in any of the five bands as our two high variability amplitude samples. The two multiband variability samples are represented by the nine AGNs with $F_{\mathrm{var}}>0.1$ and the 13 AGNs with $\Delta mag>0.4$ in all five bands. We find that $\sim$55\%--60\% of the objects of each of the four subsamples are outliers in Case C. In contrast, Cases A and B are less affected, but again they give rise to very similar results with only $\sim$35\%--40\% of the most variable AGNs among the outliers for each of the four considered subsamples.  

\subsection{Adding GALEX/IRAC bands to the PS1 bands}
\label{sec:uvmirresults}

High quality photometric redshifts require multiband observational data with broad wavelength coverage to account for the multitude of spectral components of astronomical objects. Especially regarding spectral fitting of luminous AGNs, it is very important that along with the optical/near-infrared bands, comprising prominent emission lines like $\mathrm{H_{\alpha}}$ and $\mathrm{H_{\beta}}$, mid-infrared bands are available to describe the power law emission component towards longer wavelengths \citep{2012ApJ...748..142D}. In addition, depending on the redshift range, UV bands cover strong spectral lines, such as $\mathrm{Ly_{\alpha}}$, CIV$\,\lambda\mathrm{1549\AA}$, and MgII$\,\lambda\mathrm{2798\AA}$, the drop in flux of the "big blue bump", or the power law component towards shorter wavelengths \citep{2004ApJ...615..135S}. For this reason we add observations in the NUV and FUV GALEX bands, as well as in the IRAC1 and IRAC2 MIR band exposures of the Spitzer space telescope, to our five optical/near-infrared PS1 bands. This allows us to investigate to what extent the photo-z quality increases by applying an enlarged photometry set for our sample of variable AGNs.

\begin{figure*}
\centering
\subfloat{%
        \includegraphics[width=0.4\textwidth]{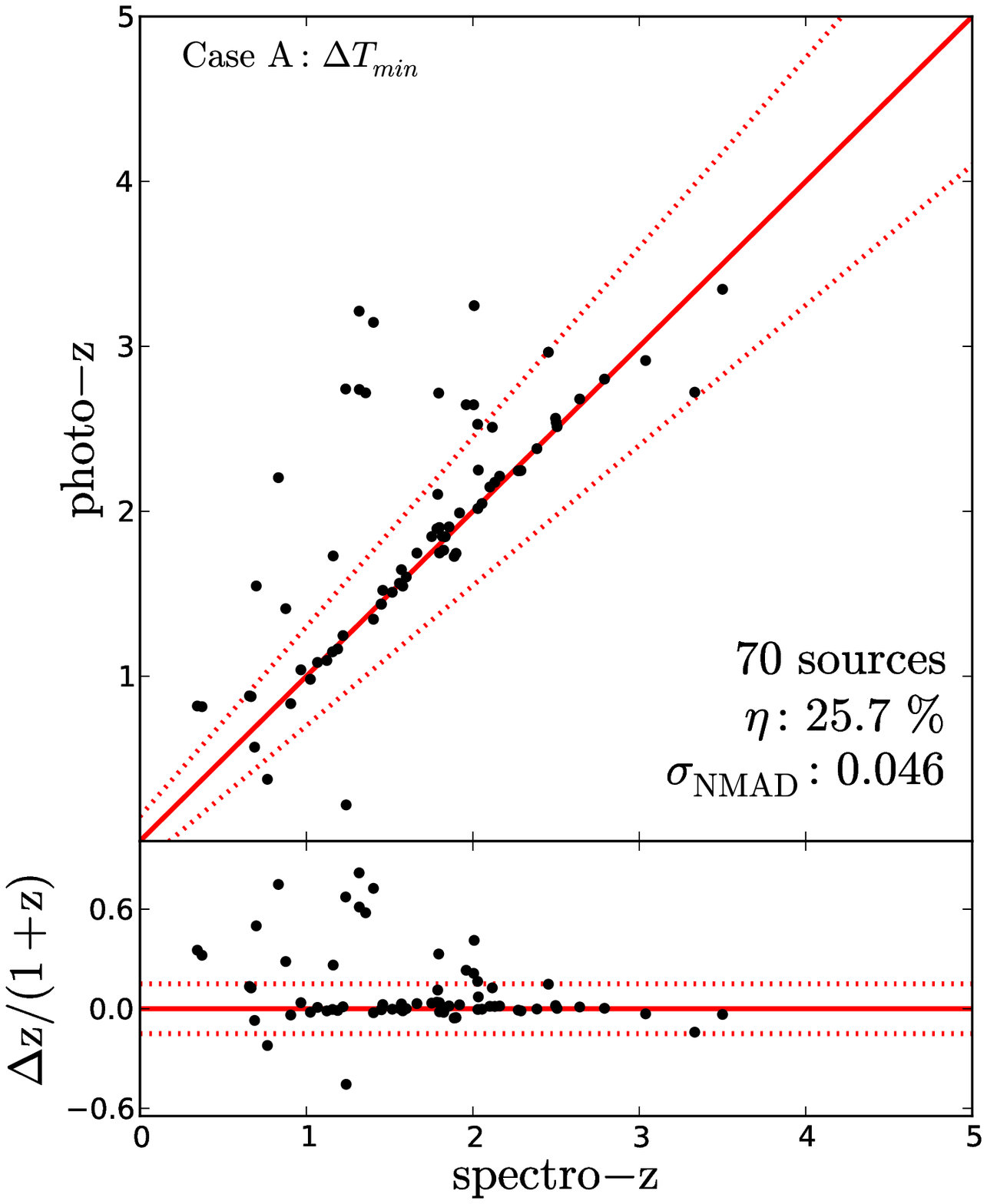}}
\quad
\subfloat{%
        \includegraphics[width=0.4\textwidth]{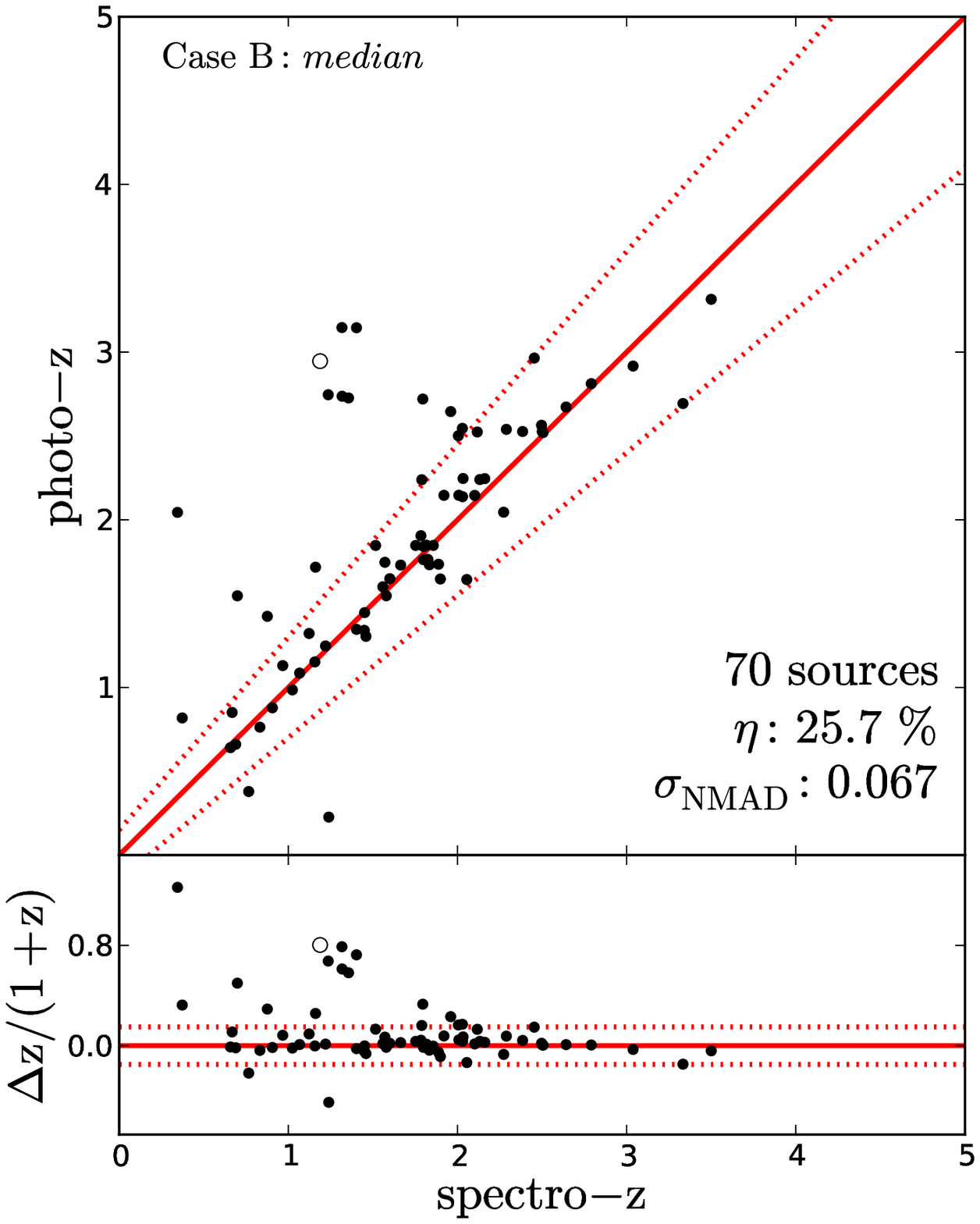}}

\subfloat{%
        \includegraphics[width=0.4\textwidth]{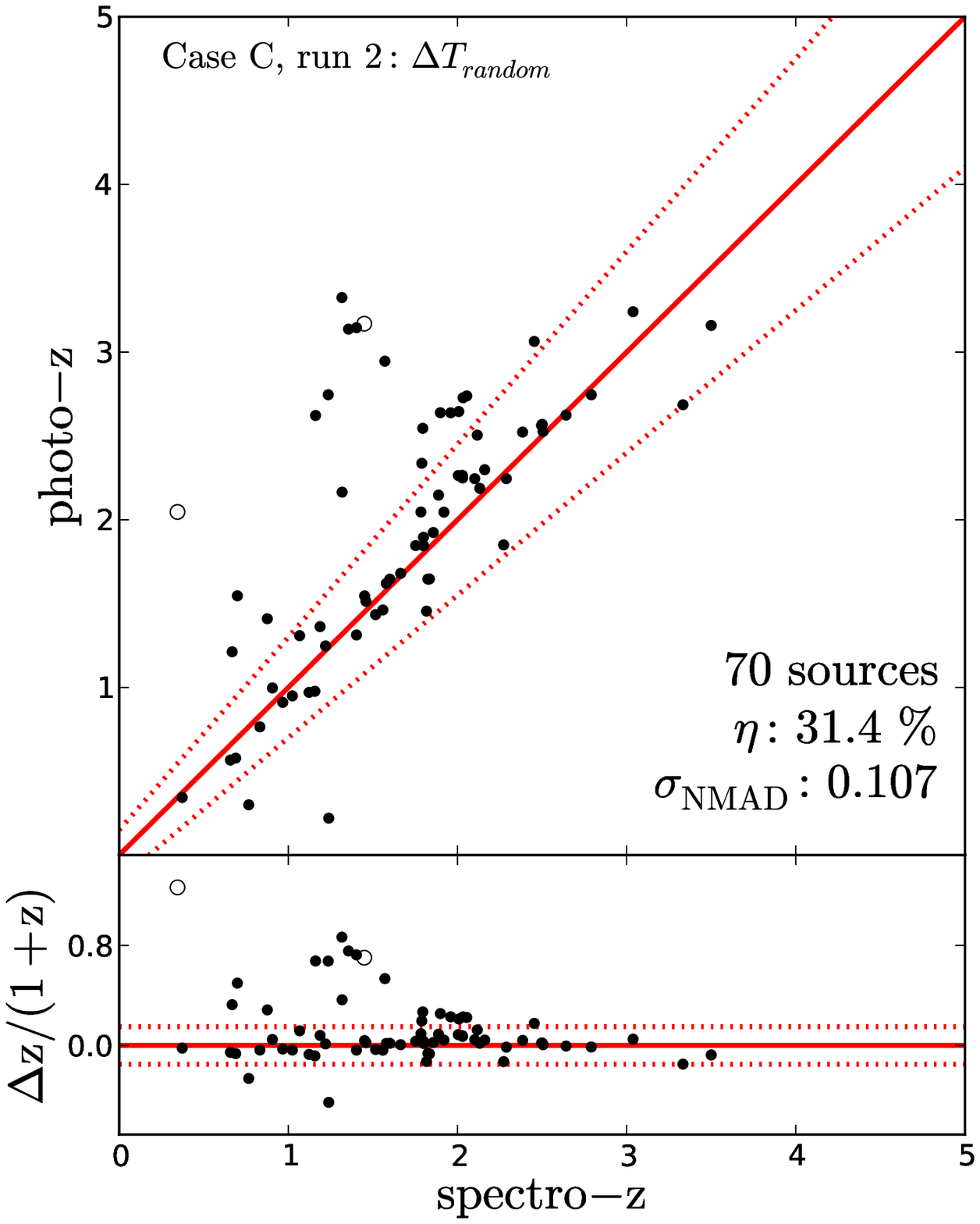}}
\quad
\subfloat{%
        \includegraphics[width=0.4\textwidth]{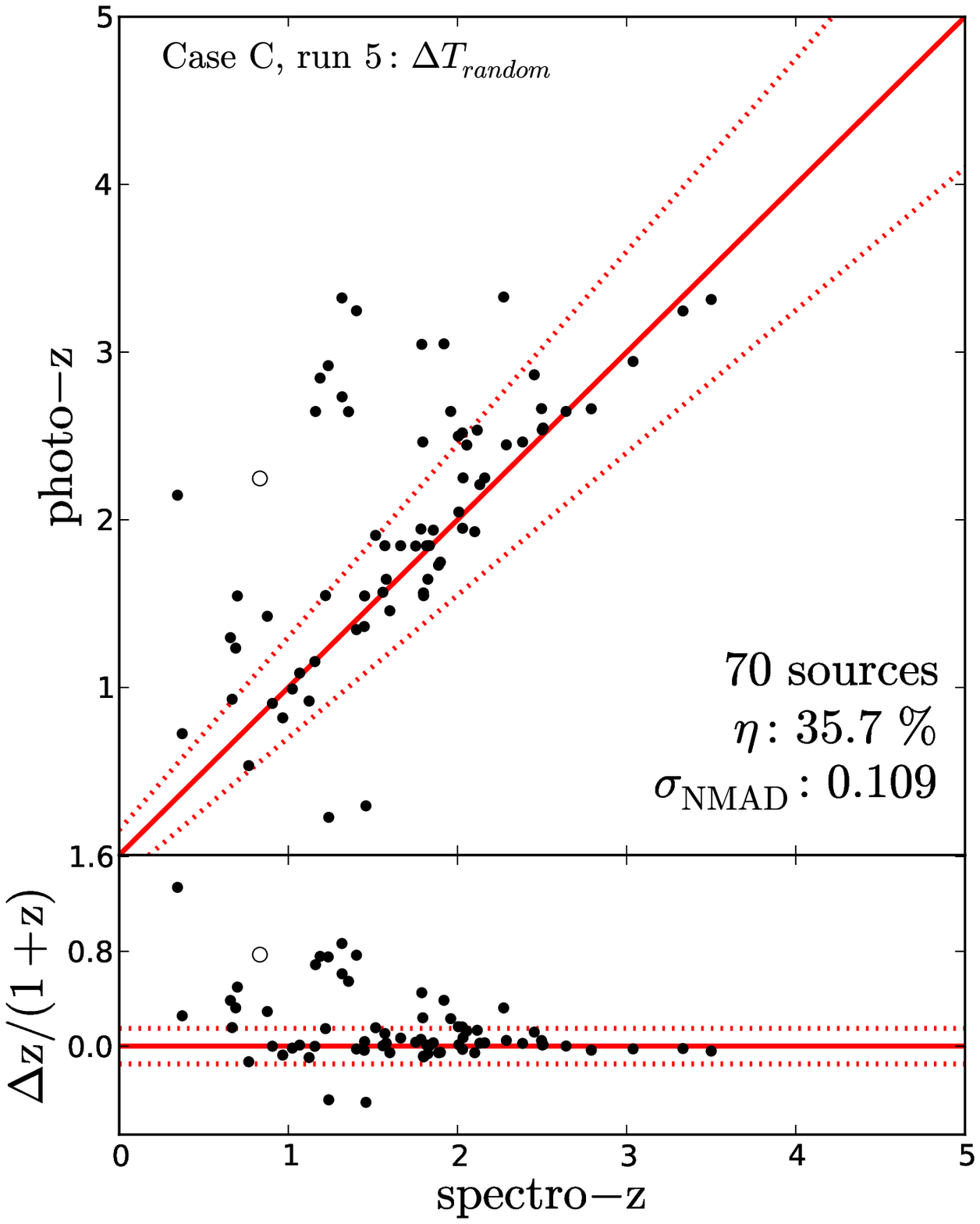}}

\caption{Comparison of photometric and spectroscopic redshifts for the 70 AGNs with GALEX photometry of the MDF04 sample. Empty circles represent sources for which the second peak of the redshift probability distribution agrees with the correct redshift. The solid line represents the one-to-one relation, and the dotted lines correspond to $z_{\mathrm{phot}}=z_{\mathrm{spec}}\pm 0.15\left(1+z_{\mathrm{spec}}\right)$.}
\label{fig:zresultuvps1ir}
\end{figure*}

Using the same PS1 input photometry sets as described in the previous section, only extended by the UV/MIR bands, we ran LePhare to derive photometric redshifts for Cases A, B, and C. Following S09, we add 0.3 mag in quadrature to the NUV/FUV photometry errors and 0.2 mag to the IRAC1/IRAC2 photometry errors. Since magnitude values higher than 25 are not credible for the NUV/FUV bands, we excluded observations fainter than this limit during the fitting procedure. We did not perform a magnitude cut in the IRAC bands, since none of our objects has IRAC magnitudes larger than 20. 
\begin{table}
\caption{Assessing photo-z quality for the MDF04 sample using GALEX, PS1, and IRAC photometry.}
\centering
\begin{tabular}{ccccccc}
\hline\hline
   & \multicolumn{2}{c}{Case A} & \multicolumn{2}{c}{Case B} & \multicolumn{2}{c}{Case C} \\

        & $\eta\,(\%)$ & $\sigma_{\mathrm{NMAD}}$ 
        & $\eta\,(\%)$ & $\sigma_{\mathrm{NMAD}}$ \\
        \hline
        & 25.7 & 0.046 & 25.7 & 0.067 & 31.9 & 0.108 \\ 
        \hline
\end{tabular}
\tablefoot{The quoted values of Case C are the average values of the ten realizations.}
\label{tab:zresultuvps1ir}
\end{table}
\begin{table}
\caption{Results for each of the ten realizations of Case C using GALEX, PS1, and IRAC photometry.}
\centering
\begin{tabular}{ccccc}
\hline\hline
 & \multicolumn{3}{c}{Case C} \\
 
    & Run ID & $\eta\,(\%)$ & $\sigma_{\mathrm{NMAD}}$ \\
    \hline
    & 1 & 32.9  &   0.143 & \\
        & 2 & 31.4  &   0.107 & \\
        & 3 & 34.3  &   0.112 & \\
        & 4 & 30.0  &   0.107 & \\
        & 5 & 35.7  &   0.109 & \\      
        & 6 & 34.3  &   0.113 & \\
        & 7 & 31.4  &   0.103 & \\
        & 8 & 25.7  &   0.113 & \\
        & 9 & 34.3  &   0.059 & \\
        & 10 & 28.6  &   0.115 & \\
    \hline
\end{tabular}
\label{tab:caseCrunsuvmir}
\end{table}
The results for each case are summarized in Fig. \ref{fig:zresultuvps1ir} and Table \ref{tab:zresultuvps1ir}. Table \ref{tab:caseCrunsuvmir} contains the results for each of the ten realizations of Case C. Obviously, adding the GALEX/IRAC bands to the five PS1 bands improves the photo-z quality for all considered cases. Especially for Case C, the fraction of outliers decreased by almost a factor of two and $\sigma_{\mathrm{NMAD}}$ by almost a factor of four. As before, Case A yields the best results in terms of accuracy and fraction of outliers; however, Case B is only outperformed by Case A by the lower $\sigma_{\mathrm{NMAD}}$ value. The superior accuracy of Case A is also clearly apparent in Fig. \ref{fig:zresultuvps1ir}, with the correctly fitted redshifts located very close to the one-to-one relation. On the whole, the differences between all considered cases are considerably reduced as compared to the results without the GALEX/IRAC bands. Even though the redshift accuracy increased for all considered cases, one might expect that adding UV/MIR bands would lead to even greater improvement of the photo-z quality. However, our sample consists of strongly varying sources, and although variability in the IRAC bands may be negligible, it is very likely that variability in the GALEX bands affects our results. Since we do not have GALEX light curves, we cannot properly correct for variability, and adding 0.3 mag in quadrature may only partly alleviate the problem. Finally all four panels of Fig. \ref{fig:zresultuvps1ir} reveal that the enhanced number of outliers in the redshift range $1.8<z<2.2$ observed in Fig. \ref{fig:zresultps1} disappears, when using the additional information provided by the UV/MIR bands. Furthermore, there are no secondary peaks in the redshift probability distribution for all objects of Case A, which is also true for 74 and 73 objects of the 75 sources of Case B and Run 2 of Case C, respectively.

\begin{figure}
\centering
\text{\hspace*{6mm}Case A: $\Delta T_{\mathrm{min}}$}
\subfloat{%
        \includegraphics[width=0.45\textwidth]{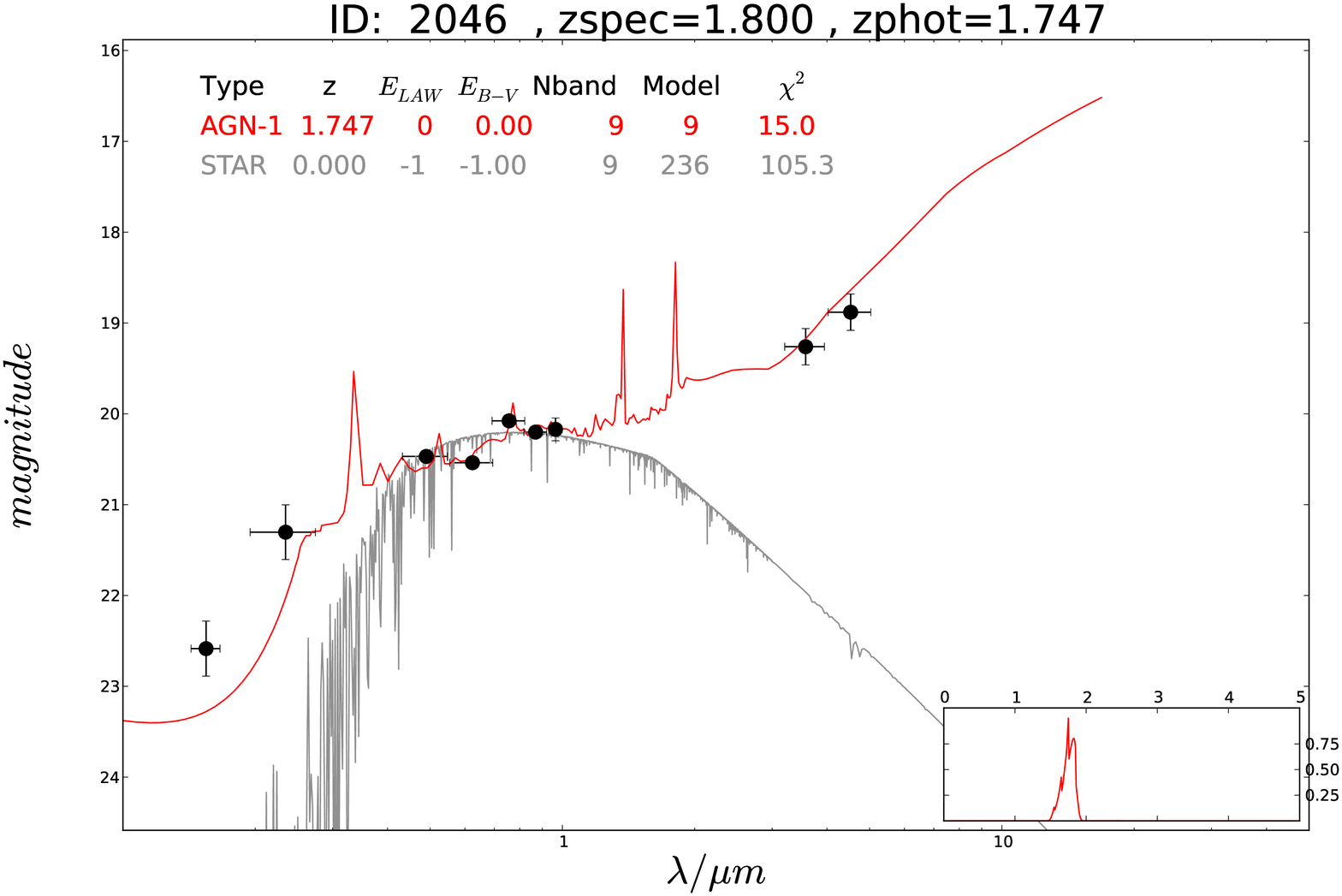}}

\text{\hspace*{6mm}Case B: median}      
\subfloat{%
        \includegraphics[width=0.45\textwidth]{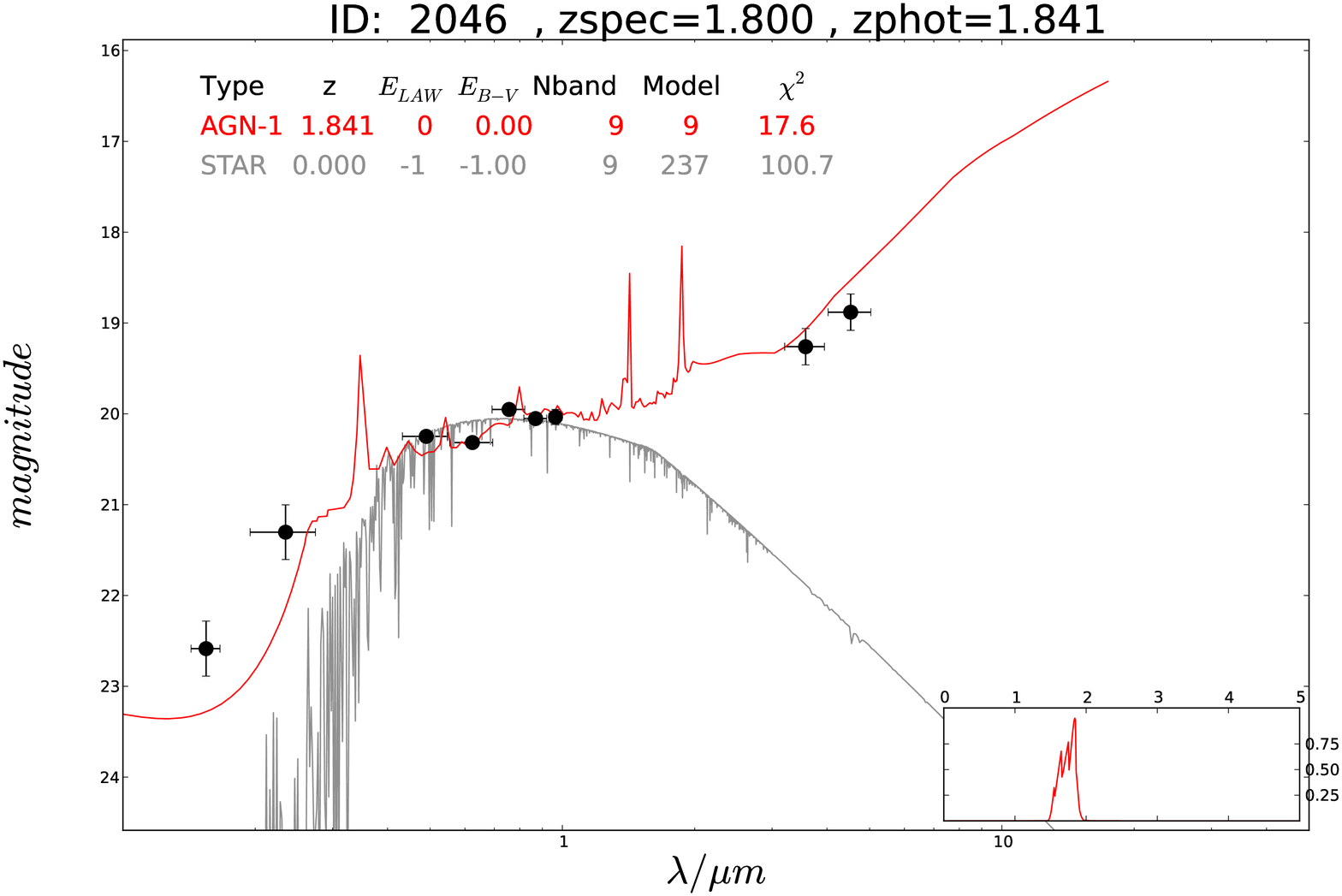}}

\text{\hspace*{6mm}Case C: $\Delta T_{\mathrm{random}}$, run 2} 
\subfloat{%
        \includegraphics[width=0.45\textwidth]{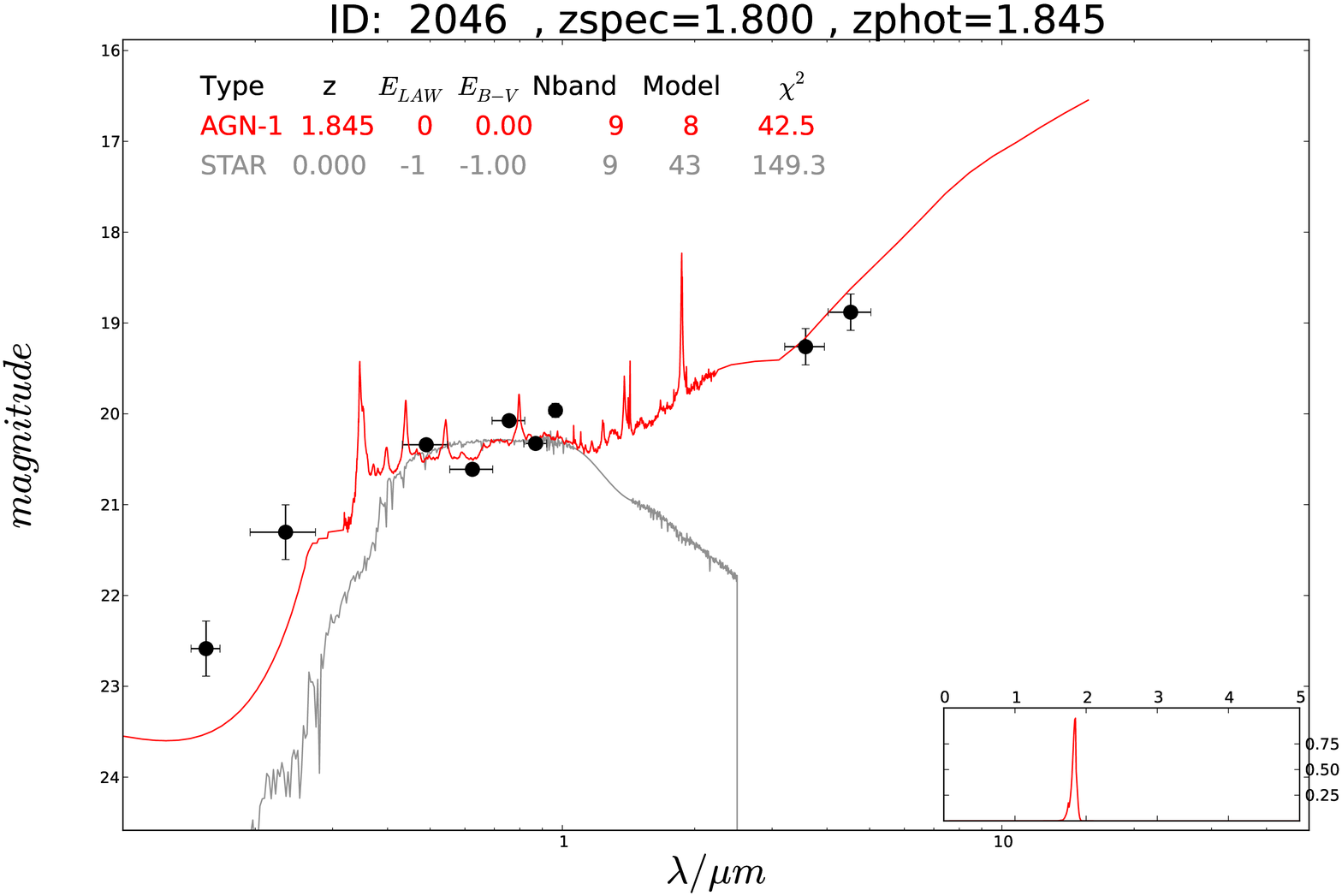}}
\caption{Same as Fig. \ref{fig:trandvstmin} for the AGN with XID 2046.}
\label{fig:uvmirtmintrand}
\end{figure}

To better understand the results quoted above, we again perform a visual inspection of the SED fits for the different input photometry cases. A representative example of what we observe for most of our objects is displayed in Fig. \ref{fig:uvmirtmintrand}. Comparing Cases A and C, we note that although the relative positions of the optical photometry of Case C are strongly affected by variability, the correct redshift is obtained for both cases. Again the median photometry of Case B does not show significant deviations from Case A, also leading to the correct redshift. The comparably small difference in the photo-z quality of all considered cases may be explained by the fact that the overall shape of the SED, implied by the relative positions of the UV, optical, and MIR bands, is predominantly determining the photometric redshift. These constraints are so strong that the influence of variability, occurring in the optical bands, is suppressed, so is less likely to bias the resulting redshift. Still in extreme cases optical variability can cause fatal errors during the fit for Case C, because variability can mimic the presence of emission lines. However, we only observe this for a few objects once the GALEX/IRAC bands are added. Still such cases give rise to the performance differences of Cases A and C, expressed in the respective outlier fractions and accuracy values listed in Table \ref{tab:zresultuvps1ir}. 

Finally we probe again whether the most variable AGNs tend to be found among the outliers for our different photometry sets, therefore we selected the two subsamples of sources with $F_{\mathrm{var}}>0.15$ (15 of the 70 AGNs) and $\Delta mag>0.8$ (12 of the 70 AGNs) in any of the five PS1 bands. Accordingly, we also defined the two subsamples of objects with $F_{\mathrm{var}}>0.1$ (8 of the 70 AGNs) and $\Delta mag>0.4$ (14 of the 70 AGNs) in all five bands simultaneously. It turns out that $\sim$30\%--40\% of the AGNs of each of the four subsamples are outliers in Case C. Regarding Cases A and B, these fractions are lower with only $\sim$10\%--20\% of the most variable AGNs among the outliers for each of the four considered subsamples. Compared to the respective values quoted in section \ref{sec:ps1only} without the UV/MIR bands, the percentages decreased by roughly a factor of two, again indicating that variability effects are now less problematic for the photo-z computation. 

\subsection{Color properties of the multiband variability}

The visual inspection of the SED fits for Cases A and B revealed that the relative positions of the selected magnitude values are very similar, whereas the Case C photometry exhibits dramatic differences to the former cases. Since the relative positions of the measurements determine the SED shape of each object, it is not surprising that Cases A and B show only minor differences in the photo-z results. However, we can think of a situation in which the selected input photometry should not be very different for these two cases. If we assume that most of our objects vary almost simultaneously in each band; i.e., if there are only negligible time lags between the light curves of the bands, then the median values of the light curves and the values with minimum temporal distance will trace very similar spectral shapes. Owing to variability, the two spectral curves will only be slightly shifted relative to each other by typically a few tenths of a magnitude or even less. This assumption requires that we are not strongly affected by time lags between the variability of prominent emission lines, which are present in one band but missing in another, and the variability of the continuum radiation. Considering that the PS1 bands are broadband filters, it is difficult to disentangle the relative contributions of line variability and continuum variability in a given band (see also \citet{2012ApJ...744..147S} for a discussion of the impact of emission lines on broadband color variability of quasars). However, the continuum variability is often found to be larger than the line variability by a factor of a few \citep{2004ApJ...613..682P}. 

\begin{figure*}
\centering
\textbf{\hspace*{12mm}AGN (XID 1)}
\subfloat{%
        \includegraphics[width=0.83\textwidth]{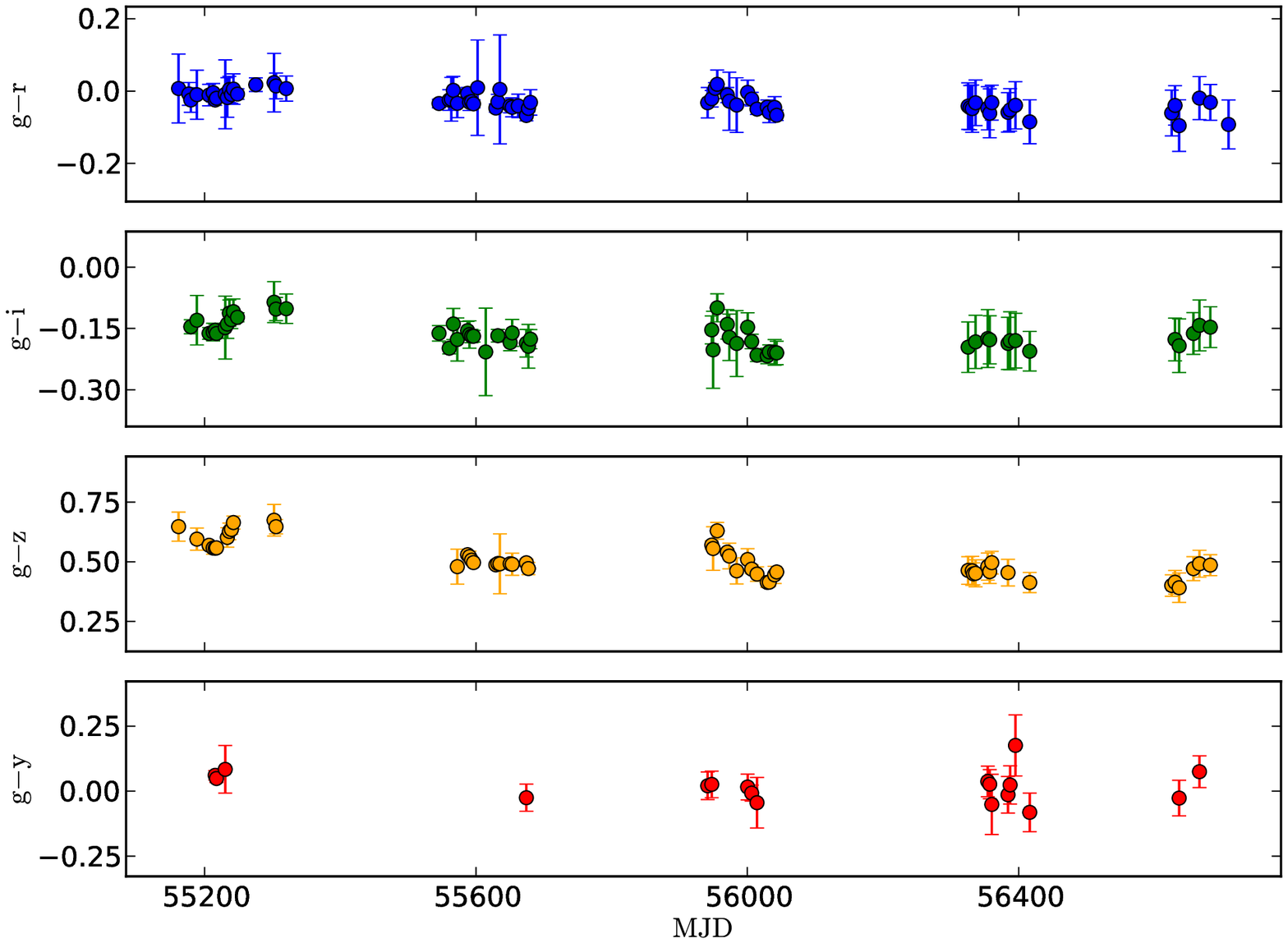}}

\textbf{\hspace*{12mm}AGN (XID 53781)}
\subfloat{%
        \includegraphics[width=0.83\textwidth]{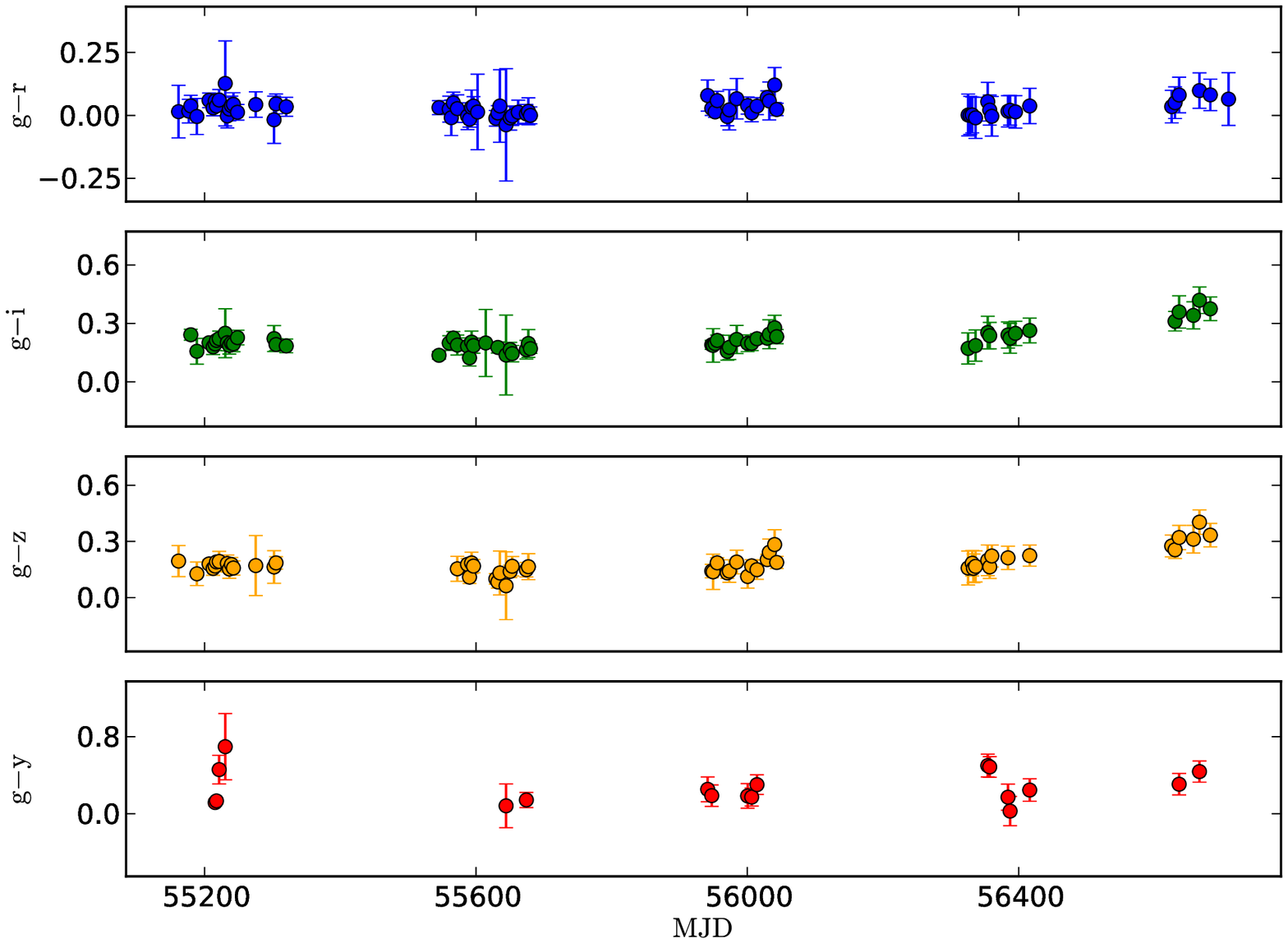}}
\caption{Interpolated MDF04 color light curves for the AGNs with XID 1 and XID 53781.}
\label{fig:colorlc1}
\end{figure*}

To validate this hypothesis we analyzed the color light curves of our sample. Although the different PS1 band observations are not simultaneous, we can obtain approximate colors by taking the observing times of one band as reference values to calculate the contemporaneous magnitude values of the other bands from linear interpolation between light curve points adjacent to the reference values. We chose the "bluest" band ($g_{\mathrm{P1}}$) as our reference band and computed the interpolated colors $g_{\mathrm{P1}}-r_{\mathrm{P1}}$, $g_{\mathrm{P1}}-i_{\mathrm{P1}}$, $g_{\mathrm{P1}}-z_{\mathrm{P1}}$, and $g_{\mathrm{P1}}-y_{\mathrm{P1}}$. We stress, however, that because of the comparably low number of points in the $y_{\mathrm{P1}}$ band light curves, the $g_{\mathrm{P1}}-y_{\mathrm{P1}}$ colors can only be interpolated with limited quality and should be considered with caution. A visual inspection of the color light curves of our sample indicates that the vast majority of our AGNs indeed have approximately constant colors. In addition we verify this observation by calculating the excess variance from the colors and color errors. Although we have computed the excess variance from the light curve fluxes and flux errors throughout this work, we also found that we obtain consistent results when calculating $\sigma^{2}_{\mathrm{rms}}$ from magnitudes and detect variability via the condition $\sigma^{2}_{\mathrm{rms}}-err\left(\sigma^{2}_{\mathrm{rms}}\right)>0$. According to the latter condition 1 ($g_{\mathrm{P1}}-r_{\mathrm{P1}}$), 10 ($g_{\mathrm{P1}}-i_{\mathrm{P1}}$), 22 ($g_{\mathrm{P1}}-z_{\mathrm{P1}}$), and 29 ($g_{\mathrm{P1}}-y_{\mathrm{P1}}$) AGNs out of the 75 sources exhibit color variability. However, of these AGNs, only 13 vary in more than one color, and the considerable $g_{\mathrm{P1}}-y_{\mathrm{P1}}$ color variability possibly stems from poor interpolation. As an example, Fig. \ref{fig:colorlc1} shows the interpolated color light curves for two AGNs of our sample. Except for the colors $g_{\mathrm{P1}}-z_{\mathrm{P1}}$ of XID 1 and $g_{\mathrm{P1}}-y_{\mathrm{P1}}$ of XID 53781, all color light curves are constant within the uncertainties. We point out that these findings strongly suggest that the emission components probed by the different broadband PS1 filters approximately vary as a unit for most AGNs of our sample. In view of these findings, it appears reasonable that the photo-z quality of Cases A and B is not significantly different, but still Case A yields slightly better results. 

\section{Conclusions}
\label{sec:conclu}

We studied AGN variability in five optical bands for a large sample of X-ray-selected point-like AGNs from the XMM-COSMOS survey, taking advantage of the multi-epoch observations provided by the PS1 3$\pi$ and Medium Deep Field surveys. To measure variability, we utilized a simple statistic that estimates the probability of variability and the normalized excess variance that quantifies the variability amplitude. With the help of these two variability parameters, we defined a sample of varying AGNs for each PS1 band of the 3$\pi$ and MDF04 survey. The samples of variable objects comprise 90 ($g_{\mathrm{P1}}$), 54 ($r_{\mathrm{P1}}$), 14 ($i_{\mathrm{P1}}$), 37 ($z_{\mathrm{P1}}$), and 8 ($y_{\mathrm{P1}}$) sources for the 3$\pi$ survey and 184 ($g_{\mathrm{P1}}$), 181 ($r_{\mathrm{P1}}$), 162 ($i_{\mathrm{P1}}$), 131 ($z_{\mathrm{P1}}$), and 74 ($y_{\mathrm{P1}}$) sources for the MDF04 survey. We find that the PS1 3$\pi$ survey allows variable sources to
be reliably selected if the intrinsic variability amplitude is large. For those objects that are detected as variable from 3$\pi$ light curves with at least 3\% fractional variability, we are able to define a pure but incomplete sample of variable sources. Therefore it is possible to detect variable objects for three-quarters of the sky with this data, which is of paramount importance for future missions like EUCLID. 

In addition we investigated the effects of variability on the computation of photometric redshifts. We did this by comparing the well known spectroscopic redshifts of our AGN sample with the photometric redshifts obtained by applying three different kinds of input photometry for our fitting procedure. For each of the five PS1 bands $g_{\mathrm{P1}}$, $r_{\mathrm{P1}}$, $i_{\mathrm{P1}}$, $z_{\mathrm{P1}}$, and $y_{\mathrm{P1}}$, we selected either the pointings with minimal temporal distance in observing time, the median magnitude values of the light curves, or randomly drawn light curve points to calculate the photometric redshift. We note that optical variability significantly limits the achievable photo-z quality. Particularly when only optical bands are used to derive the photometric redshifts, it is crucial to select the photometry with minimized distance in observing time. Omitting a correction of variability in this case leads to very large outlier percentages ($\sim$57\%) and very low accuracies ($\sim$0.400). Taking the median magnitude values as photometry yields only slightly worse results than choosing the points with minimal temporal distance. This is found to be due to the fact that the AGNs in our sample vary almost simultaneously in all optical bands. We stress that we were not able to obtain photometric redshift accuracies better than 0.07 and outlier fractions less than 33\% for our sample of variable AGNs using only the five PS1 bands of the MDF04 survey, even if we consider photometry corresponding to the closest approximation of a snapshot SED. Furthermore, we point out that secondary peaks in the redshift probability distributions should always be rated appropriately in the association of photometric redshifts with astronomical objects.

Considering the same input photometry cases using 3$\pi$ survey light curves for the five PS1 bands results in even lower photo-z quality. Owing to the sampling pattern of the 3$\pi$ survey light curves, the three studied cases give rise to very similar percentages of outliers of typically 60\%--65\% and accuracies of $\sim$0.400. Since the sparse sampling of the 3$\pi$ survey light curves does not allow for selecting photometry resembling an appropriate snapshot SED, the photo-z quality for strongly varying sources may be rather low in general as long as only the five PS1 bands are used for the redshift computation.     

Once UV/GALEX and MIR/IRAC observations are available, which do not contain variability information, the influence of variability is considerably weakened, with the introduced constraints on the overall shape of the spectral energy distribution dominating variability effects of the optical bands. Although the photometric redshift quality generally improves when adding these bands, we still obtain no less than 26\% of outliers and an accuracy of 0.05 at best using MDF04 photometry. 

When considering deep, wide-area surveys that critically rely on precise photometric redshifts, such as the upcoming EUCLID and eROSITA missions, objects showing signs of variability should receive a flag stating that their photometric redshifts may be of low quality. Outlier percentages of less than 5\% and accuracies better than 0.02 for strongly variable AGNs may only be feasible with large photometry sets that comprise broadband and narrowband filters in a wide wavelength range.   

\begin{acknowledgements}
The Pan-STARRS1 Surveys (PS1) have been made possible through contributions of the Institute for Astronomy, the University of Hawaii, the Pan-STARRS Project Office, the Max-Planck Society and its participating institutes, the Max Planck Institute for Astronomy, Heidelberg, and the Max Planck Institute for Extraterrestrial Physics, Garching, The Johns Hopkins University, Durham University, the University of Edinburgh, Queen's University Belfast, the Harvard-Smithsonian Center for Astrophysics, the Las Cumbres Observatory Global Telescope Network Incorporated, the National Central University of Taiwan, the Space Telescope Science Institute, the National Aeronautics and Space Administration under Grant No. NNX08AR22G issued through the Planetary Science Division of the NASA Science Mission Directorate, the National Science Foundation under Grant No. AST-1238877, the University of Maryland, and Eotvos Lorand University (ELTE), and the Los Alamos National Laboratory. We thank the anonymous referee for very beneficial comments. T.S. thanks Johannes Koppenh\"ofer, Audrey Galametz, Giorgio Lanzuisi, Christian Obermeier, and Gabriele Ponti for many helpful discussions and comments. T.S. gratefully acknowledges support by Paul Nandra. 
\end{acknowledgements}

\bibliographystyle{aa} 
\bibliography{corr_simm_2015a.bib} 

\begin{thebibliography}{134}
\expandafter\ifx\csname natexlab\endcsname\relax\def\natexlab#1{#1}\fi

\bibitem[{{Aird} {et~al.}(2010){Aird}, {Nandra}, {Laird}, {Georgakakis},
  {Ashby}, {Barmby}, {Coil}, {Huang}, {Koekemoer}, {Steidel}, \&
  {Willmer}}]{2010MNRAS.401.2531A}
{Aird}, J., {Nandra}, K., {Laird}, E.~S., {et~al.} 2010, \mnras, 401, 2531

\bibitem[{{Allevato} {et~al.}(2013){Allevato}, {Paolillo}, {Papadakis}, \&
  {Pinto}}]{2013ApJ...771....9A}
{Allevato}, V., {Paolillo}, M., {Papadakis}, I., \& {Pinto}, C. 2013, \apj,
  771, 9

\bibitem[{{Amendola} {et~al.}(2013){Amendola}, {Appleby}, {Bacon}, {Baker},
  {Baldi}, {Bartolo}, {Blanchard}, {Bonvin}, {Borgani}, {Branchini}, {Burrage},
  {Camera}, {Carbone}, {Casarini}, {Cropper}, {de Rham}, {Di Porto}, {Ealet},
  {Ferreira}, {Finelli}, {Garc{\'{\i}}a-Bellido}, {Giannantonio}, {Guzzo},
  {Heavens}, {Heisenberg}, {Heymans}, {Hoekstra}, {Hollenstein}, {Holmes},
  {Horst}, {Jahnke}, {Kitching}, {Koivisto}, {Kunz}, {La Vacca}, {March},
  {Majerotto}, {Markovic}, {Marsh}, {Marulli}, {Massey}, {Mellier}, {Mota},
  {Nunes}, {Percival}, {Pettorino}, {Porciani}, {Quercellini}, {Read},
  {Rinaldi}, {Sapone}, {Scaramella}, {Skordis}, {Simpson}, {Taylor}, {Thomas},
  {Trotta}, {Verde}, {Vernizzi}, {Vollmer}, {Wang}, {Weller}, \&
  {Zlosnik}}]{2013LRR....16....6A}
{Amendola}, L., {Appleby}, S., {Bacon}, D., {et~al.} 2013, Living Reviews in
  Relativity, 16, 6

\bibitem[{{Andrae} {et~al.}(2013){Andrae}, {Kim}, \&
  {Bailer-Jones}}]{2013A&A...554A.137A}
{Andrae}, R., {Kim}, D.-W., \& {Bailer-Jones}, C.~A.~L. 2013, \aap, 554, A137

\bibitem[{{Arnouts} {et~al.}(1999){Arnouts}, {Cristiani}, {Moscardini},
  {Matarrese}, {Lucchin}, {Fontana}, \& {Giallongo}}]{1999MNRAS.310..540A}
{Arnouts}, S., {Cristiani}, S., {Moscardini}, L., {et~al.} 1999, \mnras, 310,
  540

\bibitem[{{Assef} {et~al.}(2008){Assef}, {Kochanek}, {Brodwin}, {Brown},
  {Caldwell}, {Cool}, {Eisenhardt}, {Eisenstein}, {Gonzalez}, {Jannuzi},
  {Jones}, {McKenzie}, {Murray}, \& {Stern}}]{2008ApJ...676..286A}
{Assef}, R.~J., {Kochanek}, C.~S., {Brodwin}, M., {et~al.} 2008, \apj, 676, 286

\bibitem[{{Babbedge} {et~al.}(2004){Babbedge}, {Rowan-Robinson},
  {Gonzalez-Solares}, {Polletta}, {Berta}, {P{\'e}rez-Fournon}, {Oliver},
  {Salaman}, {Irwin}, \& {Weatherley}}]{2004MNRAS.353..654B}
{Babbedge}, T.~S.~R., {Rowan-Robinson}, M., {Gonzalez-Solares}, E., {et~al.}
  2004, \mnras, 353, 654

\bibitem[{{Barro} {et~al.}(2011){Barro}, {P{\'e}rez-Gonz{\'a}lez}, {Gallego},
  {Ashby}, {Kajisawa}, {Miyazaki}, {Villar}, {Yamada}, \&
  {Zamorano}}]{2011ApJS..193...30B}
{Barro}, G., {P{\'e}rez-Gonz{\'a}lez}, P.~G., {Gallego}, J., {et~al.} 2011,
  \apjs, 193, 30

\bibitem[{{Bender} {et~al.}(2001){Bender}, {Appenzeller}, {B{\"o}hm}, {Drory},
  {Fricke}, {Gabasch}, {Heidt}, {Hopp}, {J{\"a}ger}, {K{\"u}mmel}, {Mehlert},
  {M{\"o}llenhoff}, {Moorwood}, {Nicklas}, {Noll}, {Saglia}, {Seifert},
  {Seitz}, {Stahl}, {Sutorius}, {Szeifert}, {Wagner}, \&
  {Ziegler}}]{2001defi.conf...96B}
{Bender}, R., {Appenzeller}, I., {B{\"o}hm}, A., {et~al.} 2001, in Deep Fields,
  ed. S.~{Cristiani}, A.~{Renzini}, \& R.~E. {Williams}, 96

\bibitem[{{Ben{\'{\i}}tez}(2000)}]{2000ApJ...536..571B}
{Ben{\'{\i}}tez}, N. 2000, \apj, 536, 571

\bibitem[{{Bolzonella} {et~al.}(2000){Bolzonella}, {Miralles}, \&
  {Pell{\'o}}}]{2000A&A...363..476B}
{Bolzonella}, M., {Miralles}, J.-M., \& {Pell{\'o}}, R. 2000, \aap, 363, 476

\bibitem[{{Bordoloi} {et~al.}(2010){Bordoloi}, {Lilly}, \&
  {Amara}}]{2010MNRAS.406..881B}
{Bordoloi}, R., {Lilly}, S.~J., \& {Amara}, A. 2010, \mnras, 406, 881

\bibitem[{{Bordoloi} {et~al.}(2012){Bordoloi}, {Lilly}, {Amara}, {Oesch},
  {Bardelli}, {Zucca}, {Vergani}, {Nagao}, {Murayama}, {Shioya}, \&
  {Taniguchi}}]{2012MNRAS.421.1671B}
{Bordoloi}, R., {Lilly}, S.~J., {Amara}, A., {et~al.} 2012, \mnras, 421, 1671

\bibitem[{{Brammer} {et~al.}(2008){Brammer}, {van Dokkum}, \&
  {Coppi}}]{2008ApJ...686.1503B}
{Brammer}, G.~B., {van Dokkum}, P.~G., \& {Coppi}, P. 2008, \apj, 686, 1503

\bibitem[{{Brusa} {et~al.}(2010){Brusa}, {Civano}, {Comastri}, {Miyaji},
  {Salvato}, {Zamorani}, {Cappelluti}, {Fiore}, {Hasinger}, {Mainieri},
  {Merloni}, {Bongiorno}, {Capak}, {Elvis}, {Gilli}, {Hao}, {Jahnke},
  {Koekemoer}, {Ilbert}, {Le Floc'h}, {Lusso}, {Mignoli}, {Schinnerer},
  {Silverman}, {Treister}, {Trump}, {Vignali}, {Zamojski}, {Aldcroft},
  {Aussel}, {Bardelli}, {Bolzonella}, {Cappi}, {Caputi}, {Contini},
  {Finoguenov}, {Fruscione}, {Garilli}, {Impey}, {Iovino}, {Iwasawa},
  {Kampczyk}, {Kartaltepe}, {Kneib}, {Knobel}, {Kovac}, {Lamareille},
  {Leborgne}, {Le Brun}, {Le Fevre}, {Lilly}, {Maier}, {McCracken}, {Pello},
  {Peng}, {Perez-Montero}, {de Ravel}, {Sanders}, {Scodeggio}, {Scoville},
  {Tanaka}, {Taniguchi}, {Tasca}, {de la Torre}, {Tresse}, {Vergani}, \&
  {Zucca}}]{2010ApJ...716..348B}
{Brusa}, M., {Civano}, F., {Comastri}, A., {et~al.} 2010, \apj, 716, 348

\bibitem[{{Butler} \& {Bloom}(2011)}]{2011AJ....141...93B}
{Butler}, N.~R. \& {Bloom}, J.~S. 2011, \aj, 141, 93

\bibitem[{{Cappelluti} {et~al.}(2009){Cappelluti}, {Brusa}, {Hasinger},
  {Comastri}, {Zamorani}, {Finoguenov}, {Gilli}, {Puccetti}, {Miyaji},
  {Salvato}, {Vignali}, {Aldcroft}, {B{\"o}hringer}, {Brunner}, {Civano},
  {Elvis}, {Fiore}, {Fruscione}, {Griffiths}, {Guzzo}, {Iovino}, {Koekemoer},
  {Mainieri}, {Scoville}, {Shopbell}, {Silverman}, \&
  {Urry}}]{2009A&A...497..635C}
{Cappelluti}, N., {Brusa}, M., {Hasinger}, G., {et~al.} 2009, \aap, 497, 635

\bibitem[{{Cappelluti} {et~al.}(2011){Cappelluti}, {Predehl}, {B{\"o}hringer},
  {Brunner}, {Brusa}, {Burwitz}, {Churazov}, {Dennerl}, {Finoguenov},
  {Freyberg}, {Friedrich}, {Hasinger}, {Kenziorra}, {Kreykenbohm}, {Lamer},
  {Meidinger}, {M{\"u}hlegger}, {Pavlinsky}, {Robrade}, {Santangelo},
  {Schmitt}, {Schwope}, {Steinmitz}, {Str{\"u}der}, {Sunyaev}, \&
  {Tenzer}}]{2011MSAIS..17..159C}
{Cappelluti}, N., {Predehl}, P., {B{\"o}hringer}, H., {et~al.} 2011, Memorie
  della Societa Astronomica Italiana Supplementi, 17, 159

\bibitem[{{Cardamone} {et~al.}(2010){Cardamone}, {van Dokkum}, {Urry},
  {Taniguchi}, {Gawiser}, {Brammer}, {Taylor}, {Damen}, {Treister}, {Cobb},
  {Bond}, {Schawinski}, {Lira}, {Murayama}, {Saito}, \&
  {Sumikawa}}]{2010ApJS..189..270C}
{Cardamone}, C.~N., {van Dokkum}, P.~G., {Urry}, C.~M., {et~al.} 2010, \apjs,
  189, 270

\bibitem[{{Cardelli} {et~al.}(1989){Cardelli}, {Clayton}, \&
  {Mathis}}]{1989ApJ...345..245C}
{Cardelli}, J.~A., {Clayton}, G.~C., \& {Mathis}, J.~S. 1989, \apj, 345, 245

\bibitem[{{Carini} {et~al.}(2007){Carini}, {Noble}, {Taylor}, \&
  {Culler}}]{2007AJ....133..303C}
{Carini}, M.~T., {Noble}, J.~C., {Taylor}, R., \& {Culler}, R. 2007, \aj, 133,
  303

\bibitem[{{Carliles} {et~al.}(2010){Carliles}, {Budav{\'a}ri}, {Heinis},
  {Priebe}, \& {Szalay}}]{2010ApJ...712..511C}
{Carliles}, S., {Budav{\'a}ri}, T., {Heinis}, S., {Priebe}, C., \& {Szalay},
  A.~S. 2010, \apj, 712, 511

\bibitem[{{Chambers}(2014)}]{2014AAS...22311601C}
{Chambers}, K.~C. 2014, in American Astronomical Society Meeting Abstracts,
  Vol. 223, American Astronomical Society Meeting Abstracts \#223, \#116.01

\bibitem[{{Collier} \& {Peterson}(2001)}]{2001ApJ...555..775C}
{Collier}, S. \& {Peterson}, B.~M. 2001, \apj, 555, 775

\bibitem[{{Collister} \& {Lahav}(2004)}]{2004PASP..116..345C}
{Collister}, A.~A. \& {Lahav}, O. 2004, \pasp, 116, 345

\bibitem[{{Csabai} {et~al.}(2003){Csabai}, {Budav{\'a}ri}, {Connolly},
  {Szalay}, {Gy{\H o}ry}, {Ben{\'{\i}}tez}, {Annis}, {Brinkmann}, {Eisenstein},
  {Fukugita}, {Gunn}, {Kent}, {Lupton}, {Nichol}, \&
  {Stoughton}}]{2003AJ....125..580C}
{Csabai}, I., {Budav{\'a}ri}, T., {Connolly}, A.~J., {et~al.} 2003, \aj, 125,
  580

\bibitem[{{Dahlen} {et~al.}(2010){Dahlen}, {Mobasher}, {Dickinson}, {Ferguson},
  {Giavalisco}, {Grogin}, {Guo}, {Koekemoer}, {Lee}, {Lee}, {Nonino}, {Riess},
  \& {Salimbeni}}]{2010ApJ...724..425D}
{Dahlen}, T., {Mobasher}, B., {Dickinson}, M., {et~al.} 2010, \apj, 724, 425

\bibitem[{{Dahlen} {et~al.}(2013){Dahlen}, {Mobasher}, {Faber}, {Ferguson},
  {Barro}, {Finkelstein}, {Finlator}, {Fontana}, {Gruetzbauch}, {Johnson},
  {Pforr}, {Salvato}, {Wiklind}, {Wuyts}, {Acquaviva}, {Dickinson}, {Guo},
  {Huang}, {Huang}, {Newman}, {Bell}, {Conselice}, {Galametz}, {Gawiser},
  {Giavalisco}, {Grogin}, {Hathi}, {Kocevski}, {Koekemoer}, {Koo}, {Lee},
  {McGrath}, {Papovich}, {Peth}, {Ryan}, {Somerville}, {Weiner}, \&
  {Wilson}}]{2013ApJ...775...93D}
{Dahlen}, T., {Mobasher}, B., {Faber}, S.~M., {et~al.} 2013, \apj, 775, 93

\bibitem[{{De Cicco} {et~al.}(2015){De Cicco}, {Paolillo}, {Covone}, {Falocco},
  {Longo}, {Grado}, {Limatola}, {Botticella}, {Pignata}, {Cappellaro},
  {Vaccari}, {Trevese}, {Vagnetti}, {Salvato}, {Radovich}, {Brandt},
  {Capaccioli}, {Napolitano}, \& {Schipani}}]{2015A&A...574A.112D}
{De Cicco}, D., {Paolillo}, M., {Covone}, G., {et~al.} 2015, \aap, 574, A112

\bibitem[{{DePoy} {et~al.}(2008){DePoy}, {Abbott}, {Annis}, {Antonik},
  {Barcel{\'o}}, {Bernstein}, {Bigelow}, {Brooks}, {Buckley-Geer}, {Campa},
  {Cardiel}, {Castander}, {Castilla}, {Cease}, {Chappa}, {Dede}, {Derylo},
  {Diehl}, {Doel}, {DeVicente}, {Estrada}, {Finley}, {Flaugher}, {Gaztanaga},
  {Gerdes}, {Gladders}, {Guarino}, {Gutierrez}, {Hamilton}, {Haney}, {Holland},
  {Honscheid}, {Huffman}, {Karliner}, {Kau}, {Kent}, {Kozlovsky}, {Kubik},
  {Kuehn}, {Kuhlmann}, {Kuk}, {Leger}, {Lin}, {Martinez}, {Martinez},
  {Merritt}, {Mohr}, {Moore}, {Moore}, {Nord}, {Ogando}, {Olsen}, {Onal},
  {Peoples}, {Qian}, {Roe}, {Sanchez}, {Scarpine}, {Schmidt}, {Schmitt},
  {Schubnell}, {Schultz}, {Selen}, {Shaw}, {Simaitis}, {Slaughter}, {Smith},
  {Spinka}, {Stefanik}, {Stuermer}, {Talaga}, {Tarle}, {Thaler}, {Tucker},
  {Walker}, {Worswick}, \& {Zhao}}]{2008SPIE.7014E..0ED}
{DePoy}, D.~L., {Abbott}, T., {Annis}, J., {et~al.} 2008, in Society of
  Photo-Optical Instrumentation Engineers (SPIE) Conference Series, Vol. 7014,
  Society of Photo-Optical Instrumentation Engineers (SPIE) Conference Series,
  0

\bibitem[{{Donley} {et~al.}(2012){Donley}, {Koekemoer}, {Brusa}, {Capak},
  {Cardamone}, {Civano}, {Ilbert}, {Impey}, {Kartaltepe}, {Miyaji}, {Salvato},
  {Sanders}, {Trump}, \& {Zamorani}}]{2012ApJ...748..142D}
{Donley}, J.~L., {Koekemoer}, A.~M., {Brusa}, M., {et~al.} 2012, \apj, 748, 142

\bibitem[{{Edelson} \& {Nandra}(1999)}]{1999ApJ...514..682E}
{Edelson}, R. \& {Nandra}, K. 1999, \apj, 514, 682

\bibitem[{{Edelson} {et~al.}(1990){Edelson}, {Krolik}, \&
  {Pike}}]{1990ApJ...359...86E}
{Edelson}, R.~A., {Krolik}, J.~H., \& {Pike}, G.~F. 1990, \apj, 359, 86

\bibitem[{{Falocco} {et~al.}(2015){Falocco}, {Paolillo}, {Covone}, {De Cicco},
  {Longo}, {Grado}, {Limatola}, {Vaccari}, {Botticella}, {Pignata},
  {Cappellaro}, {Trevese}, {Vagnetti}, {Salvato}, {Radovich}, {Hsu},
  {Capaccioli}, {Napolitano}, {Brandt}, {Baruffolo}, {Cascone}, \&
  {Schipani}}]{2015A&A...579A.115F}
{Falocco}, S., {Paolillo}, M., {Covone}, G., {et~al.} 2015, \aap, 579, A115

\bibitem[{{Feldmann} {et~al.}(2006){Feldmann}, {Carollo}, {Porciani}, {Lilly},
  {Capak}, {Taniguchi}, {Le F{\`e}vre}, {Renzini}, {Scoville}, {Ajiki},
  {Aussel}, {Contini}, {McCracken}, {Mobasher}, {Murayama}, {Sanders},
  {Sasaki}, {Scarlata}, {Scodeggio}, {Shioya}, {Silverman}, {Takahashi},
  {Thompson}, \& {Zamorani}}]{2006MNRAS.372..565F}
{Feldmann}, R., {Carollo}, C.~M., {Porciani}, C., {et~al.} 2006, \mnras, 372,
  565

\bibitem[{{Finlator} {et~al.}(2007){Finlator}, {Dav{\'e}}, \&
  {Oppenheimer}}]{2007MNRAS.376.1861F}
{Finlator}, K., {Dav{\'e}}, R., \& {Oppenheimer}, B.~D. 2007, \mnras, 376, 1861

\bibitem[{{Gabasch} {et~al.}(2004){Gabasch}, {Bender}, {Seitz}, {Hopp},
  {Saglia}, {Feulner}, {Snigula}, {Drory}, {Appenzeller}, {Heidt}, {Mehlert},
  {Noll}, {B{\"o}hm}, {J{\"a}ger}, {Ziegler}, \&
  {Fricke}}]{2004A&A...421...41G}
{Gabasch}, A., {Bender}, R., {Seitz}, S., {et~al.} 2004, \aap, 421, 41

\bibitem[{{Gaskell} \& {Klimek}(2003)}]{2003A&AT...22..661G}
{Gaskell}, C.~M. \& {Klimek}, E.~S. 2003, Astronomical and Astrophysical
  Transactions, 22, 661

\bibitem[{{George} {et~al.}(2000){George}, {Turner}, {Yaqoob}, {Netzer},
  {Laor}, {Mushotzky}, {Nandra}, \& {Takahashi}}]{2000ApJ...531...52G}
{George}, I.~M., {Turner}, T.~J., {Yaqoob}, T., {et~al.} 2000, \apj, 531, 52

\bibitem[{{Gerdes} {et~al.}(2010){Gerdes}, {Sypniewski}, {McKay}, {Hao},
  {Weis}, {Wechsler}, \& {Busha}}]{2010ApJ...715..823G}
{Gerdes}, D.~W., {Sypniewski}, A.~J., {McKay}, T.~A., {et~al.} 2010, \apj, 715,
  823

\bibitem[{{Giallongo} {et~al.}(1998){Giallongo}, {D'Odorico}, {Fontana},
  {Cristiani}, {Egami}, {Hu}, \& {McMahon}}]{1998AJ....115.2169G}
{Giallongo}, E., {D'Odorico}, S., {Fontana}, A., {et~al.} 1998, \aj, 115, 2169

\bibitem[{{Gonz{\'a}lez-Mart{\'{\i}}n}
  {et~al.}(2011){Gonz{\'a}lez-Mart{\'{\i}}n}, {Papadakis}, {Reig}, \&
  {Zezas}}]{2011A&A...526A.132G}
{Gonz{\'a}lez-Mart{\'{\i}}n}, O., {Papadakis}, I., {Reig}, P., \& {Zezas}, A.
  2011, \aap, 526, A132

\bibitem[{{Gopal-Krishna} {et~al.}(2003){Gopal-Krishna}, {Stalin}, {Sagar}, \&
  {Wiita}}]{2003ApJ...586L..25G}
{Gopal-Krishna}, {Stalin}, C.~S., {Sagar}, R., \& {Wiita}, P.~J. 2003, \apjl,
  586, L25

\bibitem[{{Graham} {et~al.}(2014){Graham}, {Djorgovski}, {Drake}, {Mahabal},
  {Chang}, {Stern}, {Donalek}, \& {Glikman}}]{2014MNRAS.439..703G}
{Graham}, M.~J., {Djorgovski}, S.~G., {Drake}, A.~J., {et~al.} 2014, \mnras,
  439, 703

\bibitem[{{Grazian} {et~al.}(2006){Grazian}, {Fontana}, {de Santis}, {Nonino},
  {Salimbeni}, {Giallongo}, {Cristiani}, {Gallozzi}, \&
  {Vanzella}}]{2006A&A...449..951G}
{Grazian}, A., {Fontana}, A., {de Santis}, C., {et~al.} 2006, \aap, 449, 951

\bibitem[{{Gupta} \& {Joshi}(2005)}]{2005A&A...440..855G}
{Gupta}, A.~C. \& {Joshi}, U.~C. 2005, \aap, 440, 855

\bibitem[{{Hasinger} {et~al.}(2007){Hasinger}, {Cappelluti}, {Brunner},
  {Brusa}, {Comastri}, {Elvis}, {Finoguenov}, {Fiore}, {Franceschini}, {Gilli},
  {Griffiths}, {Lehmann}, {Mainieri}, {Matt}, {Matute}, {Miyaji}, {Molendi},
  {Paltani}, {Sanders}, {Scoville}, {Tresse}, {Urry}, {Vettolani}, \&
  {Zamorani}}]{2007ApJS..172...29H}
{Hasinger}, G., {Cappelluti}, N., {Brunner}, H., {et~al.} 2007, \apjs, 172, 29

\bibitem[{{Heasley}(2008)}]{2008AIPC.1082..352H}
{Heasley}, J.~N. 2008, in American Institute of Physics Conference Series, Vol.
  1082, American Institute of Physics Conference Series, ed. C.~A.~L.
  {Bailer-Jones}, 352--358

\bibitem[{{Hoaglin} {et~al.}(1983){Hoaglin}, {Mosteller}, \&
  {Tukey}}]{1983ured.book.....H}
{Hoaglin}, D.~C., {Mosteller}, F., \& {Tukey}, J.~W. 1983, {Understanding
  robust and exploratory data anlysis}

\bibitem[{{Hodapp} {et~al.}(2004){Hodapp}, {Siegmund}, {Kaiser}, {Chambers},
  {Laux}, {Morgan}, \& {Mannery}}]{2004SPIE.5489..667H}
{Hodapp}, K.~W., {Siegmund}, W.~A., {Kaiser}, N., {et~al.} 2004, in Society of
  Photo-Optical Instrumentation Engineers (SPIE) Conference Series, Vol. 5489,
  Ground-based Telescopes, ed. J.~M. {Oschmann}, Jr., 667--678

\bibitem[{{Hsu} {et~al.}(2014){Hsu}, {Salvato}, {Nandra}, {Brusa}, {Bender},
  {Buchner}, {Donley}, {Kocevski}, {Guo}, {Hathi}, {Rangel}, {Willner},
  {Brightman}, {Georgakakis}, {Budav{\'a}ri}, {Szalay}, {Ashby}, {Barro},
  {Dahlen}, {Faber}, {Ferguson}, {Galametz}, {Grazian}, {Grogin}, {Huang},
  {Koekemoer}, {Lucas}, {McGrath}, {Mobasher}, {Peth}, {Rosario}, \&
  {Trump}}]{2014ApJ...796...60H}
{Hsu}, L.-T., {Salvato}, M., {Nandra}, K., {et~al.} 2014, \apj, 796, 60

\bibitem[{{Ilbert} {et~al.}(2006){Ilbert}, {Arnouts}, {McCracken},
  {Bolzonella}, {Bertin}, {Le F{\`e}vre}, {Mellier}, {Zamorani}, {Pell{\`o}},
  {Iovino}, {Tresse}, {Le Brun}, {Bottini}, {Garilli}, {Maccagni}, {Picat},
  {Scaramella}, {Scodeggio}, {Vettolani}, {Zanichelli}, {Adami}, {Bardelli},
  {Cappi}, {Charlot}, {Ciliegi}, {Contini}, {Cucciati}, {Foucaud}, {Franzetti},
  {Gavignaud}, {Guzzo}, {Marano}, {Marinoni}, {Mazure}, {Meneux}, {Merighi},
  {Paltani}, {Pollo}, {Pozzetti}, {Radovich}, {Zucca}, {Bondi}, {Bongiorno},
  {Busarello}, {de La Torre}, {Gregorini}, {Lamareille}, {Mathez}, {Merluzzi},
  {Ripepi}, {Rizzo}, \& {Vergani}}]{2006A&A...457..841I}
{Ilbert}, O., {Arnouts}, S., {McCracken}, H.~J., {et~al.} 2006, \aap, 457, 841

\bibitem[{{Ilbert} {et~al.}(2009){Ilbert}, {Capak}, {Salvato}, {Aussel},
  {McCracken}, {Sanders}, {Scoville}, {Kartaltepe}, {Arnouts}, {Le Floc'h},
  {Mobasher}, {Taniguchi}, {Lamareille}, {Leauthaud}, {Sasaki}, {Thompson},
  {Zamojski}, {Zamorani}, {Bardelli}, {Bolzonella}, {Bongiorno}, {Brusa},
  {Caputi}, {Carollo}, {Contini}, {Cook}, {Coppa}, {Cucciati}, {de la Torre},
  {de Ravel}, {Franzetti}, {Garilli}, {Hasinger}, {Iovino}, {Kampczyk},
  {Kneib}, {Knobel}, {Kovac}, {Le Borgne}, {Le Brun}, {F{\`e}vre}, {Lilly},
  {Looper}, {Maier}, {Mainieri}, {Mellier}, {Mignoli}, {Murayama}, {Pell{\`o}},
  {Peng}, {P{\'e}rez-Montero}, {Renzini}, {Ricciardelli}, {Schiminovich},
  {Scodeggio}, {Shioya}, {Silverman}, {Surace}, {Tanaka}, {Tasca}, {Tresse},
  {Vergani}, \& {Zucca}}]{2009ApJ...690.1236I}
{Ilbert}, O., {Capak}, P., {Salvato}, M., {et~al.} 2009, \apj, 690, 1236

\bibitem[{{Ilbert} {et~al.}(2010){Ilbert}, {Salvato}, {Le Floc'h}, {Aussel},
  {Capak}, {McCracken}, {Mobasher}, {Kartaltepe}, {Scoville}, {Sanders},
  {Arnouts}, {Bundy}, {Cassata}, {Kneib}, {Koekemoer}, {Le F{\`e}vre}, {Lilly},
  {Surace}, {Taniguchi}, {Tasca}, {Thompson}, {Tresse}, {Zamojski}, {Zamorani},
  \& {Zucca}}]{2010ApJ...709..644I}
{Ilbert}, O., {Salvato}, M., {Le Floc'h}, E., {et~al.} 2010, \apj, 709, 644

\bibitem[{{Ivezic} {et~al.}(2006){Ivezic}, {Tyson}, {Strauss}, {Kahn},
  {Stubbs}, {Pinto}, {Cook}, \& {LSST Collaboration}}]{2006AAS...209.8602I}
{Ivezic}, Z., {Tyson}, A.~J., {Strauss}, M.~A., {et~al.} 2006, in Bulletin of
  the American Astronomical Society, Vol.~38, American Astronomical Society
  Meeting Abstracts, 1017

\bibitem[{{Kaiser} {et~al.}(2010){Kaiser}, {Burgett}, {Chambers}, {Denneau},
  {Heasley}, {Jedicke}, {Magnier}, {Morgan}, {Onaka}, \&
  {Tonry}}]{2010SPIE.7733E..12K}
{Kaiser}, N., {Burgett}, W., {Chambers}, K., {et~al.} 2010, in Society of
  Photo-Optical Instrumentation Engineers (SPIE) Conference Series, Vol. 7733,
  Society of Photo-Optical Instrumentation Engineers (SPIE) Conference Series

\bibitem[{{Kelly} {et~al.}(2009){Kelly}, {Bechtold}, \&
  {Siemiginowska}}]{2009ApJ...698..895K}
{Kelly}, B.~C., {Bechtold}, J., \& {Siemiginowska}, A. 2009, \apj, 698, 895

\bibitem[{{Kelly} {et~al.}(2011){Kelly}, {Sobolewska}, \&
  {Siemiginowska}}]{2011ApJ...730...52K}
{Kelly}, B.~C., {Sobolewska}, M., \& {Siemiginowska}, A. 2011, \apj, 730, 52

\bibitem[{{Kelly} {et~al.}(2013){Kelly}, {Treu}, {Malkan}, {Pancoast}, \&
  {Woo}}]{2013ApJ...779..187K}
{Kelly}, B.~C., {Treu}, T., {Malkan}, M., {Pancoast}, A., \& {Woo}, J.-H. 2013,
  \apj, 779, 187

\bibitem[{{Kim} {et~al.}(2011){Kim}, {Protopapas}, {Byun}, {Alcock}, {Khardon},
  \& {Trichas}}]{2011ApJ...735...68K}
{Kim}, D.-W., {Protopapas}, P., {Byun}, Y.-I., {et~al.} 2011, \apj, 735, 68

\bibitem[{{Kokubo}(2015)}]{2015MNRAS.449...94K}
{Kokubo}, M. 2015, \mnras, 449, 94

\bibitem[{{Kokubo} {et~al.}(2014){Kokubo}, {Morokuma}, {Minezaki}, {Doi},
  {Kawaguchi}, {Sameshima}, \& {Koshida}}]{2014ApJ...783...46K}
{Kokubo}, M., {Morokuma}, T., {Minezaki}, T., {et~al.} 2014, \apj, 783, 46

\bibitem[{{Koppenhoefer} {et~al.}(2013){Koppenhoefer}, {Saglia}, \&
  {Riffeser}}]{2013ExA....35..329K}
{Koppenhoefer}, J., {Saglia}, R.~P., \& {Riffeser}, A. 2013, Experimental
  Astronomy, 35, 329

\bibitem[{{Kotulla} {et~al.}(2009){Kotulla}, {Fritze}, {Weilbacher}, \&
  {Anders}}]{2009MNRAS.396..462K}
{Kotulla}, R., {Fritze}, U., {Weilbacher}, P., \& {Anders}, P. 2009, \mnras,
  396, 462

\bibitem[{{Koz{\l}owski} {et~al.}(2012){Koz{\l}owski}, {Kochanek}, {Jacyszyn},
  {Udalski}, {Szyma{\'n}ski}, {Poleski}, {Kubiak}, {Soszy{\'n}ski},
  {Pietrzy{\'n}ski}, {Wyrzykowski}, {Ulaczyk}, \&
  {Pietrukowicz}}]{2012ApJ...746...27K}
{Koz{\l}owski}, S., {Kochanek}, C.~S., {Jacyszyn}, A.~M., {et~al.} 2012, \apj,
  746, 27

\bibitem[{{Koz{\l}owski} {et~al.}(2011){Koz{\l}owski}, {Kochanek}, \&
  {Udalski}}]{2011ApJS..194...22K}
{Koz{\l}owski}, S., {Kochanek}, C.~S., \& {Udalski}, A. 2011, \apjs, 194, 22

\bibitem[{{Koz{\l}owski} {et~al.}(2010){Koz{\l}owski}, {Kochanek}, {Udalski},
  {Wyrzykowski}, {Soszy{\'n}ski}, {Szyma{\'n}ski}, {Kubiak}, {Pietrzy{\'n}ski},
  {Szewczyk}, {Ulaczyk}, {Poleski}, \& {OGLE
  Collaboration}}]{2010ApJ...708..927K}
{Koz{\l}owski}, S., {Kochanek}, C.~S., {Udalski}, A., {et~al.} 2010, \apj, 708,
  927

\bibitem[{{Koz{\l}owski} {et~al.}(2013){Koz{\l}owski}, {Onken}, {Kochanek},
  {Udalski}, {Szyma{\'n}ski}, {Kubiak}, {Pietrzy{\'n}ski}, {Soszy{\'n}ski},
  {Wyrzykowski}, {Ulaczyk}, {Poleski}, {Pietrukowicz}, {Skowron}, {OGLE
  Collaboration}, {Meixner}, \& {Bonanos}}]{2013ApJ...775...92K}
{Koz{\l}owski}, S., {Onken}, C.~A., {Kochanek}, C.~S., {et~al.} 2013, \apj,
  775, 92

\bibitem[{{Kron}(1980)}]{1980ApJS...43..305K}
{Kron}, R.~G. 1980, \apjs, 43, 305

\bibitem[{{Lanzuisi} {et~al.}(2014){Lanzuisi}, {Ponti}, {Salvato}, {Hasinger},
  {Cappelluti}, {Bongiorno}, {Brusa}, {Lusso}, {Nandra}, {Merloni},
  {Silverman}, {Trump}, {Vignali}, {Comastri}, {Gilli}, {Schramm},
  {Steinhardt}, {Sanders}, {Kartaltepe}, {Rosario}, \&
  {Trakhtenbrot}}]{2014ApJ...781..105L}
{Lanzuisi}, G., {Ponti}, G., {Salvato}, M., {et~al.} 2014, \apj, 781, 105

\bibitem[{{Laureijs} {et~al.}(2011){Laureijs}, {Amiaux}, {Arduini},
  {Augu{\`e}res}, {Brinchmann}, {Cole}, {Cropper}, {Dabin}, {Duvet}, {Ealet},
  \& et~al.}]{2011arXiv1110.3193L}
{Laureijs}, R., {Amiaux}, J., {Arduini}, S., {et~al.} 2011, ArXiv e-prints
  [\eprint[arXiv]{1110.3193}]

\bibitem[{{Lawrence} \& {Papadakis}(1993)}]{1993ApJ...414L..85L}
{Lawrence}, A. \& {Papadakis}, I. 1993, \apjl, 414, L85

\bibitem[{{Leauthaud} {et~al.}(2007){Leauthaud}, {Massey}, {Kneib}, {Rhodes},
  {Johnston}, {Capak}, {Heymans}, {Ellis}, {Koekemoer}, {Le F{\`e}vre},
  {Mellier}, {R{\'e}fr{\'e}gier}, {Robin}, {Scoville}, {Tasca}, {Taylor}, \&
  {Van Waerbeke}}]{2007ApJS..172..219L}
{Leauthaud}, A., {Massey}, R., {Kneib}, J.-P., {et~al.} 2007, \apjs, 172, 219

\bibitem[{{Leighly}(1999)}]{1999ApJS..125..297L}
{Leighly}, K.~M. 1999, \apjs, 125, 297

\bibitem[{{Lilly} {et~al.}(2009){Lilly}, {Le Brun}, {Maier}, {Mainieri},
  {Mignoli}, {Scodeggio}, {Zamorani}, {Carollo}, {Contini}, {Kneib}, {Le
  F{\`e}vre}, {Renzini}, {Bardelli}, {Bolzonella}, {Bongiorno}, {Caputi},
  {Coppa}, {Cucciati}, {de la Torre}, {de Ravel}, {Franzetti}, {Garilli},
  {Iovino}, {Kampczyk}, {Kovac}, {Knobel}, {Lamareille}, {Le Borgne}, {Pello},
  {Peng}, {P{\'e}rez-Montero}, {Ricciardelli}, {Silverman}, {Tanaka}, {Tasca},
  {Tresse}, {Vergani}, {Zucca}, {Ilbert}, {Salvato}, {Oesch}, {Abbas},
  {Bottini}, {Capak}, {Cappi}, {Cassata}, {Cimatti}, {Elvis}, {Fumana},
  {Guzzo}, {Hasinger}, {Koekemoer}, {Leauthaud}, {Maccagni}, {Marinoni},
  {McCracken}, {Memeo}, {Meneux}, {Porciani}, {Pozzetti}, {Sanders},
  {Scaramella}, {Scarlata}, {Scoville}, {Shopbell}, \&
  {Taniguchi}}]{2009ApJS..184..218L}
{Lilly}, S.~J., {Le Brun}, V., {Maier}, C., {et~al.} 2009, \apjs, 184, 218

\bibitem[{{Luo} {et~al.}(2010){Luo}, {Brandt}, {Xue}, {Brusa}, {Alexander},
  {Bauer}, {Comastri}, {Koekemoer}, {Lehmer}, {Mainieri}, {Rafferty},
  {Schneider}, {Silverman}, \& {Vignali}}]{2010ApJS..187..560L}
{Luo}, B., {Brandt}, W.~N., {Xue}, Y.~Q., {et~al.} 2010, \apjs, 187, 560

\bibitem[{{MacLeod} {et~al.}(2011){MacLeod}, {Brooks}, {Ivezi{\'c}},
  {Kochanek}, {Gibson}, {Meisner}, {Koz{\l}owski}, {Sesar}, {Becker}, \& {de
  Vries}}]{2011ApJ...728...26M}
{MacLeod}, C.~L., {Brooks}, K., {Ivezi{\'c}}, {\v Z}., {et~al.} 2011, \apj,
  728, 26

\bibitem[{{MacLeod} {et~al.}(2010){MacLeod}, {Ivezi{\'c}}, {Kochanek},
  {Koz{\l}owski}, {Kelly}, {Bullock}, {Kimball}, {Sesar}, {Westman}, {Brooks},
  {Gibson}, {Becker}, \& {de Vries}}]{2010ApJ...721.1014M}
{MacLeod}, C.~L., {Ivezi{\'c}}, {\v Z}., {Kochanek}, C.~S., {et~al.} 2010,
  \apj, 721, 1014

\bibitem[{{MacLeod} {et~al.}(2012){MacLeod}, {Ivezi{\'c}}, {Sesar}, {de Vries},
  {Kochanek}, {Kelly}, {Becker}, {Lupton}, {Hall}, {Richards}, {Anderson}, \&
  {Schneider}}]{2012ApJ...753..106M}
{MacLeod}, C.~L., {Ivezi{\'c}}, {\v Z}., {Sesar}, B., {et~al.} 2012, \apj, 753,
  106

\bibitem[{{Madau}(1995)}]{1995ApJ...441...18M}
{Madau}, P. 1995, \apj, 441, 18

\bibitem[{{Magnier} {et~al.}(2013){Magnier}, {Schlafly}, {Finkbeiner}, {Juric},
  {Tonry}, {Burgett}, {Chambers}, {Flewelling}, {Kaiser}, {Kudritzki},
  {Morgan}, {Price}, {Sweeney}, \& {Stubbs}}]{2013ApJS..205...20M}
{Magnier}, E.~A., {Schlafly}, E., {Finkbeiner}, D., {et~al.} 2013, \apjs, 205,
  20

\bibitem[{{Markowitz} {et~al.}(2003){Markowitz}, {Edelson}, {Vaughan},
  {Uttley}, {George}, {Griffiths}, {Kaspi}, {Lawrence}, {McHardy}, {Nandra},
  {Pounds}, {Reeves}, {Schurch}, \& {Warwick}}]{2003ApJ...593...96M}
{Markowitz}, A., {Edelson}, R., {Vaughan}, S., {et~al.} 2003, \apj, 593, 96

\bibitem[{{McHardy} {et~al.}(2004){McHardy}, {Papadakis}, {Uttley}, {Page}, \&
  {Mason}}]{2004MNRAS.348..783M}
{McHardy}, I.~M., {Papadakis}, I.~E., {Uttley}, P., {Page}, M.~J., \& {Mason},
  K.~O. 2004, \mnras, 348, 783

\bibitem[{{McLaughlin} {et~al.}(1996){McLaughlin}, {Mattox}, {Cordes}, \&
  {Thompson}}]{1996ApJ...473..763M}
{McLaughlin}, M.~A., {Mattox}, J.~R., {Cordes}, J.~M., \& {Thompson}, D.~J.
  1996, \apj, 473, 763

\bibitem[{{Merloni} {et~al.}(2012){Merloni}, {Predehl}, {Becker},
  {B{\"o}hringer}, {Boller}, {Brunner}, {Brusa}, {Dennerl}, {Freyberg},
  {Friedrich}, {Georgakakis}, {Haberl}, {Hasinger}, {Meidinger}, {Mohr},
  {Nandra}, {Rau}, {Reiprich}, {Robrade}, {Salvato}, {Santangelo}, {Sasaki},
  {Schwope}, {Wilms}, \& {German eROSITA Consortium}}]{2012arXiv1209.3114M}
{Merloni}, A., {Predehl}, P., {Becker}, W., {et~al.} 2012, ArXiv e-prints
  [\eprint[arXiv]{1209.3114}]

\bibitem[{{Metcalfe} {et~al.}(2013){Metcalfe}, {Farrow}, {Cole}, {Draper},
  {Norberg}, {Burgett}, {Chambers}, {Denneau}, {Flewelling}, {Kaiser},
  {Kudritzki}, {Magnier}, {Morgan}, {Price}, {Sweeney}, {Tonry}, {Wainscoat},
  \& {Waters}}]{2013MNRAS.435.1825M}
{Metcalfe}, N., {Farrow}, D.~J., {Cole}, S., {et~al.} 2013, \mnras, 435, 1825

\bibitem[{{Mobasher} {et~al.}(2007){Mobasher}, {Capak}, {Scoville}, {Dahlen},
  {Salvato}, {Aussel}, {Thompson}, {Feldmann}, {Tasca}, {Le Fevre}, {Lilly},
  {Carollo}, {Kartaltepe}, {McCracken}, {Mould}, {Renzini}, {Sanders},
  {Shopbell}, {Taniguchi}, {Ajiki}, {Shioya}, {Contini}, {Giavalisco},
  {Ilbert}, {Iovino}, {Le Brun}, {Mainieri}, {Mignoli}, \&
  {Scodeggio}}]{2007ApJS..172..117M}
{Mobasher}, B., {Capak}, P., {Scoville}, N.~Z., {et~al.} 2007, \apjs, 172, 117

\bibitem[{{Mohr} {et~al.}(2008){Mohr}, {Adams}, {Barkhouse}, {Beldica},
  {Bertin}, {Cai}, {da Costa}, {Darnell}, {Daues}, {Jarvis}, {Gower}, {Lin},
  {Martelli}, {Neilsen}, {Ngeow}, {Ogando}, {Parga}, {Sheldon}, {Tucker},
  {Kuropatkin}, \& {Stoughton}}]{2008SPIE.7016E..0LM}
{Mohr}, J.~J., {Adams}, D., {Barkhouse}, W., {et~al.} 2008, in Society of
  Photo-Optical Instrumentation Engineers (SPIE) Conference Series, Vol. 7016,
  Society of Photo-Optical Instrumentation Engineers (SPIE) Conference Series,
  0

\bibitem[{{Morganson} {et~al.}(2014){Morganson}, {Burgett}, {Chambers},
  {Green}, {Kaiser}, {Magnier}, {Marshall}, {Morgan}, {Price}, {Rix},
  {Schlafly}, {Tonry}, \& {Walter}}]{2014ApJ...784...92M}
{Morganson}, E., {Burgett}, W.~S., {Chambers}, K.~C., {et~al.} 2014, \apj, 784,
  92

\bibitem[{{Nandra} {et~al.}(1997){Nandra}, {George}, {Mushotzky}, {Turner}, \&
  {Yaqoob}}]{1997ApJ...476...70N}
{Nandra}, K., {George}, I.~M., {Mushotzky}, R.~F., {Turner}, T.~J., \&
  {Yaqoob}, T. 1997, \apj, 476, 70

\bibitem[{{Oke} \& {Gunn}(1983)}]{1983ApJ...266..713O}
{Oke}, J.~B. \& {Gunn}, J.~E. 1983, \apj, 266, 713

\bibitem[{{O'Neill} {et~al.}(2005){O'Neill}, {Nandra}, {Papadakis}, \&
  {Turner}}]{2005MNRAS.358.1405O}
{O'Neill}, P.~M., {Nandra}, K., {Papadakis}, I.~E., \& {Turner}, T.~J. 2005,
  \mnras, 358, 1405

\bibitem[{{Padmanabhan} {et~al.}(2005){Padmanabhan}, {Budav{\'a}ri},
  {Schlegel}, {Bridges}, {Brinkmann}, {Cannon}, {Connolly}, {Croom}, {Csabai},
  {Drinkwater}, {Eisenstein}, {Hewett}, {Loveday}, {Nichol}, {Pimbblet}, {De
  Propris}, {Schneider}, {Scranton}, {Seljak}, {Shanks}, {Szapudi}, {Szalay},
  \& {Wake}}]{2005MNRAS.359..237P}
{Padmanabhan}, N., {Budav{\'a}ri}, T., {Schlegel}, D.~J., {et~al.} 2005,
  \mnras, 359, 237

\bibitem[{{Palanque-Delabrouille} {et~al.}(2011){Palanque-Delabrouille},
  {Yeche}, {Myers}, {Petitjean}, {Ross}, {Sheldon}, {Aubourg}, {Delubac}, {Le
  Goff}, {P{\^a}ris}, {Rich}, {Dawson}, {Schneider}, \&
  {Weaver}}]{2011A&A...530A.122P}
{Palanque-Delabrouille}, N., {Yeche}, C., {Myers}, A.~D., {et~al.} 2011, \aap,
  530, A122

\bibitem[{{Paolillo} {et~al.}(2004){Paolillo}, {Schreier}, {Giacconi},
  {Koekemoer}, \& {Grogin}}]{2004ApJ...611...93P}
{Paolillo}, M., {Schreier}, E.~J., {Giacconi}, R., {Koekemoer}, A.~M., \&
  {Grogin}, N.~A. 2004, \apj, 611, 93

\bibitem[{{Papadakis}(2004)}]{2004MNRAS.348..207P}
{Papadakis}, I.~E. 2004, \mnras, 348, 207

\bibitem[{{Pell{\'o}} {et~al.}(2009){Pell{\'o}}, {Rudnick}, {De Lucia},
  {Simard}, {Clowe}, {Jablonka}, {Milvang-Jensen}, {Saglia}, {White},
  {Arag{\'o}n-Salamanca}, {Halliday}, {Poggianti}, {Best}, {Dalcanton},
  {Dantel-Fort}, {Fort}, {von der Linden}, {Mellier}, {Rottgering}, \&
  {Zaritsky}}]{2009A&A...508.1173P}
{Pell{\'o}}, R., {Rudnick}, G., {De Lucia}, G., {et~al.} 2009, \aap, 508, 1173

\bibitem[{{Peterson} {et~al.}(2004){Peterson}, {Ferrarese}, {Gilbert}, {Kaspi},
  {Malkan}, {Maoz}, {Merritt}, {Netzer}, {Onken}, {Pogge}, {Vestergaard}, \&
  {Wandel}}]{2004ApJ...613..682P}
{Peterson}, B.~M., {Ferrarese}, L., {Gilbert}, K.~M., {et~al.} 2004, \apj, 613,
  682

\bibitem[{{Ponti} {et~al.}(2012){Ponti}, {Papadakis}, {Bianchi}, {Guainazzi},
  {Matt}, {Uttley}, \& {Bonilla}}]{2012A&A...542A..83P}
{Ponti}, G., {Papadakis}, I., {Bianchi}, S., {et~al.} 2012, \aap, 542, A83

\bibitem[{{Predehl} {et~al.}(2007){Predehl}, {Andritschke}, {Bornemann},
  {Br{\"a}uninger}, {Briel}, {Brunner}, {Burkert}, {Dennerl}, {Eder},
  {Freyberg}, {Friedrich}, {F{\"u}rmetz}, {Hartmann}, {Hartner}, {Hasinger},
  {Herrmann}, {Holl}, {Huber}, {Kendziorra}, {Kink}, {Meidinger}, {M{\"u}ller},
  {Pavlinsky}, {Pfeffermann}, {Roh{\'e}}, {Santangelo}, {Schmitt}, {Schwope},
  {Steinmetz}, {Str{\"u}der}, {Sunyaev}, {Tiedemann}, {Vongehr}, {Wilms},
  {Erhard}, {Gutruf}, {Jugler}, {Kampf}, {Graue}, {Citterio}, {Valsecci},
  {Vernani}, \& {Zimmerman}}]{2007SPIE.6686E..17P}
{Predehl}, P., {Andritschke}, R., {Bornemann}, W., {et~al.} 2007, in Society of
  Photo-Optical Instrumentation Engineers (SPIE) Conference Series, Vol. 6686,
  Society of Photo-Optical Instrumentation Engineers (SPIE) Conference Series,
  17

\bibitem[{{Prevot} {et~al.}(1984){Prevot}, {Lequeux}, {Prevot}, {Maurice}, \&
  {Rocca-Volmerange}}]{1984A&A...132..389P}
{Prevot}, M.~L., {Lequeux}, J., {Prevot}, L., {Maurice}, E., \&
  {Rocca-Volmerange}, B. 1984, \aap, 132, 389

\bibitem[{{Richards} {et~al.}(2002){Richards}, {Fan}, {Newberg}, {Strauss},
  {Vanden Berk}, {Schneider}, {Yanny}, {Boucher}, {Burles}, {Frieman}, {Gunn},
  {Hall}, {Ivezi{\'c}}, {Kent}, {Loveday}, {Lupton}, {Rockosi}, {Schlegel},
  {Stoughton}, {SubbaRao}, \& {York}}]{2002AJ....123.2945R}
{Richards}, G.~T., {Fan}, X., {Newberg}, H.~J., {et~al.} 2002, \aj, 123, 2945

\bibitem[{{Rosario} {et~al.}(2013){Rosario}, {Mozena}, {Wuyts}, {Nandra},
  {Koekemoer}, {McGrath}, {Hathi}, {Dekel}, {Donley}, {Dunlop}, {Faber},
  {Ferguson}, {Giavalisco}, {Grogin}, {Guo}, {Kocevski}, {Koo}, {Laird},
  {Newman}, {Rangel}, \& {Somerville}}]{2013ApJ...763...59R}
{Rosario}, D.~J., {Mozena}, M., {Wuyts}, S., {et~al.} 2013, \apj, 763, 59

\bibitem[{{Ruan} {et~al.}(2012){Ruan}, {Anderson}, {MacLeod}, {Becker},
  {Burnett}, {Davenport}, {Ivezi{\'c}}, {Kochanek}, {Plotkin}, {Sesar}, \&
  {Stuart}}]{2012ApJ...760...51R}
{Ruan}, J.~J., {Anderson}, S.~F., {MacLeod}, C.~L., {et~al.} 2012, \apj, 760,
  51

\bibitem[{{Saglia} {et~al.}(2012){Saglia}, {Tonry}, {Bender}, {Greisel},
  {Seitz}, {Senger}, {Snigula}, {Phleps}, {Wilman}, {Bailer-Jones}, {Klement},
  {Rix}, {Smith}, {Green}, {Burgett}, {Chambers}, {Heasley}, {Kaiser},
  {Magnier}, {Morgan}, {Price}, {Stubbs}, \& {Wainscoat}}]{2012ApJ...746..128S}
{Saglia}, R.~P., {Tonry}, J.~L., {Bender}, R., {et~al.} 2012, \apj, 746, 128

\bibitem[{{Salvato} {et~al.}(2009){Salvato}, {Hasinger}, {Ilbert}, {Zamorani},
  {Brusa}, {Scoville}, {Rau}, {Capak}, {Arnouts}, {Aussel}, {Bolzonella},
  {Buongiorno}, {Cappelluti}, {Caputi}, {Civano}, {Cook}, {Elvis}, {Gilli},
  {Jahnke}, {Kartaltepe}, {Impey}, {Lamareille}, {Le Floc'h}, {Lilly},
  {Mainieri}, {McCarthy}, {McCracken}, {Mignoli}, {Mobasher}, {Murayama},
  {Sasaki}, {Sanders}, {Schiminovich}, {Shioya}, {Shopbell}, {Silverman},
  {Smol{\v c}i{\'c}}, {Surace}, {Taniguchi}, {Thompson}, {Trump}, {Urry}, \&
  {Zamojski}}]{2009ApJ...690.1250S}
{Salvato}, M., {Hasinger}, G., {Ilbert}, O., {et~al.} 2009, \apj, 690, 1250

\bibitem[{{Salvato} {et~al.}(2011){Salvato}, {Ilbert}, {Hasinger}, {Rau},
  {Civano}, {Zamorani}, {Brusa}, {Elvis}, {Vignali}, {Aussel}, {Comastri},
  {Fiore}, {Le Floc'h}, {Mainieri}, {Bardelli}, {Bolzonella}, {Bongiorno},
  {Capak}, {Caputi}, {Cappelluti}, {Carollo}, {Contini}, {Garilli}, {Iovino},
  {Fotopoulou}, {Fruscione}, {Gilli}, {Halliday}, {Kneib}, {Kakazu},
  {Kartaltepe}, {Koekemoer}, {Kovac}, {Ideue}, {Ikeda}, {Impey}, {Le Fevre},
  {Lamareille}, {Lanzuisi}, {Le Borgne}, {Le Brun}, {Lilly}, {Maier},
  {Manohar}, {Masters}, {McCracken}, {Messias}, {Mignoli}, {Mobasher}, {Nagao},
  {Pello}, {Puccetti}, {Perez-Montero}, {Renzini}, {Sargent}, {Sanders},
  {Scodeggio}, {Scoville}, {Shopbell}, {Silvermann}, {Taniguchi}, {Tasca},
  {Tresse}, {Trump}, \& {Zucca}}]{2011ApJ...742...61S}
{Salvato}, M., {Ilbert}, O., {Hasinger}, G., {et~al.} 2011, \apj, 742, 61

\bibitem[{{Sanders} {et~al.}(2007){Sanders}, {Salvato}, {Aussel}, {Ilbert},
  {Scoville}, {Surace}, {Frayer}, {Sheth}, {Helou}, {Brooke}, {Bhattacharya},
  {Yan}, {Kartaltepe}, {Barnes}, {Blain}, {Calzetti}, {Capak}, {Carilli},
  {Carollo}, {Comastri}, {Daddi}, {Ellis}, {Elvis}, {Fall}, {Franceschini},
  {Giavalisco}, {Hasinger}, {Impey}, {Koekemoer}, {Le F{\`e}vre}, {Lilly},
  {Liu}, {McCracken}, {Mobasher}, {Renzini}, {Rich}, {Schinnerer}, {Shopbell},
  {Taniguchi}, {Thompson}, {Urry}, \& {Williams}}]{2007ApJS..172...86S}
{Sanders}, D.~B., {Salvato}, M., {Aussel}, H., {et~al.} 2007, \apjs, 172, 86

\bibitem[{{Schlafly} \& {Finkbeiner}(2011)}]{2011ApJ...737..103S}
{Schlafly}, E.~F. \& {Finkbeiner}, D.~P. 2011, \apj, 737, 103

\bibitem[{{Schlafly} {et~al.}(2012){Schlafly}, {Finkbeiner}, {Juri{\'c}},
  {Magnier}, {Burgett}, {Chambers}, {Grav}, {Hodapp}, {Kaiser}, {Kudritzki},
  {Martin}, {Morgan}, {Price}, {Rix}, {Stubbs}, {Tonry}, \&
  {Wainscoat}}]{2012ApJ...756..158S}
{Schlafly}, E.~F., {Finkbeiner}, D.~P., {Juri{\'c}}, M., {et~al.} 2012, \apj,
  756, 158

\bibitem[{{Schmidt} {et~al.}(2010){Schmidt}, {Marshall}, {Rix}, {Jester},
  {Hennawi}, \& {Dobler}}]{2010ApJ...714.1194S}
{Schmidt}, K.~B., {Marshall}, P.~J., {Rix}, H.-W., {et~al.} 2010, \apj, 714,
  1194

\bibitem[{{Schmidt} {et~al.}(2012){Schmidt}, {Rix}, {Shields}, {Knecht},
  {Hogg}, {Maoz}, \& {Bovy}}]{2012ApJ...744..147S}
{Schmidt}, K.~B., {Rix}, H.-W., {Shields}, J.~C., {et~al.} 2012, \apj, 744, 147

\bibitem[{{Scott} {et~al.}(2004){Scott}, {Kriss}, {Brotherton}, {Green},
  {Hutchings}, {Shull}, \& {Zheng}}]{2004ApJ...615..135S}
{Scott}, J.~E., {Kriss}, G.~A., {Brotherton}, M., {et~al.} 2004, \apj, 615, 135

\bibitem[{{Stalin} {et~al.}(2004){Stalin}, {Gopal-Krishna}, {Sagar}, \&
  {Wiita}}]{2004MNRAS.350..175S}
{Stalin}, C.~S., {Gopal-Krishna}, {Sagar}, R., \& {Wiita}, P.~J. 2004, \mnras,
  350, 175

\bibitem[{{Stalin} {et~al.}(2005){Stalin}, {Gupta}, {Gopal-Krishna}, {Wiita},
  \& {Sagar}}]{2005MNRAS.356..607S}
{Stalin}, C.~S., {Gupta}, A.~C., {Gopal-Krishna}, {Wiita}, P.~J., \& {Sagar},
  R. 2005, \mnras, 356, 607

\bibitem[{{Stubbs} {et~al.}(2010){Stubbs}, {Doherty}, {Cramer}, {Narayan},
  {Brown}, {Lykke}, {Woodward}, \& {Tonry}}]{2010ApJS..191..376S}
{Stubbs}, C.~W., {Doherty}, P., {Cramer}, C., {et~al.} 2010, \apjs, 191, 376

\bibitem[{{Sun} {et~al.}(2014){Sun}, {Wang}, {Chen}, \&
  {Zheng}}]{2014ApJ...792...54S}
{Sun}, Y.-H., {Wang}, J.-X., {Chen}, X.-Y., \& {Zheng}, Z.-Y. 2014, \apj, 792,
  54

\bibitem[{{Surace} {et~al.}(2005){Surace}, {Shupe}, {Fang}, {Evans}, {Alexov},
  {Frayer}, {Lonsdale}, \& {SWIRE Team}}]{2005AAS...207.6301S}
{Surace}, J.~A., {Shupe}, D.~L., {Fang}, F., {et~al.} 2005, in Bulletin of the
  American Astronomical Society, Vol.~37, American Astronomical Society Meeting
  Abstracts, 1246

\bibitem[{{Taniguchi} {et~al.}(2007){Taniguchi}, {Scoville}, {Murayama},
  {Sanders}, {Mobasher}, {Aussel}, {Capak}, {Ajiki}, {Miyazaki}, {Komiyama},
  {Shioya}, {Nagao}, {Sasaki}, {Koda}, {Carilli}, {Giavalisco}, {Guzzo},
  {Hasinger}, {Impey}, {LeFevre}, {Lilly}, {Renzini}, {Rich}, {Schinnerer},
  {Shopbell}, {Kaifu}, {Karoji}, {Arimoto}, {Okamura}, \&
  {Ohta}}]{2007ApJS..172....9T}
{Taniguchi}, Y., {Scoville}, N., {Murayama}, T., {et~al.} 2007, \apjs, 172, 9

\bibitem[{{Tonry} \& {Onaka}(2009)}]{2009amos.confE..40T}
{Tonry}, J. \& {Onaka}, P. 2009, in Advanced Maui Optical and Space
  Surveillance Technologies Conference

\bibitem[{{Tonry} {et~al.}(2012{\natexlab{a}}){Tonry}, {Stubbs}, {Kilic},
  {Flewelling}, {Deacon}, {Chornock}, {Berger}, {Burgett}, {Chambers},
  {Kaiser}, {Kudritzki}, {Hodapp}, {Magnier}, {Morgan}, {Price}, \&
  {Wainscoat}}]{2012ApJ...745...42T}
{Tonry}, J.~L., {Stubbs}, C.~W., {Kilic}, M., {et~al.} 2012{\natexlab{a}},
  \apj, 745, 42

\bibitem[{{Tonry} {et~al.}(2012{\natexlab{b}}){Tonry}, {Stubbs}, {Lykke},
  {Doherty}, {Shivvers}, {Burgett}, {Chambers}, {Hodapp}, {Kaiser},
  {Kudritzki}, {Magnier}, {Morgan}, {Price}, \&
  {Wainscoat}}]{2012ApJ...750...99T}
{Tonry}, J.~L., {Stubbs}, C.~W., {Lykke}, K.~R., {et~al.} 2012{\natexlab{b}},
  \apj, 750, 99

\bibitem[{{Trump} {et~al.}(2007){Trump}, {Impey}, {McCarthy}, {Elvis},
  {Huchra}, {Brusa}, {Hasinger}, {Schinnerer}, {Capak}, {Lilly}, \&
  {Scoville}}]{2007ApJS..172..383T}
{Trump}, J.~R., {Impey}, C.~D., {McCarthy}, P.~J., {et~al.} 2007, \apjs, 172,
  383

\bibitem[{{Turner} {et~al.}(1999){Turner}, {George}, {Nandra}, \&
  {Turcan}}]{1999ApJ...524..667T}
{Turner}, T.~J., {George}, I.~M., {Nandra}, K., \& {Turcan}, D. 1999, \apj,
  524, 667

\bibitem[{{Ulrich} {et~al.}(1997){Ulrich}, {Maraschi}, \&
  {Urry}}]{1997ARA&A..35..445U}
{Ulrich}, M.-H., {Maraschi}, L., \& {Urry}, C.~M. 1997, \araa, 35, 445

\bibitem[{{Vanzella} {et~al.}(2004){Vanzella}, {Cristiani}, {Fontana},
  {Nonino}, {Arnouts}, {Giallongo}, {Grazian}, {Fasano}, {Popesso}, {Saracco},
  \& {Zaggia}}]{2004A&A...423..761V}
{Vanzella}, E., {Cristiani}, S., {Fontana}, A., {et~al.} 2004, \aap, 423, 761

\bibitem[{{Vaughan} {et~al.}(2003){Vaughan}, {Edelson}, {Warwick}, \&
  {Uttley}}]{2003MNRAS.345.1271V}
{Vaughan}, S., {Edelson}, R., {Warwick}, R.~S., \& {Uttley}, P. 2003, \mnras,
  345, 1271

\bibitem[{{Wolf}(2009)}]{2009MNRAS.397..520W}
{Wolf}, C. 2009, \mnras, 397, 520

\bibitem[{{Wolf} {et~al.}(2004){Wolf}, {Meisenheimer}, {Kleinheinrich},
  {Borch}, {Dye}, {Gray}, {Wisotzki}, {Bell}, {Rix}, {Cimatti}, {Hasinger}, \&
  {Szokoly}}]{2004A&A...421..913W}
{Wolf}, C., {Meisenheimer}, K., {Kleinheinrich}, M., {et~al.} 2004, \aap, 421,
  913

\bibitem[{{Young} {et~al.}(2012){Young}, {Brandt}, {Xue}, {Paolillo},
  {Alexander}, {Bauer}, {Lehmer}, {Luo}, {Shemmer}, {Schneider}, \&
  {Vignali}}]{2012ApJ...748..124Y}
{Young}, M., {Brandt}, W.~N., {Xue}, Y.~Q., {et~al.} 2012, \apj, 748, 124

\bibitem[{{Zamojski} {et~al.}(2007){Zamojski}, {Schiminovich}, {Rich},
  {Mobasher}, {Koekemoer}, {Capak}, {Taniguchi}, {Sasaki}, {McCracken},
  {Mellier}, {Bertin}, {Aussel}, {Sanders}, {Le F{\`e}vre}, {Ilbert},
  {Salvato}, {Thompson}, {Kartaltepe}, {Scoville}, {Barlow}, {Forster},
  {Friedman}, {Martin}, {Morrissey}, {Neff}, {Seibert}, {Small}, {Wyder},
  {Bianchi}, {Donas}, {Heckman}, {Lee}, {Madore}, {Milliard}, {Szalay},
  {Welsh}, \& {Yi}}]{2007ApJS..172..468Z}
{Zamojski}, M.~A., {Schiminovich}, D., {Rich}, R.~M., {et~al.} 2007, \apjs,
  172, 468

\bibitem[{{Zhou} {et~al.}(2010){Zhou}, {Zhang}, {Wang}, \&
  {Zhu}}]{2010ApJ...710...16Z}
{Zhou}, X.-L., {Zhang}, S.-N., {Wang}, D.-X., \& {Zhu}, L. 2010, \apj, 710, 16

\bibitem[{{Zu} {et~al.}(2013){Zu}, {Kochanek}, {Koz{\l}owski}, \&
  {Udalski}}]{2013ApJ...765..106Z}
{Zu}, Y., {Kochanek}, C.~S., {Koz{\l}owski}, S., \& {Udalski}, A. 2013, \apj,
  765, 106

\bibitem[{{Zuo} {et~al.}(2012){Zuo}, {Wu}, {Liu}, \&
  {Jiao}}]{2012ApJ...758..104Z}
{Zuo}, W., {Wu}, X.-B., {Liu}, Y.-Q., \& {Jiao}, C.-L. 2012, \apj, 758, 104

\end{thebibliography}

\newpage
\begin{appendix}
\section{On the uncertainty of the excess variance}
\label{sec:appendixa}

In addition to the Poisson noise error (equation \ref{eq:errnev}) used in this work, more uncertainties exist that are related to an excess variance measurement, and they are connected to the stochastic nature of the variations and the sampling pattern of the light curve. These error sources were studied in detail by \citet{2013ApJ...771....9A}. They applied Monte Carlo methods to create 5000 different light curves drawn from a power spectral density (PSD) with logarithmic slopes between $-1$ and $-3$ and measured the excess variance of these light curves adopting different sampling patterns (continuous, uniform, sparse). These investigations proved that the excess variance is a biased estimator of the intrinsic ("true") variance, which itself arises from the underlying physical process related to variability. The bias factor associated with an individual $\sigma_{\mathrm{rms}}^{2}$ measurement is shown to depend on the PSD logarithmic slope, the sampling pattern, and the signal-to-noise of the light curve, at least as long as $S/N < 3$. Since the excess variance is defined to measure the integral of the PSD over the temporal frequencies probed by a light curve, the actual value of $\sigma_{\mathrm{rms}}^{2}$ is affected by the functional form of the PSD. The optical power spectra of AGNs are usually characterized by a "red noise" PSD, i.e. a power law $\mathrm{PSD\left(\nu\right)}\propto\nu^{\gamma}$ with $\gamma<-1$. It is very likely that optical PSDs exhibit a break frequency, separating the low frequency part with $\gamma\sim -1$ from the high frequency part with $\gamma\sim -2$, which is similar to what was observed in many X-ray variability studies \citep{1993ApJ...414L..85L,1999ApJ...514..682E,2003ApJ...593...96M,2004MNRAS.348..783M}. The actual value of the optical break timescale may strongly depend on the physical parameters of each source, such as the black hole mass and luminosity, and typical values between 10--100 days, but even up to $\sim$10 years have been reported \citep{2001ApJ...555..775C,2009ApJ...698..895K}. Considering the timescales encompassed by the PS1 3$\pi$ and MDF light curves (shortest timescale $\sim$1 day, longest timescale $\sim$4 years), it is therefore unclear whether our $\sigma_{\mathrm{rms}}^{2}$ measurements predominantly integrate the PSD in the low or high frequency parts. Nevertheless, for both surveys, the light curve sampling pattern is closer to the sparse case than to the continuous or uniform ones. For these reasons the value of the bias factor $b_{\mathrm{sparse}}$ for the $\sigma_{\mathrm{rms}}^{2}$ measurements of this work is expected to lie somewhere between $b_{\mathrm{sparse}}=1.2$ (for $\gamma\sim -1$), $b_{\mathrm{sparse}}=1.0$ (for $\gamma\sim -1.5$), or $b_{\mathrm{sparse}}=0.6$ (for $\gamma\sim -2$) according to Table 2 in \citet{2013ApJ...771....9A} and is therefore negligible.

\begin{figure*}
\centering
\subfloat{%
        \includegraphics[width=.48\textwidth]{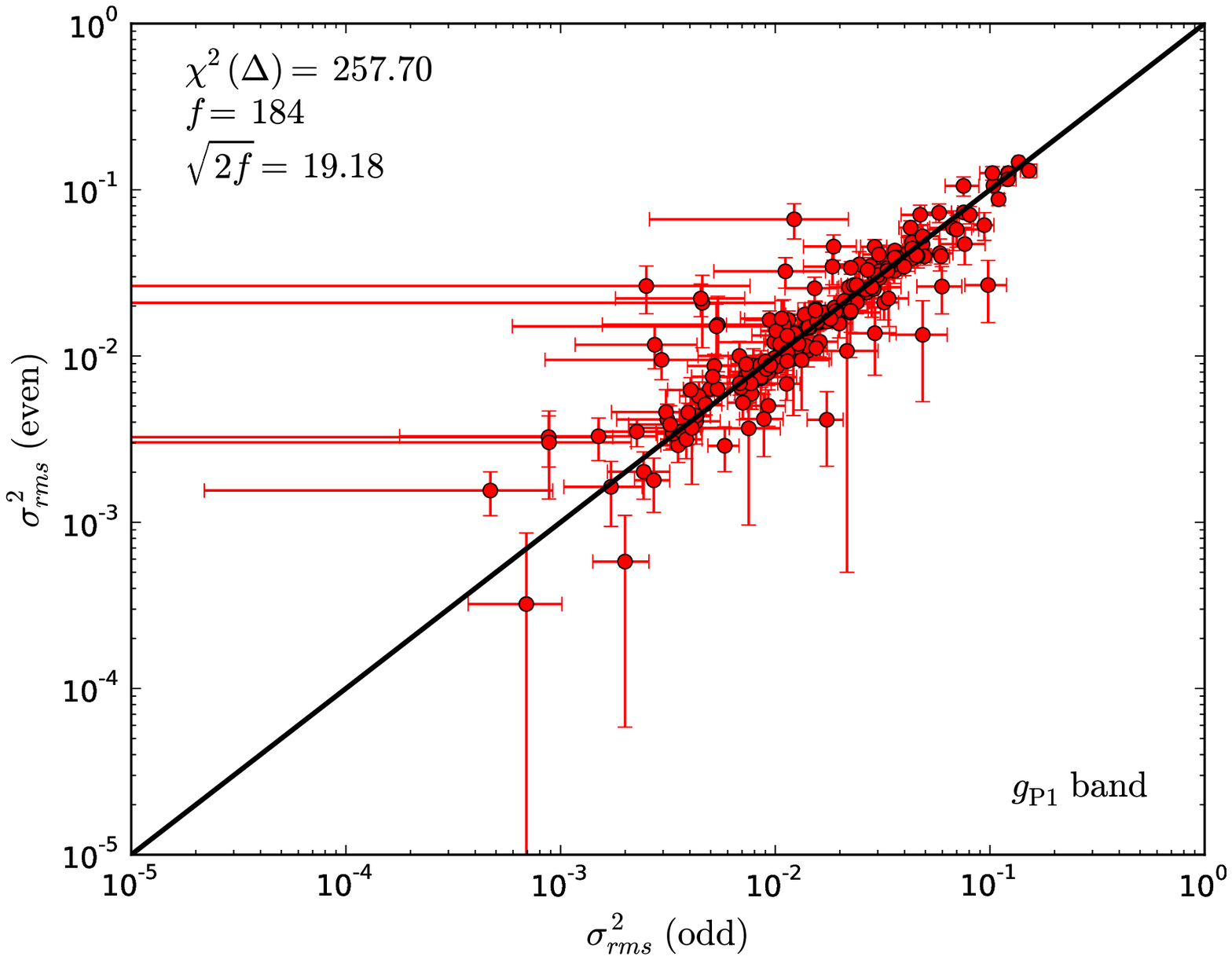}}
\quad
\subfloat{%
        \includegraphics[width=.48\textwidth]{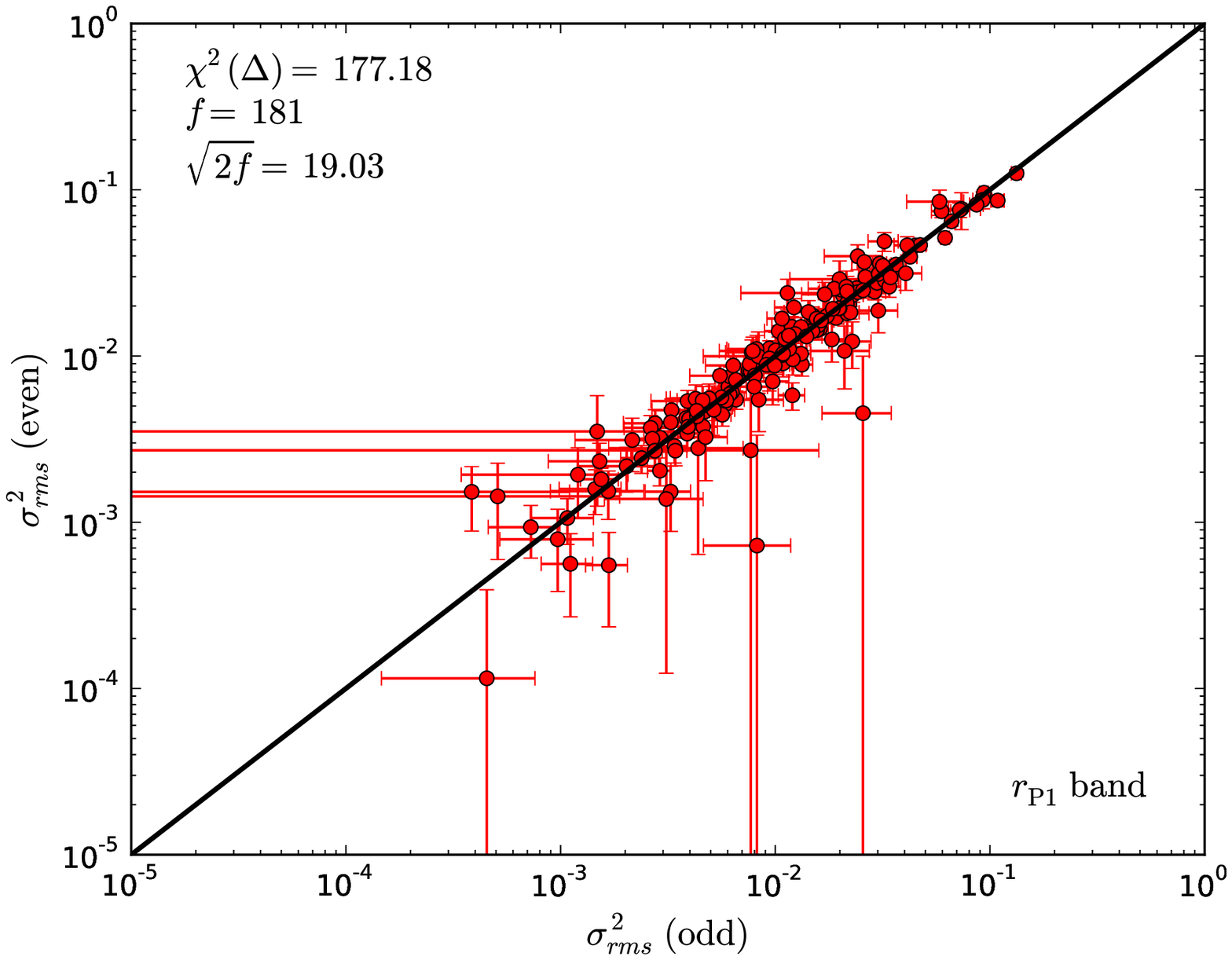}}

\subfloat{%
        \includegraphics[width=.48\textwidth]{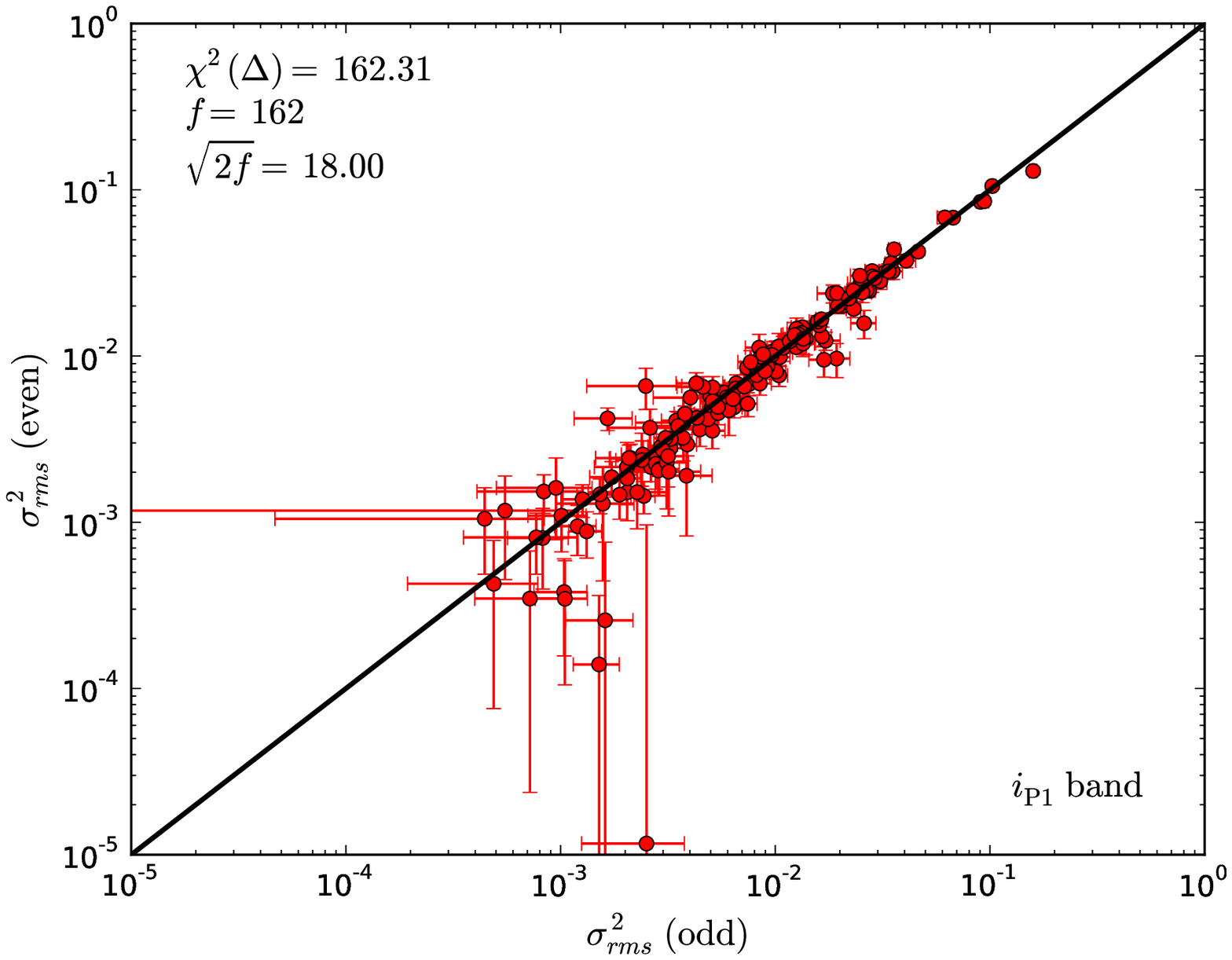}}
\quad
\subfloat{%
        \includegraphics[width=.48\textwidth]{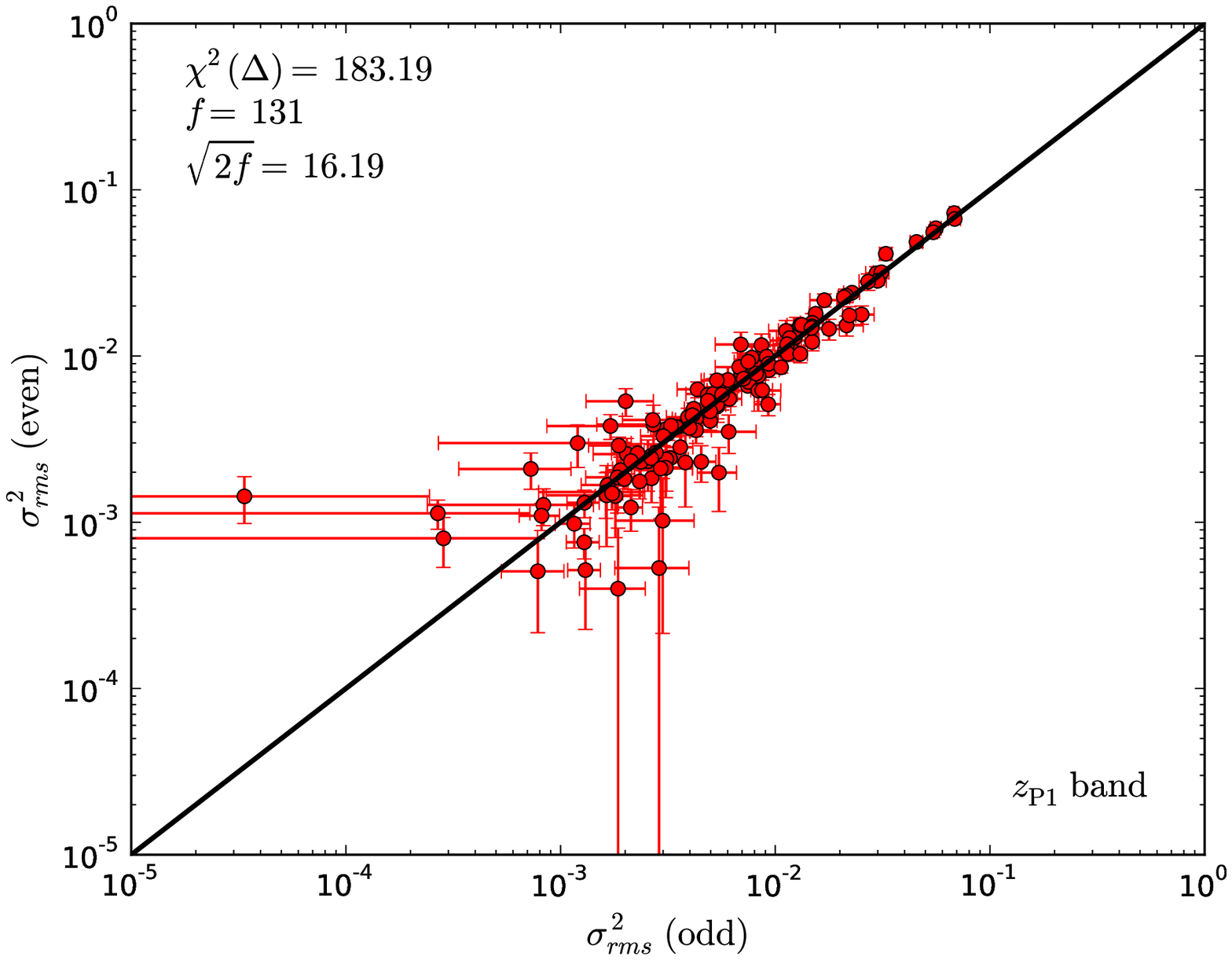}}

\subfloat{%
        \includegraphics[width=.48\textwidth]{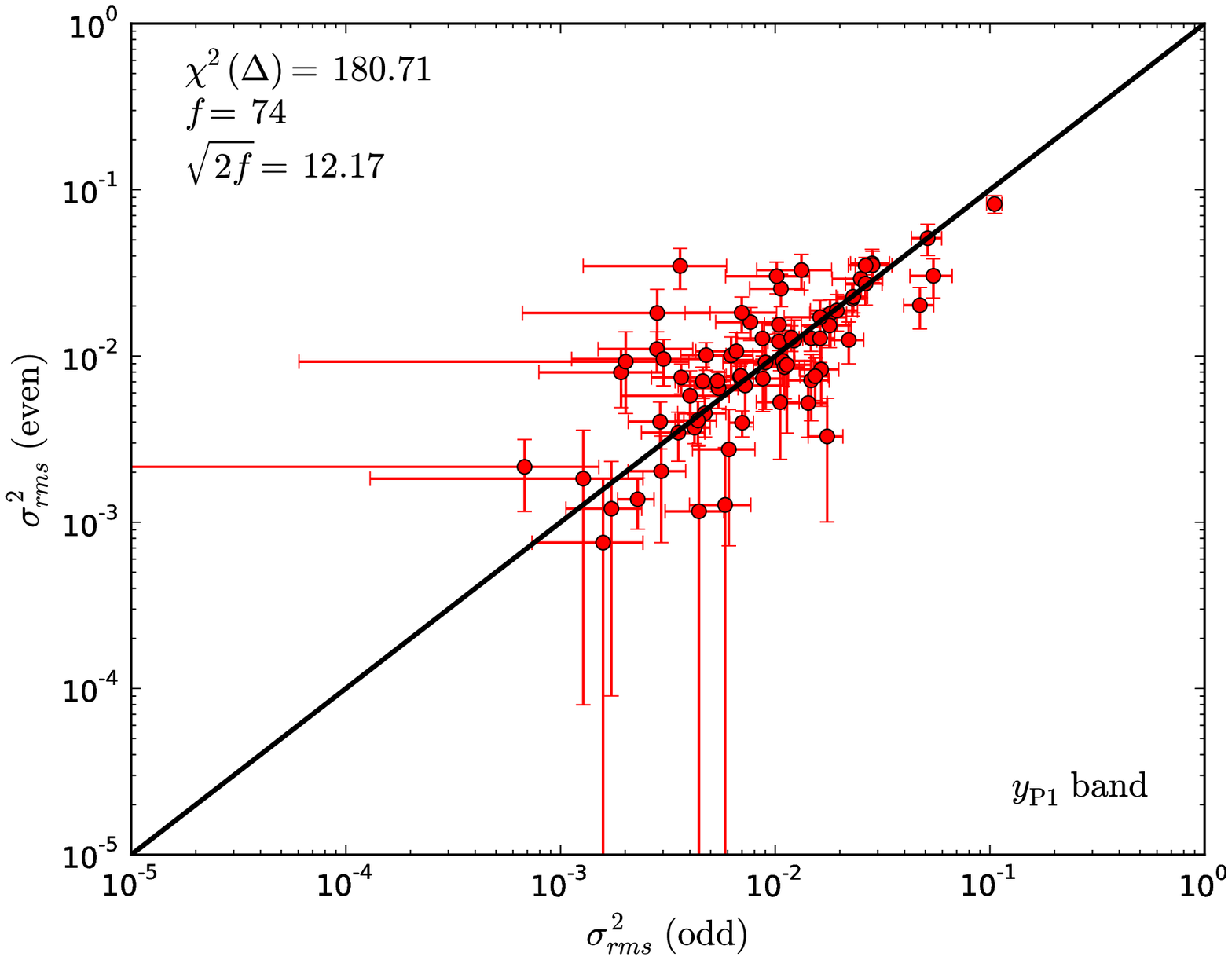}}
\caption{Excess variances of all PS1 bands calculated from only the even (y-axis) and only the odd (x-axis) light curve points for all variable AGNs of the MDF04 sample. Denoted is the $\chi^{2}$ of $\Delta=\sigma_{\mathrm{rms}}^{2}(\mathrm{even})-\sigma_{\mathrm{rms}}^{2}(\mathrm{odd}),$ together with its expectation value and standard deviation. The black line corresponds to the one-to-one relation.}
        \label{fig:evenodd}
\end{figure*}

To assess the quality of the excess variance measurements for our MDF04 sample and to check whether the assumed error is reasonable according to equation \ref{eq:errnev}, we perform a simple test by comparing the $\sigma_{\mathrm{rms}}^{2}$ values obtained from two different realizations of each light curve by calculating the excess variance once from only the even light curve points and once from only the odd light curve points. The uncertainty corresponding to these two measurements is then contrasted to the individual errors $err\left(\sigma_{\mathrm{rms}}^{2}\right)$ assigned to each variability measurement. We do this by calculating
\begin{flalign}
        \label{eq:deltanev}
        \chi^{2}\left(\Delta\right)=\sum_{i=1}^{f}\frac{\left(\Delta_{i}-\bar{\Delta}\right)^{2}}{err\left(\Delta_{i}\right)^{2}}
\end{flalign}   
with the difference $\Delta=\sigma_{\mathrm{rms}}^{2}[\mathrm{even}]-\sigma_{\mathrm{rms}}^{2}[\mathrm{odd}]$ and squared error $err\left(\Delta\right)^{2}=err(\sigma_{\mathrm{rms}}^{2})[\mathrm{even}]^{2}+err(\sigma_{\mathrm{rms}}^{2})[\mathrm{odd}]^{2}$. The computed $\chi^{2}\left(\Delta\right)$, together with its expectation value $E\left(\chi^{2}\left(\Delta\right)\right)=f$ and standard deviation $\sigma\left(\chi^{2}\left(\Delta\right)\right)=\sqrt{2f}$, is quoted for all variable AGNs of the MDF04 sample in Fig. \ref{fig:evenodd}. We note that the quality of the $\sigma_{\mathrm{rms}}^{2}$ measurements is generally high for the $g_{\mathrm{P1}}$, $r_{\mathrm{P1}}$, $i_{\mathrm{P1}}$, and $z_{\mathrm{P1}}$ bands, because most values lie very close to the one to one relation. Since the $\chi^{2}\left(\Delta\right)$ values are very close to the respective expectation value for the $r_{\mathrm{P1}}$, $i_{\mathrm{P1}}$ bands and only deviate by a factor of $\sim$1.4 for the $g_{\mathrm{P1}}$ and $z_{\mathrm{P1}}$ bands, the Poisson noise error estimate of equation \ref{eq:errnev} represents an appropriate measurement uncertainty of the excess variance in our light curves. Only the $\sigma_{\mathrm{rms}}^{2}$ values of the $y_{\mathrm{P1}}$ band show significantly less accuracy than for the other PS1 bands, which is due to the fact that the $y_{\mathrm{P1}}$ band light curves contain fewer data points. 

\section{Method used to select the Case A photometry}
\label{sec:appendixb}

To identify the epochs minimizing the temporal distance of the five PS1 band light curves, we employ a combinatoric procedure. For each of our objects we have to consider five different light curves out of  set $B=\{g_{\mathrm{P1}}, r_{\mathrm{P1}}, i_{\mathrm{P1}}, z_{\mathrm{P1}}, y_{\mathrm{P1}}\}$, consisting of $N_{k}$ magnitude values measured at times $t_{ki}$, with $k\in B$ and $i=1, 2,..., N_{k}$. We start by taking each light curve point $i$ of the band $k=g_{\mathrm{P1}}$ and find the four light curve points $j_{\mathrm{min}}$ with minimal temporal distance to point $i$, out of the $j=1, 2,..., N_{l}$ points of all other bands $l\neq k$. Denoting the temporal distance by $\Delta t_{ki,lj}=|t_{ki}-t_{lj}|$ gives us four values $\Delta t_{ki,lj_{\mathrm{min}}}=\mathrm{min}\{\Delta t_{ki,l1},...,\Delta t_{ki,lN_{l}}\}$ for each band $l=r_{\mathrm{P1}}, i_{\mathrm{P1}}, z_{\mathrm{P1}}, y_{\mathrm{P1}}$. Then we compute the sum 
\begin{flalign}
        \label{eq:sumTg}
        \Delta T_{g_{\mathrm{P1}}i}=\Delta t_{g_{\mathrm{P1}}i,r_{\mathrm{P1}}j_{\mathrm{min}}}+\Delta t_{g_{\mathrm{P1}}i,i_{\mathrm{P1}}j_{\mathrm{min}}}+\Delta t_{g_{\mathrm{P1}}i,z_{\mathrm{P1}}j_{\mathrm{min}}}+\Delta t_{g_{\mathrm{P1}}i,y_{\mathrm{P1}}j_{\mathrm{min}}}
\end{flalign}     
and find its minimum value $\Delta T_{g_{\mathrm{P1}},\mathrm{min}}=\mathrm{min}\{\Delta T_{g_{\mathrm{P1}}1},..., \Delta T_{g_{\mathrm{P1}}N_{g_{\mathrm{P1}}}}\}$ out of the $N_{g_{\mathrm{P1}}}$ points. This gives us the minimum total time interval of the different filter observations with respect to the $g_{\mathrm{P1}}$ band light curve points. However, other combinations might exist, leading to a shorter total time interval, by taking the light curve points of another band as reference values while calculating the differences $|t_{ki}-t_{lj}|$ to the light curve points of the remaining bands. Therefore we perform the same procedure, with the reference band $k$ running through all elements of the set $B$, by calculating        
\begin{flalign}
        \label{eq:sumT}
        \Delta T_{ki}=\sum_{\substack{l\in B \\ l\neq k}}\Delta t_{ki,lj_{\mathrm{min}}}
\end{flalign}    
for each light curve point $i$ of reference band $k$. For each reference band $k$, we then determine the minimal total time interval $\Delta T_{k,\mathrm{min}}=\mathrm{min}\{\Delta T_{k1},..., \Delta T_{kN_{k}}\}$. The set of input photometry with minimum relative temporal distance is finally obtained by selecting those five light curve points that give rise to    
\begin{flalign}
        \label{eq:Tmin}
        \Delta T_{\mathrm{min}}=\mathrm{min}\{\Delta T_{g_{\mathrm{P1}},\mathrm{min}}, \Delta T_{r_{\mathrm{P1}},\mathrm{min}}, \Delta T_{i_{\mathrm{P1}},\mathrm{min}}, \Delta T_{z_{\mathrm{P1}},\mathrm{min}}, \Delta T_{y_{\mathrm{P1}},\mathrm{min}}\}.
\end{flalign} 
These five magnitude values are stored in the input catalogue for our fitting routines, together with their respective individual uncertainties $err\left(mag\right)$.

\begin{figure*}
\centering
\text{\hspace*{8mm}Case A (3$\pi$): $\Delta T_{\mathrm{min}}$\hspace*{63mm}Case C (3$\pi$): $\Delta T_{\mathrm{random}}$}
\subfloat{%
        \includegraphics[width=.48\textwidth]{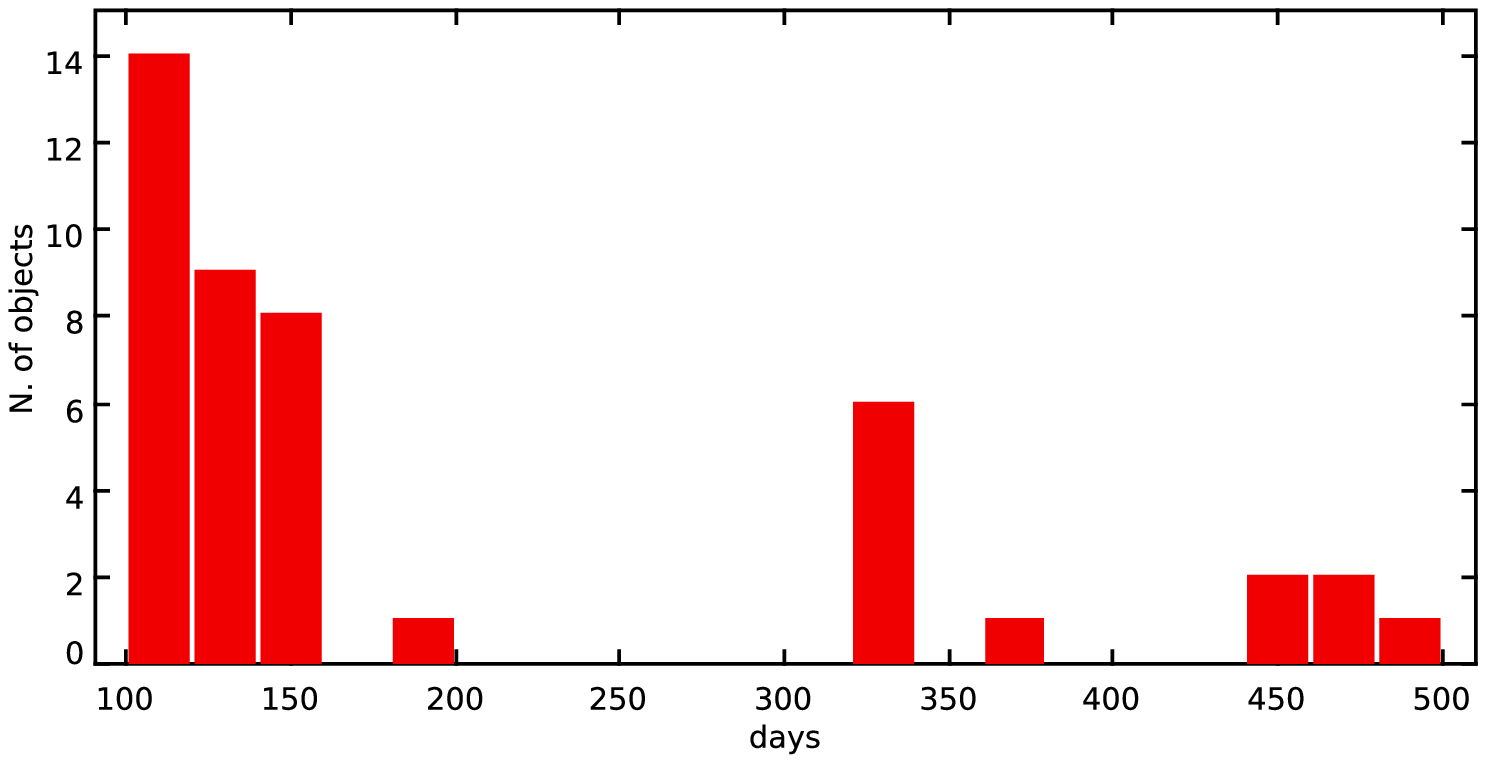}}
\quad
\subfloat{%
        \includegraphics[width=.48\textwidth]{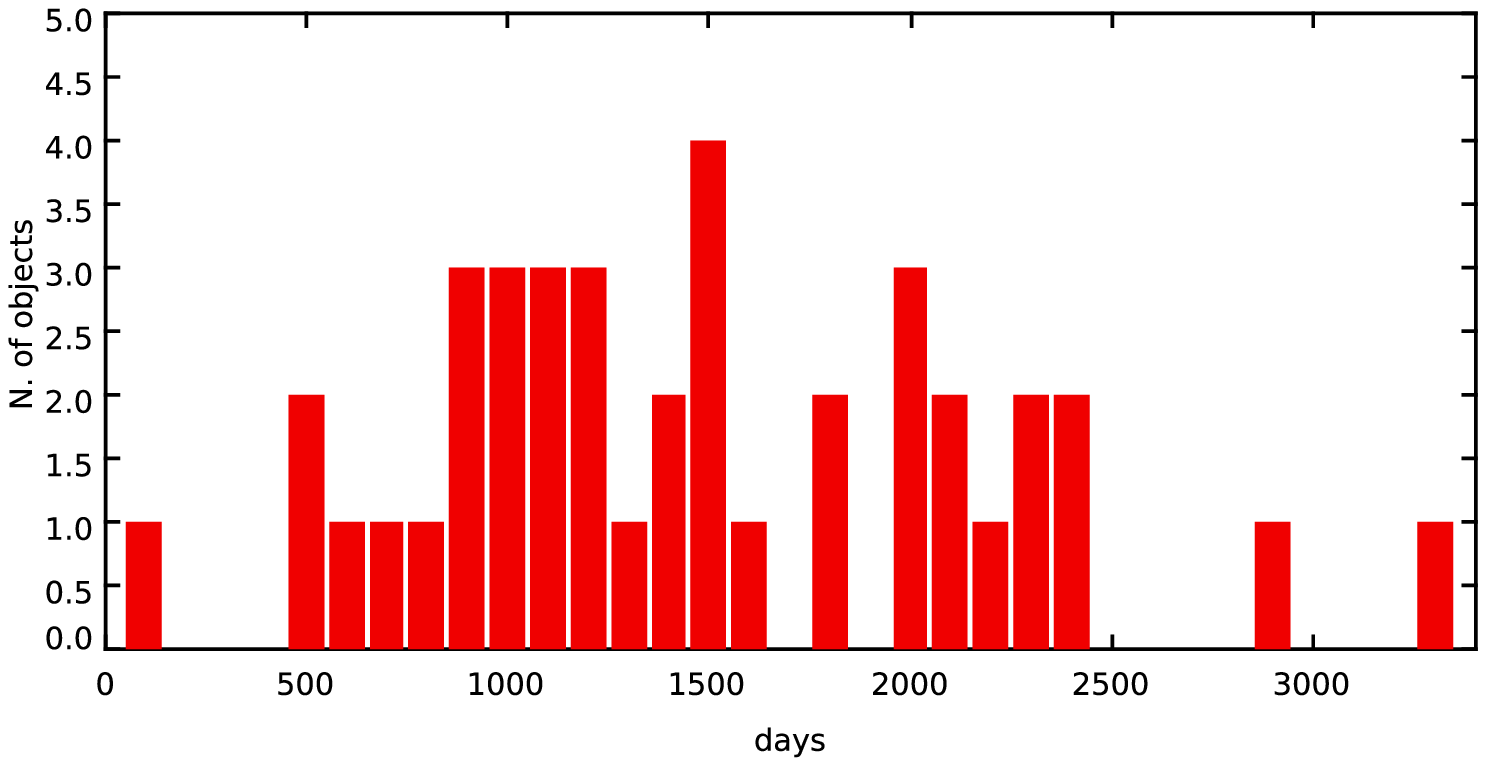}}

\text{\hspace*{8mm}Case A (MDF04): $\Delta T_{\mathrm{min}}$\hspace*{58mm}Case C (MDF04): $\Delta T_{\mathrm{random}}$}
\subfloat{%
        \includegraphics[width=.48\textwidth]{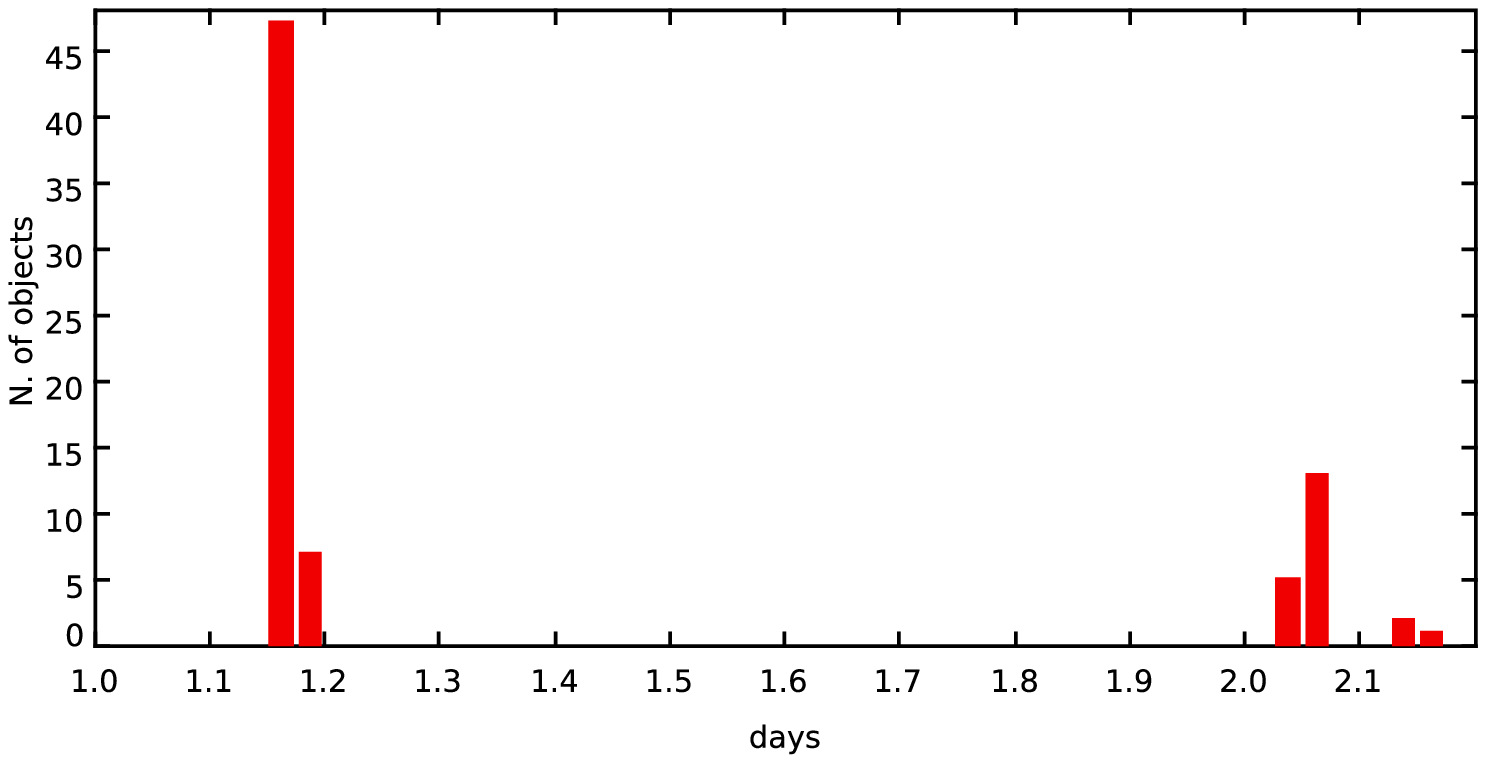}}
\quad
\subfloat{%
        \includegraphics[width=.48\textwidth]{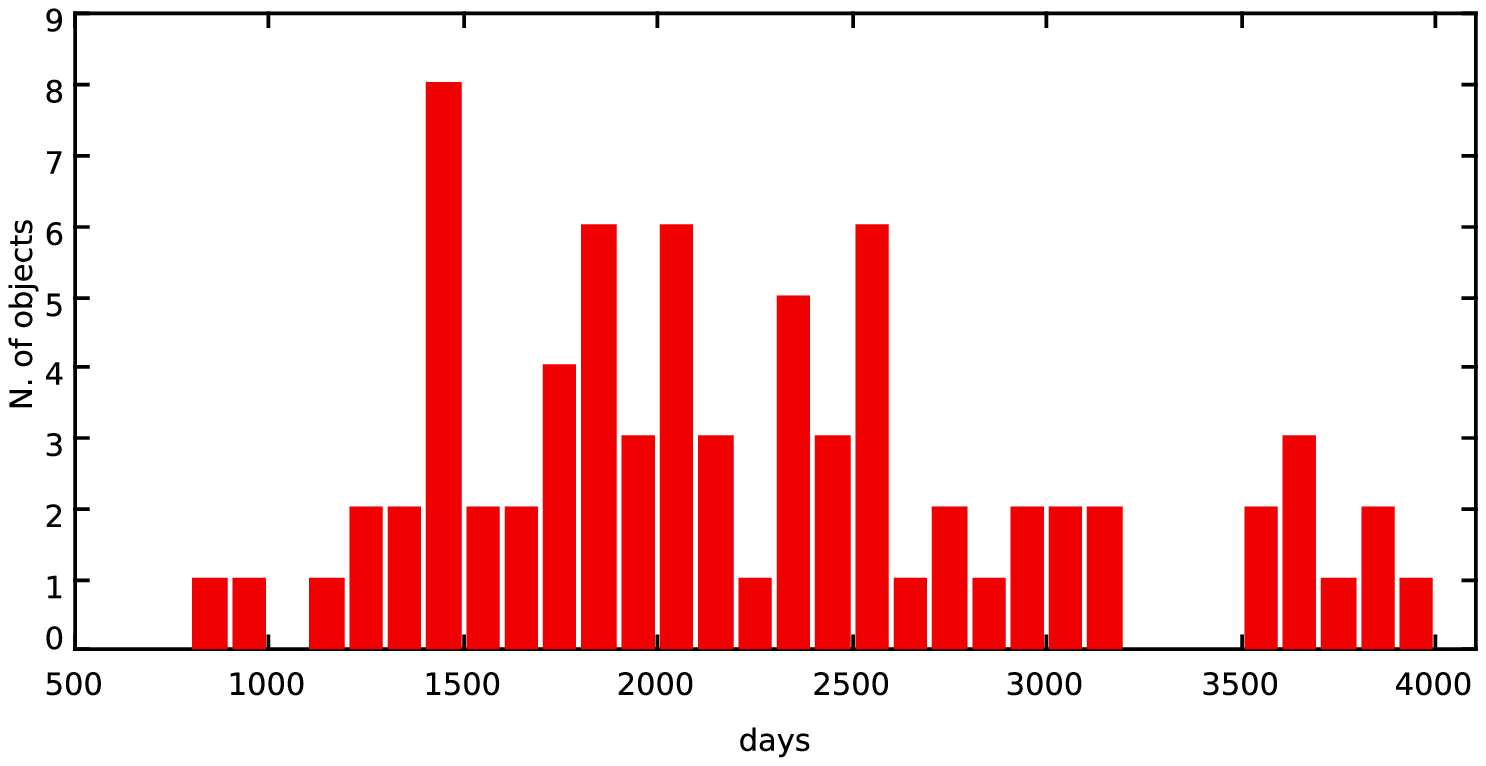}}
\caption{Distribution of $\Delta T_{\mathrm{min}}$ for the 40 AGNs of the 3$\pi$ sample and the 75 AGNs of the MDF04 sample. The distribution of $\Delta T_{\mathrm{random}}$ for one of the ten random realizations of Case C is shown for both samples in the right column.}
        \label{fig:tmintrandhist}
\end{figure*}

Owing to the different sampling patterns of the 3$\pi$ and MDF04 light curves, the minimized values $\Delta T_{\mathrm{min}}$ differ a lot for our two samples. As shown in Fig. \ref{fig:tmintrandhist} each of the 75 AGNs from the MDF04 sample has a value of $\Delta T_{\mathrm{min}}<2.5$ days. In contrast, the corresponding values for the 40 AGNs of the 3$\pi$ sample range between $\Delta T_{\mathrm{min}}=$110  and 500 days, giving a very poor approximation of a snapshot SED. For comparison, Fig. \ref{fig:tmintrandhist} also displays the histograms of the total time interval $\Delta T_{\mathrm{random}}$ for one of the ten realizations of Case C. The interval $\Delta T_{\mathrm{random}}$ is calculated after equation \ref{eq:sumT}, taking the randomly chosen $g_{\mathrm{P1}}$ band point $i$ as reference value $t_{ki}$ in the individual addends $|t_{ki}-t_{lj}|$, with $t_{lj}$ given by the other four randomly chosen light curve points $j$ of the remaining bands $l\neq g_{\mathrm{P1}}$. As intended, the distributions of $\Delta T_{\mathrm{random}}$ encompass much higher values, typically between 500 and 4000 days, than the respective $\Delta T_{\mathrm{min}}$ distributions for both the 3$\pi$ and MDF04 samples. 

\section{Catalogues and light curves of variable AGNs}
\label{sec:appendixc}

The catalogues of variable AGNs described in section \ref{sec:varsample} are provided in the online journal for every PS1 band and for both the 3$\pi$ and MDF04 surveys. Part of one of these tables is shown in Table \ref{tab:varcatalog}. The tables (ASCII format) list basic information like the identifier number, coordinates, number of light curve points, and light curve median, as well as the variability parameters defined in section \ref{sec:varmethod}. In addition, the nightly-averaged light curves for these sources, cleaned from outlier measurements as outlined in section \ref{sec:sampledef}, are available on request for every PS1 band.    
\begin{table*}
\caption{MDF04 survey catalogue of variable AGNs ($g_{\mathrm{P1}}$ band).}
\centering
\begin{tabular}{cccccccccc}
\hline\hline
XID & R.A. & Dec. & N & V & $\sigma_{\mathrm{rms}}^{2}$ & $err\left(\sigma_{\mathrm{rms}}^{2}\right)$ & median$\left(mag\right)$ & median$\left(err\left(mag\right)\right)$ & $\Delta mag$  \\
& (deg) & (deg) &  & & & & (AB) & (AB) & (AB) \\
\hline
  74 & 150.449615 & 2.246419 & 75 & 240.24 & 0.0097 & 0.0011 & 20.67 & 0.03 & 0.42 \\
  84 & 150.299744 & 2.506903 & 72 & 66.41 & 0.0070 & 0.0011 & 21.32 & 0.04 & 0.59 \\
  87 & 150.101624 & 1.848332 & 73 & 136.17 & 0.0114 & 0.0012 & 20.72 & 0.04 & 0.70 \\
  89 & 150.276276 & 2.526340 & 44 & 24.50 & 0.0299 & 0.0080 & 22.44 & 0.10 & 1.09 \\
  95 & 150.028542 & 2.209917 & 64 & 95.84 & 0.0268 & 0.0040 & 21.81 & 0.09 & 0.82 \\
\hline
\end{tabular}
\tablefoot{Column 1: XMM-COSMOS identifier number (from \citet{2009A&A...497..635C}); Columns 2--3: coordinates of the optical/IR counterpart (J2000); Column 4: number of light curve points; Column 5: V index (see equation \ref{eq:V}); Column 6: excess variance (see equation \ref{eq:nev}); Column 7: error of excess variance (see equation \ref{eq:errnev}); Column 8: median magnitude of light curve; Column 9: median error of light curve points; Column 10: $\Delta mag=\mathrm{max}\left(mag\right)-\mathrm{min}\left(mag\right)$. The table (ASCII format) is available in its entirety in the online journal.}
\label{tab:varcatalog}
\end{table*}          

\end{appendix}


\end{document}